\documentclass[%
  aps,prb,reprint,amsmath,amssymb,superscriptaddress,longbibliography,floatfix
]{revtex4-2}

\usepackage{graphicx}
\usepackage{dcolumn}
\usepackage{bm}
\usepackage{physics}
\usepackage{hyperref}
\hypersetup{hidelinks}
\usepackage{xcolor}
\usepackage{booktabs}
\usepackage{enumitem}

\newcommand{\PT}{\mathcal{PT}}
\newcommand{\Hss}{H_{\rm ss}}
\newcommand{\Htilde}{\tilde{H}}
\newcommand{\teps}{\tilde{\epsilon}_{\xi}}
\newcommand{\keff}{\kappa_{\rm imp}}
\newcommand{\Eep}{E_{\rm EP}}
\newcommand{\Gless}{G^{<}}
\newcommand{\Gret}{G^{R}}
\newcommand{\Gadv}{G^{A}}
\newcommand{\tb}{\tilde{\beta}}          
\newcommand{\gb}{\beta_0}               
\newcommand{\Gpt}{\Gamma_{\mathrm{PT}}}   
\newcommand{\dSig}{\delta_{\Sigma}}          
\newcommand{\Deff}{\Delta_{\rm eff}}     
\newcommand{\Dzero}{\Delta_0}              
\newcommand{\Dcoh}{\Delta_{\rm coh}}          
\newcommand{\Veff}{V_{\rm eff}}           

\begin{document}

\title{Emergent $\PT$ Symmetry and Exceptional Points
       in a Driven Dirac Impurity}

\author{Vinayak M.~Kulkarni}
\email{vmkphysimath@gmail.com}
\affiliation{Theoretical Sciences Unit,
  Jawaharlal Nehru Centre for Advanced Scientific Research,
  Bangalore 560064, India}

\date{August 11, 2026}

\begin{abstract}
Periodic driving can produce passive non-Hermitian impurity dynamics without
microscopic gain.  We derive this mechanism for an inversion-asymmetric
spin--orbit-coupled (SOC) Dirac impurity.  Off-shell angular harmonics generate a
real spin-odd shift, which the retarded matrix bath converts into relative
spin-dependent decay.  After common damping is removed, the resulting kernel
contains the parity--time ($\PT$) core
$\Delta_{\rm eff}\sigma_x+i\Gamma_{\rm PT}\sigma_z$ together with its causal
Kramers--Kronig detuning.  The normalized constant-core benchmark becomes an
avoided coalescence when its full finite-band frequency dependence is
restored.  By contrast, an exactly stationary rotating-drive construction,
evaluated with the full radial momentum integral and both helicity cuts,
supports a continuous family of nonlinear causal pole exceptional points
(EPs), certified by a double-zero condition, local winding, and pole
exchange.  Zero-temperature complete-basis calculations on four
$z$-shifted $N=10$ and $12$ matrix block-Wilson chains continue a
representative EP to $U=0.0025$.  On one common parameter contour both
chains have the same winding and pole exchange, and a Rouch\'e ratio
$0.678<1$ preserves the enclosed zero count between them.  This is a
controlled finite-chain result, not a numerical renormalization group (NRG)
result in the thermodynamic limit or a strong-coupling theorem.  We also show that divergent
biorthogonal projectors cancel in complete propagators and in the
particle--hole-symmetric finite-$U$ charge resolvent, so they do not generate
a universal screening enhancement.  Finally, on an equal-velocity
branchwise-linearized submanifold, the exact finite-$U$ Anderson contact
matrix is rational in a dressed rapidity and $GL(2,\mathbb C)$ covariant.
Its Yang--Baxter structure survives the non-unitary similarity and becomes a
unipotent boundary twist at the EP.  This integrability statement does not
extend to the full curved, frequency-dependent driven kernel.
\end{abstract}

\maketitle

\section{Introduction}
\label{sec:intro}

Driven quantum systems provide a versatile route to steady states
and dynamical phases that have no equilibrium analogs.
When coherent drive, dissipation, and strong correlations coexist,
the resulting many-body dynamics can generate fixed points that
reflect both microscopic symmetry and emergent non-Hermitian
structure.
Recent work has shown that minimizing a driven bosonic field within
a Keldysh mean-field framework can reproduce renormalization-group
(RG)-like fixed points~\cite{kamenev2023field,popruzhenko2014keldysh,maghrebi2016nonequilibrium},
pointing to a deeper connection between nonequilibrium stability
and non-Hermitian criticality~\cite{kulkscipo,Claeys2022RGdissipative}.

Exceptional points (EPs) are non-Hermitian degeneracies where
eigenvalues and eigenvectors coalesce, playing a central role in
such settings.
They reshape spectral topology and enable non-reciprocal transport,
sensing enhancement, and dynamical control.
EPs have been linked to the non-Hermitian skin
effect~\cite{YaoWang2018}, higher-order
degeneracies~\cite{higherorder,PhysRevLett.126.110404},
and ultrafast projective protocols in both quantum and classical
experiments~\cite{wan2023evidence,mentink2015ultrafast}.
Their influence extends beyond spectra: non-Hermitian topology
underpins phenomena in nonlinear
photonics~\cite{xia2021nonlinear,smirnova2020nonlinear},
topological acoustics~\cite{xue2022topological},
engineered metamaterials~\cite{RevModPhys.96.015002};
coherent light--matter
coupling and engineered dissipation can stabilize correlated states
such as topological Kondo phases~\cite{hu2025ceco2p2}, and
nonlinearities can generate instabilities captured by dynamical
mean-field theory~\cite{aoki2014nonequilibrium,PhysRevX.7.031013}.

Impurity systems are particularly sensitive to anisotropy and
dissipation.
Boundary-induced openness can mimic potential
scattering~\cite{kattel2024kondo,kattel2024spin,kattel2024dissipation},
while nonlinear perturbations modify inelastic processes and impurity
coherence.
Non-Hermitian phase transitions have recently been observed in
condensed-matter
experiments~\cite{li2024discoverynonhermitianphasetransition},
and nonlinear optical platforms provide precise control of
non-Hermitian
topology~\cite{dai2024non,PhysRevLett.132.156901,khanikaev2024topological}.
In correlated impurity models, non-Hermitian features manifest
through sign reversals in RG flows and causality violations in both
interacting and non-interacting
settings~\cite{PhysRevB.111.125157,han2023complex,
PhysRevB.111.045124,kulkscipo}.
Non-Hermitian Kondo physics has been explored via numerical
renormalization group, cold-atom platforms, and Floquet-engineered
conduction bands~\cite{burke2025non,PhysRevB.108.155120,SBkeld}.
Recent work on a non-Hermitian Anderson model with complex hybridization has
also established a genuine ground-state Kondo-breakdown transition by
self-consistent slave-boson and interacting Bethe-Ansatz
methods~\cite{yamamoto2026complex}.  The present problem is different: its
non-Hermitian structure is a passive retarded kernel derived from a Hermitian
driven system, and its RG result is deliberately restricted to a weak-flow
pre-EP diagnostic.
An exact limiting precedent for an emergent non-Hermitian impurity generator
was given by Oguri and Sakano~\cite{oguri2013highbias}: at infinite bias (or
in the corresponding high-temperature limit) the Anderson impurity maps to a
finite Liouville--Fock Hamiltonian, while physical Green functions require the
complete doubled-space initial/final-state construction.  That high-energy
limit is distinct from the present adiabatic passive kernel, but it reinforces
the need to retain complete retarded, advanced, and lesser objects rather than
assigning observability to an individual left or right eigenvector residue.
Integrability of the Kondo model with time-dependent couplings
has been established via quantum Knizhnik--Zamolodchikov
equations~\cite{pasnoori2025tdkondo}, and a $\PT$-protected
Kondo effect has been observed in
CeCo$_2$P$_2$~\cite{hu2025ceco2p2}.
Benchmark studies of the two-dimensional Hubbard model have
established the regime of validity of large-$N$ slave-boson and
mean-field treatments of correlated
electrons~\cite{simons2015hubbard}, directly motivating the
approach employed in Sec.~\ref{sec:mf}.

The exact algebraic question is separated from the microscopic derivation.
The present article derives the driven retarded kernel and identifies the
controlled low-energy submanifold on which its contact problem reduces to a
finite-$U$ Anderson model.  A companion mathematical work
~\cite{KulkarniYB2026} develops that reduced model independently: it proves
the finite-volume gauge equivalence, Yang--Baxter equation (YBE), $RLL$, and
$RTT$ construction, unipotent
boundary-twist limit, the resolution of the nonzero transverse
pseudospin-ladder sector, the resulting many-body Jordan structure, and the
Laguerre-scaled contraction of descendant spin roots at
\(\lambda=\infty\).  This
division of labor is important.  The exact result below is not a claim that
curvature, unequal branch velocities, or the full frequency-dependent
self-energy of the driven model are Yang--Baxter integrable.

Here we show that a driven impurity embedded in a Dirac-like bath
naturally develops a passive non-Hermitian kernel once higher angular
harmonics and the SOC matrix bath are treated together.
The off-shell auxiliary harmonics generate a real spin-odd coherent
shift.  The retarded Dirac propagator, through its energy and
spin-orbit matrix structure, converts this shift into a spin-dependent decay
rate.  After subtracting the common bath damping, the remaining
co-decaying kernel contains a dynamically induced passive
$\PT$-symmetric core, together with a Kramers--Kronig-related coherent
detuning that bounds the approach to exact coalescence.
We demonstrate that the resulting finite-drive near-EP structure is compatible
with the bath-controlled thermal closure---despite arising without any
explicit non-Hermitian term in the microscopic Hamiltonian.  The principal
spectral figures are normalized snapshots of the compensated constant core;
they test its algebra and dynamics but do not establish that the microscopic
finite-band drive reaches the benchmark EP.  A separate low-temperature
$U\to\infty$ saddle verifies projected-channel scaling and charge continuity,
while no finite-$r$ branch is found at the reference temperature
$T_{\rm b}=0.1$; no finite-temperature enhancement of the condensate is
claimed.
\emph{In contrast to earlier studies where non-Hermiticity is
imposed phenomenologically or emerges through renormalization-group
flow~\cite{PhysRevLett.121.203001,han2023complex,
PhysRevB.111.045124}, we demonstrate a complementary microscopic
route in which a strictly Hermitian, periodically driven Dirac impurity
produces a passive non-Hermitian relative kernel after coarse graining
and removal of the common decay.}
Dirac baths provide a natural setting because spin--momentum
locking, angular-momentum selection rules, and an energy-dependent
spectral width make the passive $\PT$ channel symmetry constrained.

The paper is organized as follows.
Section~\ref{sec:model} defines the full Hermitian Hamiltonian.
Section~\ref{sec:coarse} performs the bilinear Gaussian coarse graining
and establishes the emergent $\PT$ symmetry.
Section~\ref{sec:Hss} derives the steady-state Hamiltonian and
the Hadamard rotation that makes hybridization diagonal.
Section~\ref{sec:mf} presents the slave-boson mean field and
the resulting shifted resonance condition.
Sections~\ref{sec:spectra}--\ref{sec:FDR} state the spectral and
fluctuation--dissipation checks, whose full numerical panels are collected in
the Supplemental Material.  Section~\ref{sec:dynamics} discusses real-time
dynamics in the passive co-decaying frame.  Section~\ref{sec:scattering}
derives the Jordan response and full scattering eigenphases, then obtains
the exact finite-$U$ contact matrix on the controlled branchwise-linearized
submanifold and states its metric-covariant Yang--Baxter structure.
Section~\ref{sec:kondo} separates the analytic frozen estimate from the
supporting weak-coupling flow and from any thermodynamic Kondo-temperature
claim.
Section~\ref{sec:conc} concludes.

\section{Model and Hamiltonian}
\label{sec:model}

The full Hermitian Hamiltonian is
$H=H_{\rm bath}+H_{\rm imp}+H_{\rm imp\text{-}bath}+H_O+H_{\rm hyb}[\phi(t)]$.
Each term has a distinct physical role.
$H_{\rm bath}$ describes the two-dimensional Dirac conduction band
with spin-orbit coupling.
$H_{\rm imp}$ is the correlated Anderson impurity.
$H_{\rm imp\text{-}bath}$ is the bare hybridization between the
impurity and the $m=0$ bath channel.
$H_O$ represents auxiliary angular-momentum modes generated by
the trigonal crystal anisotropy.
$H_{\rm hyb}$ couples the physical bath to the angular harmonics
and is modulated by the periodic drive.
We define all symbols at first use below.  To prevent notation clashes,
we reserve $t_{\rm av}=(t+t')/2$ and $\tau=t-t'$ for the Wigner center and
relative times, respectively, and use $T_{\rm b}$ only for the reservoir
temperature.  The chiral bath index is $\eta=\pm$, whereas
$\zeta_\sigma=+1$ ($-1$) for $\sigma=\uparrow$ ($\downarrow$).
The SOC strength is $\lambda$; $\beta_0$ is the signed frozen drive-control
coordinate; and $\tilde\beta=|\beta_0||b_c|$ is its nonnegative
slave-boson mean-field (SBMF)-renormalized projection.  The normalized
compensated-core figures use
$k=0$ and contain no physical density of states (DOS) or cutoff.  The independent
causal and reachability audits use a radial cutoff
$k_{\rm max}=\pi/4$ and, for the sharp-cutoff Dirac transform only, the
separate ultraviolet energy cutoff $D_{\rm uv}=\pi/4$.
We denote the analytic compensated-core EP by
$\beta_{0,{\rm core}}^{\rm EP}=0.50$.  On the path used below,
$\Delta_{\rm eff}^{(0)}=0.10$ and
$\Gamma_{\rm PT}^{(0)}=0.20\beta_0$, so $\beta_0$ is not itself a decay
rate.  At the SOC-degenerate $k=0$ point the bath-dressed double-root
condition is identical to this core condition.  The finite-$k$ root search
is boundary-limited at the upper edge $\beta_0=0.55$ of the declared scan and
does not establish a separate interior projected EP.

\subsection{Bath}

We consider a two-dimensional (2D) host crystal with trigonal ($C_3$)
anisotropy.
The conduction electrons form a Dirac bath with spin-orbit coupling (SOC):
\begin{equation}
H_{\rm bath}
= \sum_{k,\sigma}\epsilon_k\,c^\dagger_{k\sigma}c_{k\sigma}
+ \lambda\sum_k k
  \bigl(c^\dagger_{k\uparrow}c_{k\downarrow}+\mathrm{h.c.}\bigr),
\label{eq:Hbath}
\end{equation}
where $c^\dagger_{k\sigma}$ creates a bath electron with radial
momentum $k$ (in units of inverse lattice constant) and spin
$\sigma\in\{\uparrow,\downarrow\}$;
$\epsilon_k=\alpha_{\rm b}k^2$ ($\alpha_{\rm b}>0$), with radial cutoff
$0\le k\le k_{\rm max}=\pi/4$; and $\lambda>0$ is the spin-orbit
coupling.  The separate energy cutoff $D_{\rm uv}$ enters only logarithmic
and scaling formulas below.
The chiral rotation
$c_{k\pm}=(c_{k\uparrow}\pm c_{k\downarrow})/\sqrt{2}$
diagonalizes $H_{\rm bath}$ exactly, yielding SOC-split chiral
dispersions $\varepsilon_{k\pm}=\alpha_{\rm b}k^2\pm\lambda k$ with
$\varepsilon_{k+}\neq\varepsilon_{k-}$ for all $k\neq0$.
For the full radial dispersions, the two positive-energy roots give
\begin{equation}
\begin{aligned}
\rho_+(\omega)&=\frac{\Delta-\lambda}{4\pi\alpha_{\rm b}\Delta},&
\rho_-(\omega)&=\frac{\Delta+\lambda}{4\pi\alpha_{\rm b}\Delta},\\
\Delta&=\sqrt{\lambda^2+4\alpha_{\rm b}\omega}.&&
\end{aligned}
\label{eq:exact_radial_dos}
\end{equation}
whenever both roots lie inside the radial cutoff.  Hence
$\rho_++\rho_-=1/(2\pi\alpha_{\rm b})$ for $\omega\to0^+$: the complete
quadratic-plus-SOC radial bath is not pseudogapped.  The familiar
$|\omega|/\lambda^2$ law describes only the small-$k$ cone patch of one
branch.  Below zero energy the lower branch instead has a band minimum at
$-\lambda^2/(4\alpha_{\rm b})$ and the associated radial van Hove edge.
The constant \emph{sum} does not eliminate the drive-induced relative decay:
the channel difference and the SOC matrix term $g_x^2$ remain nonzero, as
shown explicitly in Eq.~(\ref{eq:matrix_gamma_pt}).
The full-$k$ calculation below retains this exact structure; the symmetric
linear-DOS form is used only as an explicitly idealized analytic benchmark.

\subsection{Impurity}

The correlated impurity is described by the Anderson Hamiltonian:
\begin{equation}
H_{\rm imp}
=\sum_\sigma\epsilon_\xi\,d^\dagger_\sigma d_\sigma
+ U\,n_\uparrow n_\downarrow,
\label{eq:Himp}
\end{equation}
where $d^\dagger_\sigma$ creates an impurity electron with spin
$\sigma$, $n_\sigma=d^\dagger_\sigma d_\sigma$ is the occupation,
$\epsilon_\xi$ is the bare impurity level before the projected
bath-induced detuning is included, and $U>0$ is the on-site Coulomb
repulsion.

\subsection{Impurity--Bath Hybridization}

The impurity couples directly to the $m=0$ (s-wave) bath channel via
\begin{equation}
H_{\rm imp\text{-}bath}
=\sum_{k,\sigma}
\left(W_k\,c^\dagger_{k,0,\sigma} d_\sigma + \mathrm{h.c.}\right),
\label{eq:Hcd}
\end{equation}
where $W_k$ is the bare hybridization amplitude.
This term defines the physical coupling between the impurity and the
conduction bath before the angular-momentum harmonics are introduced.
Importantly, $W_k$ is time independent and is not proportional to the
frozen drive control $\beta_0$.  At the slave-boson saddle the physical
spinon--bath vertex is $rW_k$, with $r=|b_c|$.

In the presence of the driven angular-momentum sector
[Eqs.~(\ref{eq:HO})--(\ref{eq:Hhyb})], the $m=0$ bath propagator
becomes dressed by virtual processes involving the auxiliary modes
$m\neq0$: the auxiliary modes generate a retarded self-energy
$\Sigma_\sigma^{\rm bath}$ on the $m=0$ bath channel
(Sec.~\ref{sec:coarse}).
The impurity self-energy then arises from this dressed bath
propagator through Eq.~(\ref{eq:Hcd}):
it is the auxiliary dressing of the bath propagator, communicated
to the impurity via the bare coupling $W_k$, that produces the
effective non-Hermitian self-energy on the impurity sector.
This two-step mechanism---auxiliary modes dress the bath, bath
dresses the impurity---is made explicit in Sec.~\ref{sec:coarse}
and Supplemental Material.

\subsection{Auxiliary Angular-Momentum Modes}

The trigonal anisotropy $\beta_3(k)k^3\cos3\theta$ of the host crystal
(where $\beta_3(k)$ is the momentum-dependent trigonal crystal-field strength,
distinct from the hybridization amplitude $\gb$ introduced below;
and $\theta$ is the in-plane momentum angle) couples the
$m=0$ (s-wave) bath channel to higher angular-momentum harmonics
$m=\pm3,\pm6,\ldots$ via the $C_3$ selection rule $\Delta m=\pm3$.
These higher harmonics are represented by auxiliary fermions
$O_{k,m,\sigma}$ with angular momentum index $m\neq0$:
\begin{equation}
\begin{split}
    H_O=\sum_{k,m\neq0,\sigma}
\epsilon_O^{(m)}\,&O^\dagger_{k,m,\sigma}O_{k,m,\sigma}\\
&+\sum_{k,m,\sigma}\beta_3(k)k^3
O^\dagger_{k,m+3,\sigma}O_{k,m,\sigma}\\
&\quad+\mathrm{h.c.},
\label{eq:HO}
\end{split}
\end{equation}
where the mode energies satisfy $\epsilon_O^{(m)}=\epsilon_O^{(-m)}$
by mirror/time-reversal symmetry of the $C_{3v}$ crystal field.  The fields $O_{k,m,\sigma}$
should be understood as a Feshbach, or Loewdin-partitioned, representation
of remote higher-angular-harmonic crystal-field channels rather than as
an additional low-energy bath.  Their energies include the local orbital
or crystal-field splitting of these nonresonant harmonics relative to
the active $m=0$ Dirac channel.  In the low-energy limit
$\epsilon_O^{(m)}\gg D_{\rm uv}$, the auxiliary modes are therefore far off-shell
and can be integrated out exactly within the quadratic sector; their
main effect is a principal-value self-energy dressing the $m=0$ bath
propagator, which in turn induces a self-energy in the impurity sector
through the bare hybridization Eq.~(\ref{eq:Hcd})
(see Sec.~\ref{sec:coarse} and Supplemental Material).  If an
auxiliary harmonic approaches the conduction-band window, the same
formalism gives an ordinary on-shell bath damping contribution; this
spin-even damping is not identified with the passive relative $\PT$ channel.

\subsection{Periodic Drive and Hybridization}

The periodic drive couples the physical bath ($m=0$) to the remote
higher angular harmonics.  We write the Hermitian bath--auxiliary
hybridization as
\begin{equation}
\begin{split}
H_{\rm hyb}[\phi(t)]
=&\sum_{k,\theta,\sigma,m\neq0}
\bigl[
V_m(k)e^{i3m\theta}e^{i\zeta_\sigma\phi(t)} \\
&\hspace{1.2cm}\times
c^\dagger_{k,\theta,\sigma}O_{k,m,\sigma}
+\mathrm{h.c.}\bigr] .
\end{split}
\label{eq:Hhyb}
\end{equation}
Here $\zeta_\uparrow=+1$ and $\zeta_\downarrow=-1$ label the spin-helicity
channel selected by the spin-orbit-split Dirac bath.  The normalized
frozen-kernel scans use the slow oscillatory protocol
$\phi(t)=(\pi/4)\sin(\Omega t)$.  In that protocol the passive
control is the instantaneous angular velocity
$\dot\phi(t_{\rm av})=(\pi\Omega/4)\cos(\Omega t_{\rm av})$; the numerical parameter
$\gb$ used below should be read as the corresponding effective
frozen-kernel amplitude, not as the value of the phase itself.  Hermiticity
fixes the second vertex to carry the complex-conjugate phase.  Therefore
the normal Gaussian contraction of a number-conserving auxiliary sector
contains phase differences, not phase sums.  This is important: the
off-shell auxiliary sector does not by itself generate anti-Hermitian
gain or loss.

Equivalently, the phase can be removed from Eq.~(\ref{eq:Hhyb}) by the
time-dependent unitary rotation
\begin{equation}
O_{k,m,\sigma}\rightarrow e^{-i\zeta_\sigma\phi(t)}O_{k,m,\sigma} .
\label{eq:aux_gauge_rotation}
\end{equation}
The vertices then become phase-free, while the auxiliary Hamiltonian
acquires the Berry term
\begin{equation}
\epsilon_O^{(m)}\rightarrow
\epsilon_O^{(m)}-\zeta_\sigma\dot\phi(t) .
\label{eq:aux_level_shift}
\end{equation}
Thus a gapped Hermitian auxiliary sector produces a real spin-odd coherent
shift.  For the oscillatory protocol this statement is an adiabatic
low-energy projection.  For the stationary rotating protocol
$\phi(t)=\Omega t$, however, Eq.~(\ref{eq:aux_level_shift}) is constant and
the rotated quadratic Hamiltonian is exactly time independent; no freezing
or gradient expansion in $\Omega$ is then required.  The passive
non-Hermitian component appears only in the second step, when this
shifted channel is embedded in the energy-dependent retarded Dirac
bath.
For the oscillatory Gaussian projection the relevant off-shell hierarchy is
$|V|/\epsilon_O^{(m)}\ll1$ and $\Omega\ll\epsilon_O^{(m)}$; a quasistatic
interpretation additionally requires the drive to be slow on the retained
impurity-relaxation scale.  The reachability audit below shows that the
normalized EP benchmark does not satisfy these inequalities simultaneously.
Because $H_O$ is
\emph{bilinear} in $O_{k,m,\sigma}$, the modes $m\neq0$ can be integrated
out at the Gaussian level without approximation within the quadratic
sector; the interaction $U$ enters only through the slave-boson mean field
on the impurity level (Sec.~\ref{sec:Hss}).  Since
$\dot\phi(t_{\rm av})\propto\cos(\Omega t_{\rm av})$, a full cycle sweeps
through the two signs of the frozen control.  The present construction is an
instantaneous adiabatic description of that sweep; nonadiabatic EP traversal
and energy pumping are outside its scope.  The stationary rotating protocol
is used below only to provide a separate microscopic reachability witness.

\section{Coarse Graining and Emergent $\PT$ Symmetry}
\label{sec:coarse}

This section isolates the bilinear coarse graining from the later
slave-boson mean-field treatment of the local interaction.  The raw
retarded self-energy generated by the auxiliary angular modes is, in
general, a complex retarded function; it is not assumed to be positive.
The $\PT$-active low-energy kernel is obtained by projecting the
spin-dependent impurity self-energy onto a spin-conjugate pair
$z,z^*$.  This formulation keeps the microscopic origin of the passive
relative gain--loss channel explicit while avoiding any assumption that the raw
retarded envelope is real, positive, or sign definite.

\subsection{Keldysh Path Integral and Angular-Mode Integration}

We work within the Keldysh formalism~\cite{kamenev2023field}, writing
the partition function as a coherent-state path integral on the closed
time contour.  The bath and auxiliary angular modes enter quadratically
and can be eliminated at the Gaussian level.  The local Coulomb
interaction on the impurity is not treated in this Gaussian step; it is
incorporated later through the slave-boson mean-field approximation in
Sec.~\ref{sec:mf}.

The Grassmann integration can be written explicitly.  For fixed
$(k,\theta,\sigma)$, collect the auxiliary fields into a vector
$O=(\ldots,O_{m},\ldots)^T$ and write the quadratic contour action as
\begin{equation}
\begin{split}
S_O+S_{cO}=&
\int_{\mathcal C} dt dt'\,\bar O(t) A_O(t,t')O(t') \\
&+\int_{\mathcal C} dt\,
\left[\bar c(t)V_\sigma(t)O(t)+
\bar O(t)V_\sigma^\dagger(t)c(t)\right].
\end{split}
\label{eq:grassmann_aux_action}
\end{equation}
The Gaussian identity for Grassmann variables gives
\begin{align}
\int \mathcal D[\bar O,O]e^{i(S_O+S_{cO})}
&\propto \det A_O\,e^{iS_{\rm infl}},
\label{eq:grassmann_identity}\\
S_{\rm infl}
&=-\int_{\mathcal C}dt dt'\,
\bar c(t)\,\Xi_\sigma(t,t')\,c(t'),
\label{eq:grassmann_influence_main}\\
\Xi_\sigma(t,t')
&=V_\sigma(t)g_O(t,t')V_\sigma^\dagger(t') .
\label{eq:Xi_aux_kernel}
\end{align}
Here $g_O=A_O^{-1}$ is the contour auxiliary propagator.  This is the
complete Gaussian step: because the auxiliary sector conserves particle
number, there is no anomalous $OO$ contraction.  Keeping both vertices of
Eq.~(\ref{eq:Hhyb}) explicit, the normal auxiliary contraction gives
\begin{align}
\Sigma_{\sigma}^{O,R}(t,t')
&=\sum_{m,m'\neq0}
 v_{m\sigma}(t)\,
 g_{mm'}^{O,R}(t,t')\,
 v^*_{m'\sigma}(t'),
\label{eq:normal_aux_contraction}\\
v_{m\sigma}(t)
&=V_m e^{i3m\theta}e^{i\zeta_\sigma\phi(t)} .
\label{eq:normal_aux_vertex}
\end{align}
For the diagonal terms, and also for any Hermitian off-shell auxiliary
matrix after summing conjugate paths, the phases enter as
$e^{i\zeta_\sigma[\phi(t)-\phi(t')]}$.  With
$T=t_{\rm av}$ and $\tau=t-t'$, this is
$e^{i\zeta_\sigma\tau\dot\phi(T)}$ to first adiabatic order, so the
Wigner transform shifts the stationary auxiliary self-energy as
$\Sigma_0^{O,R}(\omega)\to
\Sigma_0^{O,R}[\omega+\zeta_\sigma\dot\phi(T)]$.
Its first-order expansion gives the regular Hermitian principal-value shift
in Eq.~(\ref{eq:aux_real_shift}); the explicit transform and sign are given
in the Supplemental Material.  Equivalently, after the unitary rotation in
Eq.~(\ref{eq:aux_gauge_rotation}), the same result is obtained from the
spin-dependent level shift in Eq.~(\ref{eq:aux_level_shift}).

To leading order in the slow drive and in the off-shell expansion, the
auxiliary-dressed $m=0$ bath therefore has the form
\begin{equation}
\begin{split}
\Sigma_{\sigma}^{\rm bath,R}(\omega,t_{\rm av})
=&\,\Sigma_0^{\rm bath,R}(\omega)
+\zeta_\sigma\Delta_{\rm aux}(\omega,t_{\rm av}) \\
&+O\!\left[(\Omega/\epsilon_O)^2\right] .
\end{split}
\label{eq:aux_real_shift}
\end{equation}
Here $\Delta_{\rm aux}$ is real in the low-energy window.  Microscopically
one may write, schematically,
\begin{equation}
\begin{split}
\Delta_{\rm aux}(\omega,t_{\rm av})
=&-\dot\phi(t_{\rm av})\,\partial_{\epsilon_O}
\sum_{m\neq0}|V_m|^2 \\
&\times
\mathcal{P}\frac{1}{\omega-\epsilon_O^{(m)}} ,
\end{split}
\label{eq:Delta_aux}
\end{equation}
up to form-factor and angular-overlap coefficients.  If an auxiliary
pole enters the conduction window, the corresponding delta-function
part gives ordinary bath damping; it is not identified with the passive
$\PT$ channel.

The bare impurity--bath hybridization $W_k$ in
Eq.~(\ref{eq:Hcd}) remains time independent.  The drive reaches the
impurity through the dressed bath propagator.  In contour notation the
two Gaussian eliminations give
\begin{align}
G_{c,\sigma}^{-1}
&=g_c^{-1}-\Xi_\sigma,
\qquad
\Xi_\sigma=V_\sigma\circ g_O\circ V_\sigma^\dagger,
\label{eq:dressed_c_inverse}\\
\Sigma_{d,\sigma}
&=W^\dagger\circ G_{c,\sigma}\circ W,
\label{eq:impurity_selfenergy_dressed_bath}
\end{align}
where $\circ$ denotes contour convolution.  Thus the drive-dependent
virtual path is $d\to c\to O\to c\to d$; no direct time dependence of
$W_k$ is required.  The full Schur-complement derivation is given in the
Supplemental Material.

In the frozen off-shell window, write
$\Xi_\sigma^R=\Xi_0^R+\zeta_\sigma\delta\Xi^R$ with
$\delta\Xi^R\in\mathbb R$.  Expanding the dressed-bath resolvent gives the
general first-order correction
\begin{equation}
\delta\Sigma_d^R
=W^\dagger G_{c,0}^R\,\delta\Xi^R\,G_{c,0}^R W .
\label{eq:deltaSigma_exact}
\end{equation}
Although the auxiliary correction is real, the retarded continuum
propagators are complex, so the transmitted impurity correction has both
coherent and dissipative parts.  Defining
\begin{equation}
\Sigma_{d,0}^R(\omega)=\Lambda(\omega)-i\bar\Gamma(\omega),
\label{eq:Sigma0_Lambda_Gamma}
\end{equation}
we write
\begin{equation}
\Sigma^R_{d,\sigma}
=\Sigma_{d,0}^R+\zeta_\sigma\Delta_{\rm coh}
+i \zeta_\sigma\Gamma_{\rm PT}.
\label{eq:Delta_Gamma_PT}
\end{equation}
For the nearly rigid scalar shift used in the analytic estimates,
$\delta\Xi^R\simeq\Delta_{\rm aux}$ and the spin-even auxiliary
background varies slowly.  Then
\begin{equation}
\Sigma^R_{d,\sigma}(\omega,t_{\rm av})
\simeq
\Sigma_{d,0}^R(\omega-\zeta_\sigma\Delta_{\rm aux})
\simeq
\Sigma_{d,0}^R-\zeta_\sigma\Delta_{\rm aux}
\partial_\omega\Sigma_{d,0}^R,
\label{eq:Dirac_conversion}
\end{equation}
which gives
\begin{equation}
\Delta_{\rm coh}=-\Delta_{\rm aux}\partial_\omega\Lambda,
\qquad
\Gamma_{\rm PT}=\Delta_{\rm aux}\partial_\omega\bar\Gamma .
\label{eq:Delta_Gamma_defs}
\end{equation}
Equation~(\ref{eq:deltaSigma_exact}) is the general first-order result;
Eq.~(\ref{eq:Dirac_conversion}) is its rigid-shift limit.  All functions
are evaluated in the active low-energy window.
For an SOC matrix bath, however, the exact result need not reduce to the
rigid shift of the spin-even scalar self-energy.  In the chiral basis set
$a=(\omega^+-\varepsilon_{k+})^{-1}$,
$b=(\omega^+-\varepsilon_{k-})^{-1}$, so that
$g_0=(a+b)/2$ and $g_x=(a-b)/2$.  Then
\begin{equation}
-\partial_\omega g_0=g_0^2+g_x^2,\qquad
g_b\sigma_zg_b=(g_0^2-g_x^2)\sigma_z=ab\,\sigma_z .
\label{eq:matrix_shift_identity}
\end{equation}
With the sign convention of Eq.~(\ref{eq:Delta_aux}), the exact first-order
relative coefficient is therefore
\begin{align}
\delta\Sigma_z^R
&=\Delta_{\rm aux}\sum_k|W_k|^2a_kb_k,
\notag\\
\Gamma_{\rm PT}
&=\Delta_{\rm aux}\!\left[
\partial_\omega\bar\Gamma
-2\,\operatorname{Im}\sum_k|W_k|^2g_{x,k}^2\right].
\label{eq:matrix_gamma_pt}
\end{align}
Reversing the definition of the spin-odd auxiliary shift reverses both sides
of Eq.~(\ref{eq:matrix_gamma_pt}), without changing any rate eigenvalue.
Consequently a structureless scalar flat band ($g_x=0$) cannot generate the
relative decay at this order, whereas an SOC-split bath can do so even when
its \emph{total} width is exactly flat: for
$\partial_\omega\bar\Gamma=0$ the decay imbalance is carried entirely by
the $g_x^2$ term.  This is the operative mechanism in the full-momentum
certificate below.  The SOC split bath supplies the spin--momentum-locked
helicity channels, and in that basis
\begin{align}
\bar\Gamma_\eta(\omega)
&=\pi\sum_k |W_{k\eta}|^2
\delta\!\left(\omega-\varepsilon_{c,\eta}(k)\right),
\label{eq:helicity_width}\\
\varepsilon_{c,\eta}(k)&=\alpha_{\rm b}k^2+\eta\lambda k .
\label{eq:helicity_dispersion}
\end{align}
so the SOC parameter $\lambda$ controls the channel-resolved densities of
states and the matrix form factor entering the projected EP diagnostics.
Although their positive-energy sum is flat in the full radial model,
their difference and $g_x^2$ are not.  Without SOC-induced helicity
structure the auxiliary shift remains a coherent real field rather than a
passive relative decay channel in this flat-total-width realization.

\subsection{Passive $\PT$ Core and Causal Detuning}
\label{sec:PT_all}

The projected impurity self-energy is decomposed into a common bath part,
a passive relative decay imbalance, and a Kramers--Kronig-related coherent
detuning:
\begin{equation}
\begin{split}
\Sigma_{\rm imp}^{R}(\omega,t_{\rm av})=&
\left[\Lambda(\omega)-i\bar\Gamma(\omega)\right]\mathbb{I} \\
&+\Dcoh(\omega,t_{\rm av})\sigma_z
+i\Gamma_{\rm PT}(\omega,t_{\rm av})\sigma_z .
\end{split}
\label{eq:passive_PT_selfenergy}
\end{equation}
The first line is a common Lamb shift and common decay.  It is retained in
the bath-controlled FDR, but it does not affect the relative EP structure.
The second line contains two spin-odd terms with different symmetry status:
$i\Gamma_{\rm PT}\sigma_z$ is the passive relative decay imbalance, while
$\Dcoh\sigma_z$ is a real coherent detuning generated by the same causal
retarded bath self-energy.

The SOC bath supplies the Hermitian spin-mixing channel.  In the original
spin basis, the local bath Green function has the matrix structure
\begin{equation}
\begin{split}
g_b^R(k,\omega)
&=\left[\omega^+-\epsilon_k-\lambda k\sigma_x\right]^{-1} \\
&=g_0(k,\omega)\mathbb{I}+g_x(k,\omega)\sigma_x,
\end{split}
\label{eq:soc_bath_matrix}
\end{equation}
with
\begin{equation}
g_x(k,\omega)=
\frac{\lambda k}
{(\omega^+-\epsilon_k)^2-\lambda^2 k^2} .
\label{eq:soc_bath_gx}
\end{equation}
Equivalently, with
$\epsilon_{k\pm}=\epsilon_k\pm\lambda k$,
\begin{equation}
\begin{aligned}
g_x&=\frac12\left[
\frac{1}{\omega^+-\epsilon_{k+}}-
\frac{1}{\omega^+-\epsilon_{k-}}\right],\\
\Sigma_x&=\frac12(\Sigma_+-\Sigma_-)
=\frac{\Lambda_+-\Lambda_-}{2}
-i\frac{\Gamma_+-\Gamma_-}{2}.
\end{aligned}
\label{eq:complex_channel_odd_self_energy}
\end{equation}
The principal-value part of this SOC off-diagonal bath dressing gives a
real coherent spin-mixing coefficient,
\begin{equation}
\Deff(\omega)=
\sum_k |W_k|^2\, {\rm Re}\, g_x(k,\omega),
\label{eq:Delta_eff_soc_origin}
\end{equation}
renormalized in the interacting problem by the slave-boson saddle point.
Equation~(\ref{eq:Delta_eff_soc_origin}) names only that coherent
principal-value component in the normalized Markov core.  The full causal
kernel retains the imaginary part in
Eq.~(\ref{eq:complex_channel_odd_self_energy}) together with its complete
frequency dependence.
The drive-induced shift cannot leak into this mixing channel at leading
order: expanding the dressed bath propagator to first order in the
spin-odd shift $\Delta_{\rm aux}\sigma_z$ gives
\begin{equation}
g_b^R\,\sigma_z\,g_b^R=\left(g_0^2-g_x^2\right)\sigma_z,
\label{eq:sigma_z_purity}
\end{equation}
because $\{\sigma_x,\sigma_z\}=0$, so the first-order correction is
purely $\sigma_z$ and $\Deff$ remains a static (drive-independent)
channel at this order, corresponding to the offset $\Dzero$ of
Eq.~(\ref{eq:projected_channels}); the isolated finite-drive EP of the
generic projection is therefore the microscopically natural case.
Thus, after subtracting the common damping, the general relative two-channel
kernel is
\begin{equation}
h_{\rm rel}=
\Deff\sigma_x+
\left(\Dcoh+i\Gamma_{\rm PT}\right)\sigma_z .
\label{eq:relative_detuned_kernel}
\end{equation}
The passive-$\PT$-symmetric core is obtained at the compensated detuning
point $\Dcoh=0$:
\begin{equation}
h_{\PT}=
\Deff\sigma_x+i\Gamma_{\rm PT}\sigma_z,
\qquad
\sigma_x h_{\PT}^*\sigma_x=h_{\PT} .
\label{eq:passive_PT_core}
\end{equation}
The real term $\Dcoh\sigma_z$ is not discarded.  It is $\PT$-odd under the
same operation and therefore acts as a causal detuning floor.  The pole
splitting of Eq.~(\ref{eq:relative_detuned_kernel}) is
\begin{equation}
s^2=
\Deff^2+
\left(\Dcoh+i\Gamma_{\rm PT}\right)^2 .
\label{eq:detuned_EP_splitting}
\end{equation}
For real $\Deff$, $\Dcoh$, and $\Gamma_{\rm PT}$, exact coalescence in the
passive sector requires
\begin{equation}
\Dcoh\Gamma_{\rm PT}=0,
\qquad
\Deff^2+\Dcoh^2-\Gamma_{\rm PT}^2=0 .
\label{eq:detuned_EP_conditions}
\end{equation}
Thus, for a nonzero relative decay imbalance, the exact passive EP occurs
at the compensated point $\Dcoh=0$ and $|\Deff|=|\Gamma_{\rm PT}|$.
In the idealized scalar cone-patch estimate, compensation is a
codimension-one operating-energy condition:
$\Dcoh\propto\partial_\omega\Lambda$ changes sign and vanishes at a
finite energy of order $D_{\rm uv}/e$ for the linear Dirac width, where
$\partial_\omega\bar\Gamma$ remains finite, while near the Fermi level
the companion ratio is logarithmically enhanced,
$|\Dcoh/\Gamma_{\rm PT}|\simeq(2/\pi)\ln(D_{\rm uv}/|\omega|)$.  Away from the
compensated point, the Kramers--Kronig detuning prevents a literal pole
coalescence.  Minimizing $|s|$ over real $\Deff$ at fixed $\Dcoh$ and
$\Gamma_{\rm PT}$ gives
\begin{equation}
|s|_{\rm min}=
\begin{cases}
\sqrt{2|\Dcoh\Gamma_{\rm PT}|},
& |\Dcoh|\le|\Gamma_{\rm PT}|,\\[2pt]
\sqrt{\Dcoh^2+\Gamma_{\rm PT}^2},
& |\Dcoh|>|\Gamma_{\rm PT}|,
\end{cases}
\label{eq:KK_floor}
\end{equation}
with the square-root non-analyticity characteristic of a second-order EP.
This causal detuning should not be identified with a universal
interaction-induced splitting.  In the exact equal-velocity,
scalar-hybridization linearized reduction derived below, the standard local
interaction preserves the complexified pseudospin symmetry and does not
lift the EP~\cite{KulkarniYB2026}.  Curvature, unequal velocities, an
energy-dependent self-energy, or component-dependent vertices can instead
move or split it.  The Supplemental Material derives the minimization and
the \emph{idealized symmetric-cone} sharp-cutoff Dirac Kramers--Kronig
estimate explicitly.  For that benchmark, $D_{\rm uv}=\pi/4$ gives an exact
compensation energy
$|\omega_c|=0.24608$ (the low-energy estimate is $D_{\rm uv}/e=0.28893$).
Compensation holds only at $\omega_c$: one common half-width
$\bar\Gamma=0.12$ away, the residual splitting floor is already
$0.90$--$1.00$ times the relative channel in the audited benchmark.  The
full finite-band causal kernel therefore exhibits an avoided coalescence
over a linewidth even though the constant compensated core is exactly
defective.

This two-step structure is the microscopic origin of the kernel used in
the rest of the paper: off-shell auxiliary modes generate a real spin-odd
shift, and the retarded SOC matrix bath converts that shift into a
spin-dependent width while also supplying the Hermitian spin-mixing channel.
The numerical scans treat $\gb$ as a tunable frozen effective kernel control
parameter.  In the scalar cone-patch rigid-shift estimate,
\begin{equation}
\Gamma_{\rm PT}
\sim
\Omega\,\frac{|V|^2}{\epsilon_O^2}\,
\partial_\omega\bar\Gamma,
\label{eq:Gamma_PT_micro_scale}
\end{equation}
whereas the full radial model uses the additional SOC term in
Eq.~(\ref{eq:matrix_gamma_pt}).  Thus the construction establishes the
symmetry-allowed origin of the passive relative channel, but does not
determine the numerical magnitude of the dimensionless frozen-kernel control
$\gb$.  More explicitly, writing
$x=|V|/\epsilon_O$ and using the phase amplitude $\pi/4$, the benchmark
$\Gamma_{\rm PT}/\bar\Gamma=0.10/0.12$ requires approximately
\begin{equation}
\frac{\Omega}{\epsilon_O}\gtrsim\frac{1.061}{x^2}.
\label{eq:reachability_bound}
\end{equation}
This is incompatible with both $x\ll1$ and a quasistatic oscillatory drive.
The
passivity condition itself is satisfied
($\Gamma_{\rm PT}=0.10<\bar\Gamma=0.12$), but microscopic reachability of
the normalized EP is not established.  Figure~\ref{fig:frozen_comparison}
therefore exposes the exact constant-core geometry only.
The independent causal test in Supplemental Fig.~S5 uses the same small
width relative to the unit charge gap and quantifies the finite-band
avoided-coalescence correction.

\paragraph{Stationary full-momentum witness.}
The preceding negative statement is specific to the normalized benchmark
and the oscillatory quasistatic protocol.  For $\phi(t)=\Omega t$, the
rotating-frame auxiliary levels are shifted by the constant $\mp\Omega$.
Keeping their exact frequency dependence, the curved SOC dispersion, and
the complete radial momentum integral, define on the outgoing resonance
sheet
\begin{equation}
F(z;\epsilon_d,V_O)=
\det\!\left[(z-\epsilon_d)\mathbb I-
\Sigma_d^{\mathrm{II}}(z;V_O)\right] .
\label{eq:full_k_pole_determinant}
\end{equation}
A nonlinear pole EP is a double zero
$F=\partial_zF=0$ with $\partial_z^2F\neq0$ and a rank-one inverse kernel.
To certify its existence without evaluating a singular eigenbasis, let
$z_c(\epsilon_d,V_O)$ be the continuously branch-tracked solution of
$\partial_zF(z_c)=0$ on a closed control loop $C$, and define
\begin{equation}
{\cal D}_{\rm loc}=F(z_c;\epsilon_d,V_O),\qquad
W_C=\frac{1}{2\pi}\Delta_C\arg {\cal D}_{\rm loc}.
\label{eq:local_discriminant_main}
\end{equation}
When $\partial_z^2F$ stays nonzero, the implicit-function theorem keeps
$z_c$ on one smooth critical-point branch.  Regarding ${\cal D}_{\rm loc}$
as a map from the two real controls to $(\operatorname{Re}{\cal D}_{\rm loc},
\operatorname{Im}{\cal D}_{\rm loc})$, its boundary winding is the Brouwer
degree whenever ${\cal D}_{\rm loc}\neq0$ on $C$.  A nonzero degree forces at
least one enclosed zero (with regular zeros counted by their Jacobian sign).
At such a zero,
$F=\partial_zF=0$; the nonzero second derivative makes it a quadratic pole
coalescence, and the rank-one inverse-kernel test excludes a scalar
degeneracy.  Explicit pole exchange and the half-winding of their separation
provide the equivalent EP2 monodromy check.
The full two-cut calculation in the Supplemental Material gives such a point
for the better-conditioned representative point
\begin{equation}
\begin{gathered}
\lambda=0.05,\quad W=0.10,\quad k_{\max}=\pi/4,\\
\epsilon_O=120,\quad\Omega=60,
\end{gathered}
\label{eq:full_k_ep_parameters}
\end{equation}
at
\begin{equation}
\begin{split}
V_O&=1.21342201,\qquad \epsilon_d=0.01517035,\\
z_{\rm EP}&=0.01197283-0.00280722i .
\end{split}
\label{eq:full_k_ep_solution}
\end{equation}
Both helicity cuts are open.  The driven $k=0$ upper threshold is
$\omega_{\rm th}^{+}=0.00817959$, so the pole lies $0.00379325$, or
$1.351|\operatorname{Im}z_{\rm EP}|$, above it; this threshold proximity and
the lower-branch van Hove structure are included in the analytic continuation.
Numerically, $|F|=1.44\times10^{-21}$,
$|\partial_zF|=1.09\times10^{-15}$, and
$\partial_z^2F=1.68645-0.19034i$; the local degree is $+1$ and the two poles
exchange on a closed loop in $(\epsilon_d,V_O)$.  The physical rate-matrix
eigenvalues are $1.904\times10^{-3}$ and $3.096\times10^{-3}$, with
$\bar\Gamma=0.002500$ and passivity margin
$\Gamma_{\min}/\bar\Gamma=0.761$.
Moreover, $V_O/(\epsilon_O-\Omega)=0.02022$, while the empty-auxiliary
thermal exponent is $(\epsilon_O-\Omega)/T_{\rm b}=600$ at
$T_{\rm b}=0.1$.  Thus $\Omega/\bar\Gamma\simeq2.4\times10^4$: the
oscillatory quasistatic condition is strongly violated, whereas no such
condition is required because the rotating-frame Hamiltonian is exactly
stationary.  The EP certificate is a retarded-kernel statement and is
temperature independent.  At the displayed temperature
$T_{\rm b}/\bar\Gamma=40$, so linewidth-scale spectral or transient
signatures are thermally unresolved; observing them would instead require
$T_{\rm b}\lesssim10^{-3}$ for this witness.

Continuation in SOC, with $W$ held fixed at $0.10$ while the real gate and
the off-shell auxiliary vertex are retuned by the same double-zero equations,
gives a finite locus over
$0.02\leq\lambda\leq0.1578$.  Independent local-winding and pole-exchange
certificates pass at $\lambda=0.05$, $0.10$, and $0.15$.  The continued pole
approaches an open-channel threshold near the upper end while the rate matrix
remains passive; this threshold termination is not a passivity breakdown.
No endpoint or EP phase boundary is claimed.  This establishes a controlled microscopic
family of causal pole EPs in the full curved kernel.  It does not turn the
normalized benchmark into a causal EP, identify the nonlinear pole EP with
the nilpotent Markov-core EP, or imply full-momentum Yang--Baxter
integrability.

\paragraph{Finite-\texorpdfstring{\(U\)}{U} block-Wilson continuation.}
We next test whether the representative \(\lambda=0.05\) causal pole EP
survives a local Anderson interaction without imposing the double-zero
condition on the interacting self-energy.  The positive \(2\times2\)
matrix spectral density of the same rotating full-\(k\) bath is logarithmically
discretized with \(\Lambda=1.5\), mapped by block Lanczos to a matrix Wilson
chain, and averaged over four \(z\) shifts.  We first use 12-site chains to
locate the finite-\(U\) center and evaluate
the zero-temperature correction-vector Green functions \(G\) and \(F\) in
exact fixed-charge sectors and form the equation-of-motion self-energy
\begin{equation}
\Sigma_U(z)=U F(z)G^{-1}(z).
\label{eq:finiteU_eom_selfenergy}
\end{equation}
Solving \(F_U=\partial_zF_U=0\) for the determinant of the full analytic
two-cut kernel dressed by Eq.~(\ref{eq:finiteU_eom_selfenergy}) gives, at
\(U=0.0025\),
\begin{equation}
\begin{split}
z_{\rm EP}^{(U)}&=0.01177670-0.00225815i,\\
\epsilon_d^{(U)}&=0.01440607,\qquad W^{(U)}=0.09036305.
\end{split}
\label{eq:finiteU_fullk_center}
\end{equation}
Here \(U/D_{\rm uv}=3.18\times10^{-3}\), while
\(U/\bar\Gamma_{\rm self}=1.225\) with
\(\bar\Gamma_{\rm self}=(W^{(U)})^2/4\).  Thus the interaction is small on
the bandwidth scale but moderate on the retuned local-width scale; it is not
a deep-Kondo or strong-coupling calculation.

\begin{figure*}[t]
\centering
\includegraphics[width=0.98\textwidth]{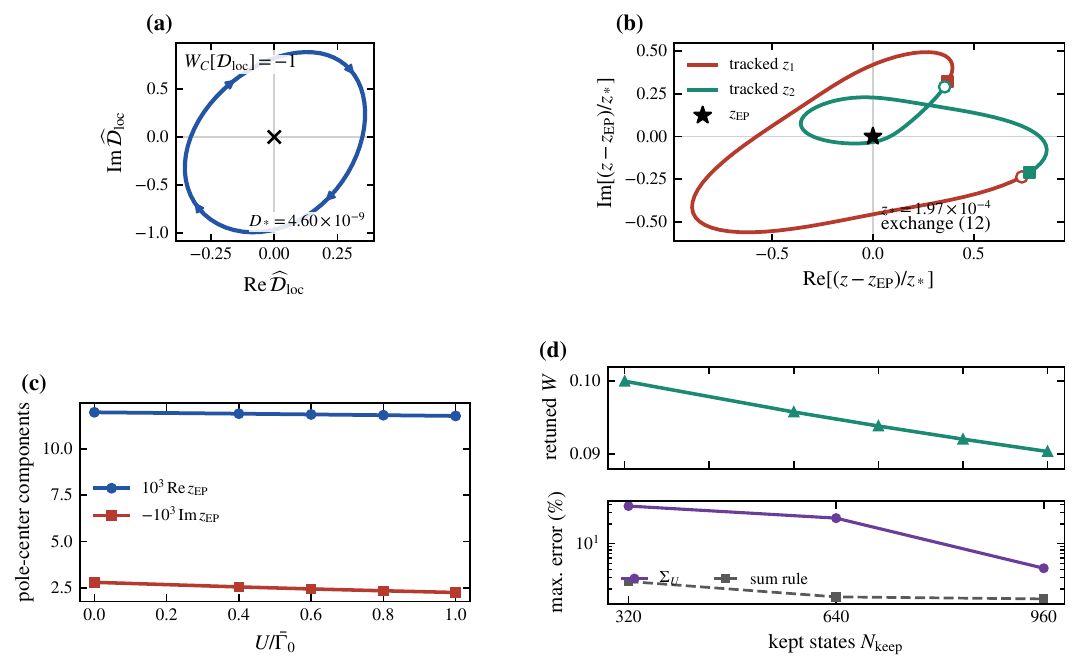}
\caption{\textbf{Finite-\(U\) continuation of the causal full-\(k\) pole EP.}
(a) The dimensionless contour
\(\widehat{\mathcal D}_{\rm loc}=\mathcal D_{\rm loc}/D_*\), with
\(D_*=\max_C|\mathcal D_{\rm loc}|=4.603\times10^{-9}\), winds once
clockwise around the origin on the interacting control ellipse at
\(U=0.0025\).  The positive rescaling removes a plotting unit and cannot
change the Brouwer degree.
(b) The two nearby poles are plotted as
\((z-z_{\rm EP})/z_*\), with
\(z_*=\max_C|z-z_{\rm EP}|=1.965\times10^{-4}\), and exchange after one
circuit; open circles and filled squares mark the initial and final tracked
labels, and the star marks Eq.~(\ref{eq:finiteU_fullk_center}).
(c) The exact finite-chain center varies smoothly from \(U=0\) to
\(U/\bar\Gamma_0=1\), where \(\bar\Gamma_0=0.0025\).
(d) The upper panel shows the retuned impurity hybridization; the lower panel
compares kept-space results with the exact response on the same four
\(z\)-shifted 12-site chains.  These errors are convergence diagnostics and are not used
to establish the winding.  The separate $N=10\to12$ common-contour gate is
given by Eq.~(\ref{eq:finiteU_fullk_rouche}) and detailed in the Supplemental
Material.}
\label{fig:finiteU_fullk_exact}
\end{figure*}

Figure~\ref{fig:finiteU_fullk_exact} gives the independent topological gate.
On an ellipse with half-widths \(10^{-4}\) in \(\epsilon_d\) and \(10^{-3}\)
in \(W\), the branch-tracked local discriminant has winding \(-1\), the two
poles exchange, and their separation has half-winding \(-1/2\).  The
quadratic, phase-step, pole-residual, and branch-assignment guards all pass
(see the Supplemental Material).  Increasing the \(z\)-average
from four to eight shifts changes the four fitted coordinates by at most
\(0.71\%\), while the exact sum rule and equation-of-motion identity hold to
roundoff.  A kept-space calculation at \(N_{\rm keep}=960\) differs from the
exact finite-chain response by at most \(4.18\%\) in the self-energy and
\(1.40\%\) in the sum rule.  These are convergence diagnostics; the
complete-basis response carries the topology claim.  No infinite-chain,
finite-temperature full-density-matrix NRG (FDM-NRG), phase-boundary, or
universal interaction-scaling
conclusion is drawn.

As an independent chain-length gate, the center is reoptimized separately
on the $N=10$ and $N=12$ complete-basis chains.  Their maximum relative
coordinate drift is $3.746\times10^{-3}$.  On a shared ellipse with
half-widths $2\times10^{-4}$ in $\epsilon_d$ and $2\times10^{-3}$ in $W$,
both chains have local-discriminant winding $-1$, endpoint pole exchange,
and pole-separation half-winding $-1/2$.  At 256 contour points,
\begin{equation}
\frac{\sup_C|D_{12}-D_{10}|}{\min_C|D_{10}|}
=\frac{1.638\times10^{-9}}{2.417\times10^{-9}}
=0.6775<1,
\label{eq:finiteU_fullk_rouche}
\end{equation}
so Rouch\'e's theorem preserves the enclosed zero count between the two
finite chains: the boundary deformation from $D_{10}$ to $D_{12}$ cannot
cross the origin.  Here $D_N$ is the finite-chain version of
Eq.~(\ref{eq:local_discriminant_main}), with its critical pole branch tracked
continuously on the shared contour.  The same gate passes at 128 points.  This establishes
finite-$N$ persistence over the tested pair, not an $N\to\infty$ limit.

\section{Steady-State Hamiltonian and Emergent Non-Hermiticity}
\label{sec:Hss}

Given the passive spin-conjugate relative self-energy from Sec.~\ref{sec:coarse}, this section derives
the explicit $4\times4$ steady-state Hamiltonian $\Htilde(k)$
and identifies the structural origin of its exceptional points.
A key step is the Hadamard rotation that diagonalizes the bare
hybridization block, revealing that each chiral spinon mode
couples directly to one bath channel.  The internal
$i\Gamma_{\rm PT}$ term still mixes those impurity modes, so the physical
two-channel scattering matrix need not be diagonal; the frozen finite-size
quantization below is therefore formulated using its exact eigenphases.

\subsection{From Self-Energy to Steady-State Hamiltonian}
\label{sec:deriveHss}

After the Gaussian coarse-graining of the bath and auxiliary angular
modes in Sec.~\ref{sec:coarse}, and after introducing the
slave-boson saddle point in Sec.~\ref{sec:mf}, the retarded spinon
action takes the form
$S_{\rm imp}^R=\sum_k\bar{f}[\omega\mathbb{I}-\Hss(k)]f$,
where the steady-state Hamiltonian $\Hss(k)$ is defined as the
inverse of the retarded Green's function at the saddle point:
$\Hss(k)\equiv[\Gret_{\rm ss}(\omega,k)]^{-1}|_{\rm saddle}$.
The gauge-invariant slave-boson amplitude is $r=|b_c|$.  The three
channels in the effective Hamiltonian have distinct microscopic origins:
the passive relative width and coherent detuning are components of the
projected retarded self-energy, whereas the impurity--bath vertex descends
from the time-independent bare coupling $W$.  Since an impurity self-energy
contains two spinon--bath vertices, the physical projected parametrization is
\begin{align}
\Gpt&=r^2\Gamma_{\rm PT}^{(0)}(\gb),&
\Deff&=r^2\Delta_0^{(0)},
\nonumber\\[-0.2em]
\Veff&=rW .
\label{eq:projected_channels}
\end{align}
Here the superscript $(0)$ denotes the frozen bath projection before
slave-boson renormalization.  The core discriminant consequently obeys
\begin{equation}
s_{\rm core}^2=r^4\left\{
\left[\Delta_0^{(0)}\right]^2
-\left[\Gamma_{\rm PT}^{(0)}(\gb)\right]^2\right\}.
\label{eq:r4_core}
\end{equation}
For $r>0$, the common factor $r^4$ cancels from the core EP condition.
The local core EP is therefore exactly invariant under changes of $r$.
A bath-dressed finite-$k$ minimum must be established independently; in the
declared scan it is boundary-limited and is not classified as an EP.
Figure~\ref{fig:frozen_comparison} uses $r=1$ and a normalized
frozen-kernel geometry path; they are intended to display the bifurcation
rather than to fix an Anderson hybridization width.  The supporting
channel-resolved RG calculation uses the conservative fixed-$W$ projection
\begin{align}
\Gamma_0&=\pi\rho W^2=0.12,\nonumber\\
\Delta_{\rm eff}^{(0)}&=0.10,\qquad
\Gamma_{\rm PT}^{(0)}=0.20\gb ,
\label{eq:smallW_path}
\end{align}
for which $\Gamma_0/\Delta_{\rm charge}=0.12$ and the core EP remains at
$\gb=0.50$.  Thus the RG test is performed in a local-moment-compatible
small-width regime, with $W$ independent of the drive coordinate.

At the compensated passive-$\PT$ point $\Dcoh=0$, substituting the projected self-energy into
the retarded Dyson equation and including the hybridization to the
chiral bath channels gives, in the spin basis
$(d_\uparrow,d_\downarrow,c_{k+},c_{k-})$,
\begin{equation}
\Hss(k)=
\begin{pmatrix}
\teps+i\Gpt & \Deff & \Veff/\sqrt{2} & \Veff/\sqrt{2}\\
\Deff & \teps-i\Gpt & \Veff/\sqrt{2} & -\Veff/\sqrt{2}\\
\Veff/\sqrt{2} & \Veff/\sqrt{2} & \varepsilon_{c,+}(k) & 0\\
\Veff/\sqrt{2} & -\Veff/\sqrt{2} & 0 & \varepsilon_{c,-}(k)
\end{pmatrix}.
\label{eq:Hss}
\end{equation}
Here $\teps=\epsilon_\xi+\delta_c+\dSig$ is the renormalized
impurity level.  The Lagrange multiplier $\delta_c$ enforces the
slave-boson constraint, while $\dSig$ is the real projected one-body
shift from the principal-value self-energy in Sec.~\ref{sec:PT_all}
(and may be absorbed into $\epsilon_\xi$ by convention).  The chiral
bath dispersions are $\varepsilon_{c,\pm}(k)=\alpha_{\rm b}k^2\pm\lambda k$.
The interaction $U$ is not an explicit two-body coupling in the
$4\times4$ Gaussian kernel; it enters through the SBMF constraint and
through the Schrieffer--Wolff and finite-\(U\) contact virtual-charge
energies used below.  For the
finite-$U$ Anderson interpretation, particle-hole symmetry is imposed
by the common impurity-level choice $\epsilon_\xi=-U/2$, not by adding
opposite $\pm U/2$ spin splittings.  A real Zeeman field, when present,
would enter as an independent spin splitting $B_{\rm imp}$ rather than
as the Hubbard interaction.

The $\PT$ symmetry of $\Hss(k)$ is explicit for real
$\Gpt$, $\Deff$, and $\Veff$: under
$P:\uparrow\leftrightarrow\downarrow$ combined with complex
conjugation, $(\teps+i\Gpt)\to(\teps-i\Gpt)^*=\teps+i\Gpt$,
while the real off-diagonal and hybridization entries are exchanged
into themselves.  Thus the $\PT$ structure is fixed by the projected
self-energy channels and the gauge-invariant amplitude $r=|b_c|$, not
by the phase of the slave-boson condensate.

\subsection{Diagonal-Hybridization Rotation}
\label{sec:rotation}

The hybridization block of $\Hss$ has Hadamard structure
$(\Veff/\sqrt{2})\bigl[\begin{smallmatrix}1&1\\1&-1
\end{smallmatrix}\bigr]$.
The unitary rotation
$U_2=\frac{1}{\sqrt{2}}\bigl[\begin{smallmatrix}
1&1\\1&-1\end{smallmatrix}\bigr]$
acting on the impurity subspace defines chiral spinon modes
$\tilde{d}_\pm=(d_\uparrow\pm d_\downarrow)/\sqrt{2}$
and diagonalizes the hybridization block exactly:
\begin{align}
U_2^{-1}h_{\rm imp}U_2
&=
\begin{pmatrix}
\teps+\Deff & i\Gpt \\
i\Gpt & \teps-\Deff
\end{pmatrix},
\label{eq:imp_rotated}\\
U_2^{-1}V_{dc}&=\Veff\,\mathbb{I},
\label{eq:V_rotated}
\end{align}
where
$h_{\rm imp}=\bigl(\begin{smallmatrix}\teps+i\Gpt&\Deff\\\Deff&\teps-i\Gpt\end{smallmatrix}\bigr)$
and
$V_{dc}=(\Veff/\sqrt{2})\bigl(\begin{smallmatrix}1&1\\1&-1\end{smallmatrix}\bigr)$.
The algebra is identical to the symmetric case:
$\frac{\Veff}{\sqrt{2}}\cdot\frac{1}{\sqrt{2}}
\bigl[\begin{smallmatrix}1&1\\1&-1\end{smallmatrix}\bigr]^2
=\frac{\Veff}{2}\cdot2\mathbb{I}=\Veff\mathbb{I}$.
The full rotated Hamiltonian in the
$(\tilde{d}_+,\tilde{d}_-,c_{k+},c_{k-})$ basis is:
\begin{equation}
\Htilde(k)=
\begin{pmatrix}
\teps+\Deff & i\Gpt & \Veff & 0\\
i\Gpt & \teps-\Deff & 0 & \Veff\\
\Veff & 0 & \varepsilon_{c,+}(k) & 0\\
0 & \Veff & 0 & \varepsilon_{c,-}(k)
\end{pmatrix}.
\label{eq:Htilde}
\end{equation}
Here $\tilde d_+$ ($\tilde d_-$) hybridizes only with $c_{k+}$ ($c_{k-}$),
while the internal $i\Gpt$ term can still convert the two asymptotic
channels.  The frozen impurity eigenvalues are
$E_\pm=\teps\pm s$, $s=\sqrt{\Deff^2-\Gpt^2}$; on the audited path
$\Deff=r^2D$, $\Gpt=r^2G$, $D=0.10$, and $G=0.20\gb$.
The plotted eigenvalues and condition numbers nevertheless use the full
$4\times4$ kernel in Eq.~(\ref{eq:Htilde}).  Eliminating its bath block gives
\begin{equation}
\mathcal P(z,k)=\mathfrak p_+(z,k)\mathfrak p_-(z,k)+\Gpt^2=0,
\label{eq:full4_schur}
\end{equation}
where
\begin{equation}
\mathfrak p_\eta(z,k)=z-\teps-\eta\Deff
-\frac{\Veff^2}{z-\varepsilon_{c,\eta}(k)},
\qquad \eta=\pm .
\label{eq:full4_schur_aeta}
\end{equation}
The full quadratic-kernel EP condition is
\begin{equation}
\mathcal P(z_{\rm EP},k)=0,
\qquad
\partial_z\mathcal P(z_{\rm EP},k)=0 .
\label{eq:full4_EP_condition}
\end{equation}
At $k=0$ the bath energies coincide, so with
$\zeta(z)=z-\teps-\Veff^2/z$ this reduces to
\begin{equation}
\mathcal P(z,0)=\zeta(z)^2-\left(\Deff^2-\Gpt^2\right).
\label{eq:k0_common_shift}
\end{equation}
Thus the impurity-related double root occurs at $\Deff^2=\Gpt^2$ in this
common-shift limit; SOC-resolved finite-$k$ asymmetry generally unfolds it.
The declared nonzero-$k$ scan finds no additional interior EP because every
minimizer lies at its upper boundary.

At $\gb=0$, $G=0$ but $D=0.10$, so the origin is a normal, coherently split
impurity doublet and the full $k=0$ eigenvector condition number is one.
The finite-drive local EP is at $|G|=|D|$, hence $|\gb|=0.50$.  In the
no-flip control, a normal-matrix guard fixes the arbitrary eigenbasis at the
ordinary origin degeneracy and prevents a false condition-number spike.

The operator-resolved finite-$U$ scalar-chain calculation is a distinct
Markov-core residue audit: it preserves the nilpotent direction in its
validated early-iteration window but supports no universal splitting exponent
or quartic coefficient (Supplemental Fig.~S7).  The interacting continuation
of the nonlinear full-$k$ pole EP is instead the matrix block-Wilson
complete-basis calculation in Fig.~\ref{fig:finiteU_fullk_exact}; the two are
not identified.  The scalar local interaction protects the EP only on the
equal-velocity, scalar-hybridization linearized submanifold; deviations in the
complete driven problem must be evaluated explicitly.

\section{Slave-Boson Mean Field and Keldysh Self-Consistency}
\label{sec:mf}

The effective impurity problem originates from a correlated Anderson
level with local repulsion $U$.  We formulate the constrained low-energy
sector using the $U\to\infty$ large-$N$ slave-boson mean-field
saddle~\cite{coleman1984mixed,read1983solution,SBkeld}.  The finite-$U=2$
Schrieffer--Wolff formulas used later are a separate frozen analytic estimate.
This section introduces the spinon operators
$f_\sigma^\dagger$ that replace the physical $d_\sigma^\dagger$
after the slave-boson decoupling, derives the self-consistent
equations for the condensate amplitude $r=|b_c|$ and Lagrange
multiplier $\delta_c$, and formulates the shifted resonance
condition in gauge-invariant language.

\subsection{Slave-Boson Representation}

To handle the Coulomb repulsion $U$ in the Keldysh framework,
we use the large-$N$ slave-boson (SBMF)
representation~\cite{coleman1984mixed,read1983solution,SBkeld}:
\begin{equation}
d^\dagger_\sigma = f^\dagger_\sigma b,
\qquad
b^\dagger b + \sum_\sigma f^\dagger_\sigma f_\sigma = Q.
\label{eq:sb_rep}
\end{equation}
Here $f^\dagger_\sigma$ is the auxiliary spinon fermion carrying
spin $\sigma$ (the physical quasiparticle in the Kondo limit),
and $b^\dagger$ is the slave boson enforcing no double occupancy.
The parameter $Q$ fixes the constraint normalization: $Q=1$ gives the
physical spin-$1/2$ infinite-$U$ constraint, while $Q=Nq_{\rm occ}$ denotes the
large-$N$ or channel-rescaled normalization.  All saddle-point
amplitudes quoted below are interpreted in the same normalization.
The constraint is imposed via a Lagrange multiplier $\delta_c$.
At the saddle point $b\to b_c\in\mathbb{C}$, the phase of
$b_c$ is an internal $U(1)$ gauge choice.  We therefore work in the
gauge-fixed convention in which observables depend on the amplitude
$r\equiv|b_c|$, and the spinon hybridization scale is $rW_k$.

The renormalized impurity level and gauge-invariant SBMF scale are
then
\begin{align}
\teps &=\epsilon_\xi+\delta_c+\dSig,
\label{eq:teps}\\
\tb &= |\gb|\,|b_c|=|\gb|\,r\qquad
\label{eq:tbeta}
\end{align}
Here $\dSig$ is the real principal-value one-body shift of the
projected impurity level; if this shift is absorbed into the bare level,
one may set $\dSig=0$ and reinterpret $\epsilon_\xi$ accordingly.
The projected channels entering $\Hss$ are defined by
Eq.~(\ref{eq:projected_channels}); in particular, the hybridization is
$\Veff=rW$ rather than $c_V\tb$.  The equations below define the
large-$N$ saddle.  For the fixed-$W$ small-width path of
Eq.~(\ref{eq:smallW_path}), a seed-independent finite-$r$ solution is found
at $T_{\rm b}=10^{-6}$ throughout $0.35\leq\gb\leq0.55$: $r$ decreases
smoothly from $0.0106078$ to $0.00949995$, with maximum constraint and
stationarity residuals $6.54\times10^{-8}$ and
$4.86\times10^{-10}$, respectively.  The maximum grid-refinement error is
$1.71\times10^{-6}$, and charge continuity is satisfied to
$3.66\times10^{-23}$.  At the manuscript reference temperature
$T_{\rm b}=0.1$, however, the same $U\to\infty$ closure has no finite-$r$
branch and terminates at the $r=0$ boundary.  The spectra displayed below
therefore use stated frozen values and are not used to claim a
finite-temperature enhancement of $r$.  The complete convergence and
seed-independence audit is given in the Supplemental Material.

\subsection{Saddle-Point Equations and Gauge-Invariant Amplitude}

The Keldysh saddle-point conditions follow by varying the complete contour
action with respect to $r=|b_c|$ and $\delta_c$ (Supplemental Material):
\begin{align}
0&=r^2+n_f-Q,\qquad
n_f=-i\int\frac{d\omega}{2\pi}{\rm Tr}_sG_{ff}^<,
\label{eq:saddle2}\\
0&=2r\delta_c+\mathcal F_W+2rD_0\,\mathcal X
+2r\mathcal I_\Gamma .
\label{eq:saddle1}
\end{align}
The hybridization force $\mathcal F_W$ is linear in $rW$; $\mathcal X$ is the
coherent-channel force associated with the unrenormalized projected
coefficient $D_0(\beta_0)$; and $\mathcal I_\Gamma$ is the full retarded/advanced/
lesser influence-kernel force.  Its Langreth form contains both
$K_\Gamma^RG^<$ and $K_\Gamma^<G^A$; dropping the latter is not a valid
thermal saddle.  In the chiral basis the relative dissipative kernel is
off-diagonal, so $G^<_{+-}$ and $G^<_{-+}$ cannot be discarded.  The
Supplemental Material gives the explicit expressions and the
no-double-counting rule.

A central point, emphasized to avoid any gauge ambiguity, is that
the physical amplitude $r=|b_c|$ is gauge-invariant, whereas the
phase of $b_c$ is an internal U(1) slave-boson gauge degree of
freedom ($b\to be^{i\theta}$, $f_\sigma\to f_\sigma e^{-i\theta}$).
Consequently no observable conclusion is drawn from the phase of
$b_c$.  All spectra, condition numbers, and scale estimates below use
the gauge-invariant amplitude $\tb=|\gb| r$ for the symmetric
finite-drive diagnostics.  If a numerical saddle is
represented with a complex $b_c$, that phase is only a convention for
the auxiliary fields and is not used as a physical source of the
resonance shift.

\subsection{Lesser Green's Function and Thermal Occupation}

The physical lesser Green's function is fixed by the thermal bath.
For a bilinear impurity problem coupled to a reservoir at temperature
$T_{\rm b}$, the Keldysh construction gives
\begin{align}
\Gless_{\rm bath}(\omega)
&=\Gret(\omega)\,\Sigma^<_{\rm bath}(\omega)\,\Gadv(\omega)
\notag\\
&=f(\omega)\,[\Gadv(\omega)-\Gret(\omega)] .
\label{eq:Gless_bath}
\end{align}
with $f(\omega)=[e^{\omega/T_{\rm b}}+1]^{-1}$.
This bath-controlled object is the $G^<$ used for the physical
occupations, with $n_f=-i\,{\rm Tr}\,G^<(t,t)$.  Equivalently,
$\Sigma^<_{\rm bath}=if\Gamma_{\rm bath}$ and
$i(G^R-G^A)=G^R\Gamma_{\rm bath}G^A$.  This sandwich form retains the
complete channel matrix and introduces no separate biorthogonal
normalization factor.

\subsection{Shifted Friedel-Type Resonance Condition}
\label{sec:FSR}

In the standard Hermitian slave-boson mean field at particle-hole
symmetry ($\epsilon_\xi=-U/2$), the Kondo resonance is pinned to the
Fermi level, $\omega_{\rm res}=0$, because the two spin occupations are
equal and the saddle point fixes $\teps=0$.
In the present driven problem the coarse-grained self-energy produces
spin-selective imaginary shifts
$x_\sigma=\zeta_\sigma\Gpt$ and therefore different effective
widths in the two spin channels.
The physical resonance position is then most conservatively described
by the gauge-invariant condition
\begin{equation}
\omega_{\rm res}\simeq \mathrm{Re}(\Eep)\simeq\mathrm{Re}(\teps),
\label{eq:omega_K}
\end{equation}
valid when the low-energy line shape remains approximately
Lorentzian.  This is a shifted resonance condition rather than a
claim that the gauge-dependent phase of $b_c$ is observable.
The shift is controlled by the renormalized level $\teps$, the
spin-dependent widths, and the self-consistent amplitude
$\tb=|\gb| |b_c|$.
In the strongly non-Lorentzian regime the
observed peak need not coincide exactly with this reference, and we
therefore treat the deviation as a line-shape effect rather than as a
violation of a sum rule.  In the frozen kernel used for the plotted
spectra, the local-moment reference has
$\mathrm{Re}(\teps)\simeq -1.0$ for
$\epsilon_\xi=-U/2=-1.0$; a fully self-consistent finite-temperature
SBMF closure, when a finite-$r$ branch exists, would replace this by its
computed saddle-point value.

\subsection{Near-EP Analytic Screening-Scale Estimate}

Near an isolated EP, the individual spectral projectors grow as the left and
right eigenvectors coalesce, while the complete resolvent remains the finite
Jordan expression derived in Sec.~\ref{sec:complete_resolvent}.  We use the
condition number $\keff$ only as a diagnostic of nonnormality and EP distance.
The screening estimate is controlled instead by the eigenvalues of the
complete channel-projected Schrieffer--Wolff matrix.
We write
\begin{align}
T_K^{\rm est}
&=D_{\rm uv}\exp\left[-\frac{1}{\chi_K j_{\rm scr}}\right],
\label{eq:TKep}\\
j_{\rm scr}
&=\max\left\{\operatorname{Re}j_\alpha:\;
 j_\alpha\in\operatorname{spec}(\rho_bJ^{\rm bio}),\;
 \operatorname{Re}j_\alpha>0\right\}.
\label{eq:j_scr_def}
\end{align}
Here $j_\alpha$ are the eigenvalues of the dimensionless biorthogonal
exchange matrix $\rho_bJ^{\rm bio}$, and only positive-real-part eigenvalues
are retained as antiferromagnetic screening channels.  The coefficient
$\chi_K>0$ fixes the conventional scaling normalization; all plotted estimates
use $\chi_K=1$.  A scale increase is possible only if the complete EP-active
channel calculation increases such a positive screening eigenvalue.  No
individual projector residue is inserted.  By contrast, a hybridization
induced EP generated by a negative scalar product $V^R V^L<0$ can
produce strong spectral nonnormality and a frozen non-Hermitian
crossover scale, but it should not automatically be identified with an
enhanced thermodynamic Kondo temperature.  Supplemental Material
gives the corresponding frozen-model scale estimate and explains why
the spin-selective sign $x_\uparrow=-x_\downarrow$ alone does not split
the scale unless the channel widths, real detunings, or exchange
eigenvalues are unequal.

\section{Spectral Properties and Physical Regimes}
\label{sec:spectra}

\begin{figure*}[t]
\centering
\includegraphics[width=0.98\textwidth]{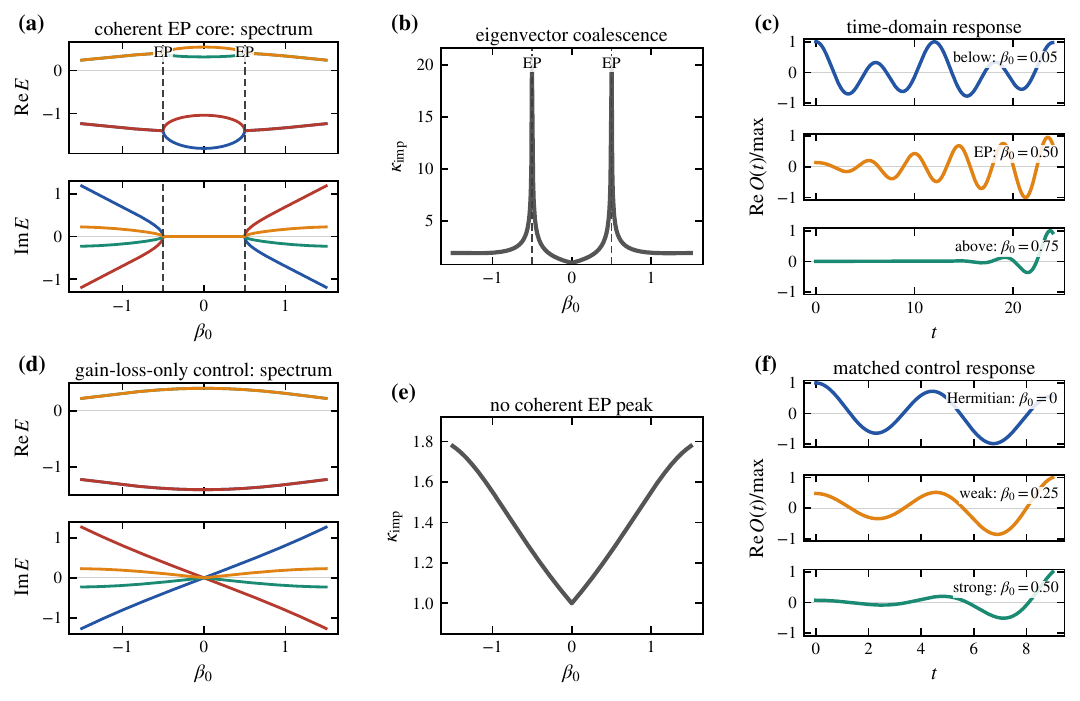}
\caption{%
\textbf{Frozen $4\times4$ compensated-core EP and matched control.}
All panels use $r=1$, $\teps=-1$, fixed scalar hybridization $W=0.75$, and
$G=\Gamma_{\rm PT}=|\beta_0|$ at $k=0$; hence no physical DOS, cutoff,
temperature, or interaction parameter enters.  Panels (a)--(c) use the
coherent splitting $D=\Delta_{\rm eff}=0.50$.  (a) Real and imaginary parts
of all four eigenvalues.  The exact local-core EPs are at
$\beta_0=\pm0.50$.  (b) The Euclidean right-eigenvector condition number.
The plotted spacing $\Delta\beta_0=0.0075$ places its nearest positive sample
at $0.5025$, where $\kappa_{\rm imp}=20.0304$; the exact origin remains
$\kappa_{\rm imp}(0)=1$.  (c) Separately stacked survival-amplitude traces
below, at, and above the EP avoid curve and label overlap.  Panels (d)--(f)
repeat the same tests after switching off the coherent splitting ($D=0$).
The smooth condition number in (e) and the matched traces in (f) form the
relative-gain--loss-only control; they are not interpreted as a coherent
impurity EP or as a Kondo-scale enhancement.
}
\label{fig:frozen_comparison}
\end{figure*}

Numerical conventions are stated separately for each calculation.  The
normalized Fig.~\ref{fig:frozen_comparison} tests only the constant
$k=0$ kernel.  The saddle uses $T_{\rm b}=10^{-6}$ (with a negative control
at $T_{\rm b}=0.1$), whereas the causal scattering/RG gate uses
$D=0.10$, $G=0.20\gb$, and $\bar\Gamma=0.12$.  No cutoff, temperature, or
interaction value from one calculation is silently transferred to another.

Figure~\ref{fig:frozen_comparison} displays the dynamical fingerprints of
the exact compensated core and its relative-gain--loss-only diagnostic limit.
The representative coefficients place a symmetric local EP pair at
$|\gb|=0.5$,
accompanied by a condition-number peak and by a crossover from
oscillatory to overdamped local dynamics.
Removing the coherent splitting channel
($\Deff=0$, Fig.~\ref{fig:frozen_comparison}(d)--(f))
removes the coherent local impurity EP mechanism.  The remaining
passive relative gain--loss dynamics is retained only as a diagnostic contrast:
it should not be interpreted as a separate positive-screening Kondo
scale enhancement.

\paragraph{Condition-number convention.}
All condition numbers shown below are computed from the Euclidean
right-eigenvector matrix of the full $4\times4$ channel-projected
kernel at the plotted $k$ slice, using a single normalization
convention for all values of \(\gb\).  We use
\(\keff\) to locate finite-\(|\gb|\) eigenvector coalescence and to
track the distance to the EP.  Its absolute magnitude is not used as a
standalone physical observable or as an independent definition of the
Kondo scale; the physical diagnostics are the spectral line shapes,
low-energy scale estimates, occupied fluctuation--dissipation-relation (FDR)-weighted spectra, and frozen
channel-projected amplitudes extracted with the same gauge-invariant
convention \(\tb=|\gb||b_c|\).

\subsection{Causal Spectral Concentration and Weak-Flow Control}

The complete causal channel density and the channel-resolved weak-coupling
flow provide supporting tests of the analytic construction; the full panels
and adverse controls are shown in Supplemental Figs.~S4 and S5.  For
$\beta_0=0.50$, the active/control ratios are $1.0567$ at $g_\ast=0.30$,
$1.0179$ at $g_\ast=0.40$, $0.9893$ at $g_\ast=0.50$, and $0.9166$ at
$g_\ast=1$.  Thus the larger leading channel-density eigenvalue and early RG
trajectory are bounded early-flow effects in the stated window, but they do not
support the stronger statement of an enhanced asymptotic Kondo temperature.
The matrix density is positive in the retained
passive window, and the result is stable under RG-step refinement.  A
rate-matched off-EP control can exceed the EP in the largest proper-delay
peak, so the Smith delay is not counted as independent evidence of
enhancement; this control is shown in the Supplemental Material.

\section{Fluctuation-Dissipation Relation and Spectral Functions}
\label{sec:FDR}

For the complete reservoir-coupled kernel, the physical lesser Green's
function is fixed by the bath self-energy.  Supplemental Fig.~S6 verifies
that the impurity and bath fluctuation--dissipation relation (FDR) ratios
follow the same thermal Fermi function on spectral support.  No
mode-by-mode occupation of the non-Hermitian eigenvectors is used in the
physical FDR, in the occupations, or in the saddle-point equations.

\paragraph{Bath-based construction (physical).}
The physical lesser Green's function is fixed by the thermal bath
self-energy.  Here this is an instantaneous-equilibrium closure at leading
adiabatic order.  Corrections are of order
$O[(\Omega/\epsilon_O)^2]$ (and of the corresponding gradient order) and
would describe drive-induced energy redistribution; no pumping statement is
made at the retained order.  For a bilinear impurity--bath coupling,
\begin{equation}
\Gless_{\rm phys}(\omega)
=
\Gret(\omega)\,\Sigma^<_{\rm bath}(\omega)\,\Gadv(\omega).
\label{eq:Gless_physical_fdr}
\end{equation}
If the bath is thermal, then
\begin{align}
\Sigma^<_{\rm bath}(\omega)
&=if(\omega)\Gamma_{\rm bath}(\omega)
\nonumber\\
&=f(\omega)
\left[\Sigma^A_{\rm bath}(\omega)-\Sigma^R_{\rm bath}(\omega)\right].
\label{eq:bath_selfenergy_fdr}
\end{align}
Combining this with the Dyson identity for the bath-induced
broadening gives
\begin{equation}
\Gless_{\rm phys}(\omega)
=
f(\omega)\,[\Gadv(\omega)-\Gret(\omega)].
\label{eq:impurity_fdr_inherited}
\end{equation}
Thus the physical impurity FDR is inherited from the reservoir and is
the $G^<$ used in the SBMF self-consistency.  Charge conservation is checked
through the steady-state collision integral
\begin{equation}
\mathcal I_N=\int\frac{d\omega}{2\pi}\,
{\rm Tr}\!\left[\Sigma^<G^>-\Sigma^>G^<\right]=0,
\label{eq:number_continuity}
\end{equation}
which vanishes for the same complete thermal bath self-energy.  This is a
nontrivial number-continuity check.

\paragraph{Impurity spectral function.}
For the complete physical kernel, $\Gadv=(\Gret)^\dagger$ and the
spectral-function matrix is
\begin{equation}
\begin{aligned}
\mathcal A(\omega)&=i[\Gret(\omega)-\Gadv(\omega)]\\
&=\Gret(\omega)\,\Gamma_{\rm tot}(\omega)\,\Gadv(\omega),
\qquad \Gamma_{\rm tot}\succeq0 .
\end{aligned}
\label{eq:DOS}
\end{equation}
With $P_{\rm imp}$ the impurity projector, the physical impurity density of
states is therefore
\begin{equation}
\rho_{\rm imp}(\omega)
=\frac{1}{2\pi}\mathrm{Tr}[P_{\rm imp}\mathcal A(\omega)]
=-\frac{1}{\pi}\mathrm{Im}\,\mathrm{Tr}[P_{\rm imp}\Gret(\omega)].
\label{eq:physical_DOS}
\end{equation}
We verify the equality of these two evaluations and positivity numerically.
A signed response of the compensated relative kernel, obtained after common
damping is subtracted, is not interpreted as a physical density of states.

\paragraph{Gauge convention for spectral functions.}
The physical self-energy from the coarse-graining is first decomposed into
a common bath shift/damping, the passive $\PT$ core of Eq.~(\ref{eq:passive_PT_core}),
and the Kramers--Kronig detuning of Eq.~(\ref{eq:relative_detuned_kernel}).
The off-shell auxiliary sector supplies the real spin-odd shift, while the
retarded SOC matrix bath converts this shift into the relative
spin-dependent width $\Gamma_{\rm PT}$ through
Eq.~(\ref{eq:matrix_gamma_pt}).  The
effective Hamiltonian is therefore written in terms of the real,
gauge-invariant amplitude $r=|b_c|$ and the projected channels
$\Gpt=r^2\Gamma_{\rm PT}^{(0)}(\gb)$,
$\Deff=r^2\Delta_0^{(0)}$, and $\Veff=rW$.
The phase of the auxiliary slave-boson field is not used separately
in spectral calculations.  With this convention, the same $\Hss$ is
used for eigenvalues, condition numbers, and spectral functions, and
no observable is attributed to the gauge-dependent phase of $b_c$.

\section{Real-Time Dynamics}
\label{sec:dynamics}

The real-time diagnostic is the complex survival amplitude
\begin{equation}
O(t)=\langle d_\uparrow|e^{-iHt}|d_\uparrow\rangle ,
\label{eq:survival_amplitude}
\end{equation}
evaluated for the same gauge-fixed frozen kernel used in the spectral
plots.  Main-text Fig.~\ref{fig:frozen_comparison}(c),(f) shows
$\mathrm{Re}\,O(t)$ as separated small multiples; the associated impurity
Bloch trajectories and norm check are retained in Supplemental Fig.~S1.
This is a non-unitary
propagator diagnostic of the local EP; it is not used as a normalized
thermal expectation value.  For Fig.~\ref{fig:frozen_comparison}(c) we use
$\teps=-1$, $D=0.50$, fixed $W=0.75$, and
$\beta_{0,{\rm core}}^{\rm EP}=0.50$.  The evolution shows a direct dynamical
signature of the EP:

\begin{itemize}
\item \emph{Below EP} ($\gb<\beta_{0,{\rm core}}^{\rm EP}$, local splitting real):
  eigenvalues are non-degenerate; the evolution shows underdamped
  oscillations with a frequency set by $\mathrm{Re}(E_+-E_-)=2\mathrm{Re}\,s$
  and a decay rate set by $\mathrm{Im}(E_\pm)$.

\item \emph{Above EP} ($\gb>\beta_{0,{\rm core}}^{\rm EP}$, local splitting imaginary):
  eigenvalues have opposite imaginary parts; the gain mode
  dominates and the evolution shows overdamped decay.

\item \emph{Without coherent splitting}
  ($\Deff=0$, Fig.~\ref{fig:frozen_comparison}(d)--(f)):
  the isolated relative gain--loss block has eigenvalues
  $\teps\pm i\Gpt$ and is therefore in the broken sector for
  $\Gpt\neq0$.  The plotted bounded trajectories are normalized or
  co-decaying diagnostics of the full bath-coupled kernel, not evidence
  for conservation of the Euclidean norm.
\end{itemize}

The contrast between the matched no-flip diagnostic and the separated
time-domain traces with coherent splitting demonstrates that both projected
impurity channels of $\Htilde$---the coherent splitting $\Deff$ and the
relative decay channel $i\Gpt$---are necessary for the full finite-drive
EP structure.  In the channel-projected parametrization, these channels are measured
relative to the common scale $\tb$ but should be viewed as distinct
projected self-energy components.  Equality of their coefficients is a
symmetric representative limit, not a symmetry requirement.

\section{Causal Scattering and Linearized Contact Algebra}
\label{sec:scattering}
\label{sec:BA}
\label{sec:complete_resolvent}

\subsection{Complete scattering and Jordan response}

We use the standard Fisher--Lee and Wigner--Smith constructions
only as causal scattering diagnostics of the projected kernel
~\cite{fisherlee1981,wigner1955,smith1960}; they are not introduced here as
new formalisms.  The basis labels must be kept explicit.  In the original
spin basis $\Psi_s=(d_\uparrow,d_\downarrow)^T$, the compensated Markov
kernel and its positive rate matrix are
\begin{align}
h_s&=\Delta_{\rm eff}\sigma_x+i\Gamma_{\rm PT}\sigma_z
-i\bar\Gamma I,\nonumber\\
\Gamma_s&=i(\Sigma_s^R-\Sigma_s^A)
=2(\bar\Gamma I-\Gamma_{\rm PT}\sigma_z)\succeq0 .
\label{eq:Gamma_bath_scattering}
\end{align}
The Hadamard matrix $U_H$ that defines the fixed chiral fields
$\Psi_\chi=U_H^\dagger\Psi_s$ gives
\begin{equation}
\begin{aligned}
h_\chi&=U_H^\dagger h_sU_H\\
&=\Delta_{\rm eff}\sigma_z+i\Gamma_{\rm PT}\sigma_x-i\bar\Gamma I,\\
\Gamma_\chi&=U_H^\dagger\Gamma_sU_H\\
&=2(\bar\Gamma I-\Gamma_{\rm PT}\sigma_x).
\end{aligned}
\label{eq:spin_chiral_scattering_map}
\end{equation}
For either representation $\nu=s,\chi$,
\begin{equation}
S_\nu(\omega)=I-i\Gamma_\nu^{1/2}G_\nu^R(\omega)
\Gamma_\nu^{1/2},
\qquad
S_\chi=U_H^\dagger S_sU_H ,
\label{eq:Fisher_Lee}
\end{equation}
and, in the passive window, $S_\nu$ is unitary.  The Wigner--Smith matrix
\begin{equation}
Q_\nu(\omega)=-iS_\nu^\dagger(\omega)\partial_\omega S_\nu(\omega)
\label{eq:Smith_Q}
\end{equation}
is Hermitian and transforms covariantly.  Thus its trace and eigenvalues,
like the eigenphases of $S$, are basis invariant.  Individual entries such
as $S_{+-}$ are representation dependent and are not used as evidence for
enhancement.  The basis-resolved Green functions and coefficients are
worked out in the Supplemental Material.

For the compensated unbroken Markov core,
$s=\sqrt{\Delta_{\rm eff}^2-\Gamma_{\rm PT}^2}\in\mathbb R$ and the
two causal scattering phases give
\begin{align}
\operatorname{Tr}Q(\teps)
&=\frac{4\bar\Gamma}{s^2+\bar\Gamma^2},\nonumber\\
\frac{\operatorname{Tr}Q_{\Gamma_{\rm PT}}(\teps)}
{\operatorname{Tr}Q_{0}(\teps)}
&=\frac{\Delta_{\rm eff}^2+\bar\Gamma^2}
{\Delta_{\rm eff}^2-\Gamma_{\rm PT}^2+\bar\Gamma^2}>1 .
\label{eq:Smith_preep_ratio_main}
\end{align}
Thus, at fixed $\Delta_{\rm eff}$ and $\bar\Gamma$, the invariant scattering
phase density is analytically concentrated at the resonance center as the
EP is approached.  Its integrated weight remains fixed, and linewidth- or
rate-matched adverse controls need not share the peak ordering.  We call
Eq.~(\ref{eq:Smith_preep_ratio_main}) path-specific spectral concentration,
not an enhanced thermodynamic Kondo temperature or a basis-independent
amplification factor.

The cancellation responsible for its regularity is an exact matrix identity.
For the traceless core
$h=\Delta_{\rm eff}\sigma_x+i\Gamma_{\rm PT}\sigma_z$, $h^2=s^2I$, write
$a=z-\epsilon_0-\Sigma_0^R(z)$.  Although the individual biorthogonal
projectors $P_\pm=(I\pm h/s)/2$ grow as $1/s$, their complete resolvent is
\begin{equation}
G^R(z)=\frac{P_+}{a-s}+\frac{P_-}{a+s}
=\frac{aI+h}{a^2-s^2}.
\label{eq:complete_resolvent}
\end{equation}
At the EP, $h_{\rm EP}^2=0$, and
\begin{equation}
(aI-h_{\rm EP})^{-1}
=\frac{I}{a}+\frac{h_{\rm EP}}{a^2},
\label{eq:jordan_resolvent_main}
\end{equation}
so the coalescence produces a finite-coefficient simple-plus-double-pole
Jordan response, with no additional $1/s$ normalization multiplier.  This
cancellation is proved by Eq.~(\ref{eq:complete_resolvent}); it is not inferred
from NRG.  An independent weak-coupling finite-$U$ biorthogonal NRG audit
resolves the
same matrix direction in branch-tracked transition residues: the channel
trace is strongly suppressed while the channel-odd and symmetric
off-diagonal components retain the nilpotent Jordan structure.  This is a
matrix-residue and physical-sum-rule consistency test, not a second proof of
the algebraic cancellation or a universal interaction-induced pole-splitting
law.  For the compensated Markovian kernel,
\begin{align}
\operatorname{Tr}Q_{\rm EP}(\omega)
&=\frac{4\bar\Gamma}
{(\omega-\operatorname{Re}E_{\rm EP})^2+\bar\Gamma^2},
\nonumber\\
\int\frac{d\omega}{2\pi}\operatorname{Tr}Q_{\rm EP}&=2 .
\label{eq:Smith_EP}
\end{align}
The numerical implementation satisfies scattering unitarity to
$1.33\times10^{-15}$ and the trace-delay/determinant identity to
$4.97\times10^{-14}$.  A rate-matched off-EP control can nevertheless have a
larger maximum proper delay, so this observable diagnoses causal pole
coalescence but is not, by itself, evidence of many-body enhancement.

The same cancellation occurs in the finite-$U$ virtual-charge sum.  At
particle--hole symmetry, with $A=U/2$ and a channel-independent bare vertex
$W$,
\begin{align}
J_{\rm SW}(s)
&=2W^2\left[(AI-h)^{-1}+(AI+h)^{-1}\right]
\notag\\
&=\frac{4AW^2}{A^2-s^2}I,\qquad
J_{\rm SW}(0)=\frac{8W^2}{U}.
\label{eq:SW_complete_charge}
\end{align}
Thus neither a Green function nor an exchange vertex acquires an independent
Petermann or condition-number multiplier.  This causal scattering statement
does not assume integrability.  We first record its exact one-particle phase
condition and then isolate the additional low-energy assumptions under which
the finite-$U$ contact problem has a genuine multiparticle Yang--Baxter
algebra.

\subsection{Low-energy contact-algebra and screening scope}
\label{sec:frozen_ba_main}

The eigenvalues $\mathsf{s}_a=e^{2i\delta_a}$ of the complete unitary
scattering matrix give the exact one-particle finite-size condition and phase
density
\begin{equation}
\det[I-e^{ik\mathcal L}S(k)]=0,
\qquad
\frac{1}{2\pi i}\partial_\omega\ln\det S
=\frac{1}{2\pi}\operatorname{Tr}Q .
\label{eq:contact_phase_compact}
\end{equation}
This is a basis-independent one-particle statement, not by itself a
multiparticle Bethe ansatz.  The Supplemental Material gives the full
contact matching.  After equal-velocity branchwise linearization, removal of
curvature, freezing of the component matrix, and restriction to scalar
momentum-independent hybridization, the finite-$U$ two-electron contact
matrix is the rational Anderson $R$ matrix in a dressed rapidity.  It obeys
the ordinary Yang--Baxter equation and is $GL(2,\mathbb C)$ covariant; the EP
then enters as an invertible unipotent boundary twist.  The bounded-ring
many-body theorem is proved separately in Ref.~\cite{KulkarniYB2026}.
Curvature, unequal velocities, causal frequency dependence, and non-scalar
projected vertices lie outside that theorem, so no full-momentum
integrability is claimed here.

For comparison with the causal weak-flow calculation, the only screening
estimate retained is the conventional frozen Schrieffer--Wolff form
\begin{equation}
\begin{aligned}
J_a^{\rm SW}&=2|W_a|^2
\left(|\tilde\epsilon_{d,a}|^{-1}
+|\tilde\epsilon_{d,a}+U|^{-1}\right),\\
T_{K,a}^{\rm SW(est)}&\simeq D_{{\rm uv},a}
\exp[-1/(\chi_K\rho_{b,a}J_a^{\rm SW})].
\end{aligned}
\label{eq:TK_SW_est}
\end{equation}
SOC can change this conditional estimate through a positive channel density,
velocity Jacobian, overlap, or ordinary charge denominator.  It does not
supply an independent $1/s$ factor.  Negative or complex exchange
eigenvalues are not identified with Kondo screening.

\section{Scope of the Near-EP Scale Estimate}
\label{sec:kondo}

Near the EP, the condition number
$\keff\sim C/|s|$ measures eigenvector nonorthogonality.  It is useful
for locating the projected-core degeneracy but is not a spectral residue,
exchange vertex, or quasiparticle weight.  The singular parts cancel in the
complete Green function [Eq.~(\ref{eq:complete_resolvent})] and in the
finite-$U$ charge sum [Eq.~(\ref{eq:SW_complete_charge})].  Consequently no
$Z^{\rm bio}$ or Petermann multiplier is inserted into a propagator,
diagrammatic vertex, or Kondo exponent.

On the exact linearized submanifold, the local \(U\) preserves the
complexified pseudospin symmetry and therefore does not cut off this
algebraic Jordan limit.  This does not require a divergent observable: the
complete resolvent has finite Jordan coefficients and the charge sum has no
standalone $1/s$ multiplier.  In the complete driven
kernel, curvature, velocity mismatch, frequency dependence, or
component-dependent vertices can move or split the EP, but their effect must
be calculated microscopically rather than represented by a universal
interaction floor.

Equation~(\ref{eq:TK_SW_est}) retains the conventional frozen
Schrieffer--Wolff scale estimate and
permits a conditional increase relative to a matched Hermitian reference
only when the projected Dirac/SOC channel increases a positive screening
width or exchange eigenvalue.  That conditional estimate is conceptually
distinct both from multiplying a Hermitian scale by a biorthogonal norm and
from the Yang--Baxter identity.

The direct numerical evidence is narrower.  The leading eigenvalue of the
complete causal channel density is larger in the chosen Fermi window, and
the largest exchange eigenvalue initially grows faster than in the matched
Hermitian control.
Supplemental Fig.~S5 shows that this ordering reverses before strong
coupling.  The low-temperature slave-boson saddle in Supplemental Fig.~S2
validates the $r^2$ scaling and charge constraint, but the same closure has no
finite-$r$ branch at $T_{\rm b}=0.1$.  We therefore describe only bounded
spectral concentration and an early-flow ordering on the displayed path.
The estimate in Eq.~(\ref{eq:TK_SW_est}) remains conditional on a positive
screening channel, and no enhanced thermodynamic Kondo temperature is
claimed.

\section{Conclusions}
\label{sec:conc}

We have shown that a strictly Hermitian, periodically driven Dirac impurity
admits an emergent passive non-Hermitian retarded kernel through two linked
steps: off-shell angular harmonics produce a real spin-odd shift, and the SOC
matrix bath converts it into relative decay.  The causal finite-band problem
also clarifies the limits of the constant-core picture.  The normalized
oscillatory-drive benchmark exhibits avoided coalescence, whereas a distinct
stationary rotating-drive construction supports a continuous family of
full-momentum causal pole EPs.  Their double zeros are protected by local
winding and pole exchange.  On the representative finite systems, the
complete-basis matrix block-Wilson calculations preserve this topology at
$U/\bar\Gamma_{\rm self}=1.225$.  Independently reoptimized $N=10$ and $12$
chains share the same pole exchange and zero count, with the common-contour
Rouch\'e ratio $0.6775<1$.  This establishes finite-interaction survival and
finite-$N$ persistence over the stated pair, not a thermodynamic-limit
full-density-matrix NRG fixed point.

At fixed nonzero spectral detuning, the complete response has no standalone
eigenvector-normalization divergence as the EP is approached.  Divergent
individual biorthogonal projectors combine into the finite-coefficient Jordan
simple-plus-double-pole resolvent, and the particle--hole-symmetric finite-$U$
charge resolvent has the same cancellation.  The physical double pole remains;
what is absent is an additional Petermann or condition-number multiplier.
The Wigner--Smith trace therefore
provides a causal pole and linewidth fingerprint rather than an independent
enhancement factor.  A separate operator-resolved NRG audit preserves the
nilpotent residue direction in its validated early-iteration window, while
the weak-flow and saddle calculations establish neither a universal
interaction law nor an enhanced thermodynamic Kondo temperature.

Finally, branchwise linearization identifies a precise integrable
submanifold.  Its finite-$U$ contact matrix is the rational Anderson matrix
in a dressed rapidity and is $GL(2,\mathbb C)$ covariant.  Thus the off-EP
metric similarity preserves the YBE, while the EP is represented by an
invertible unipotent boundary twist.  Curvature and the energy-dependent
driven self-energy lie outside this theorem.  Accordingly, neither
full-momentum Yang--Baxter integrability nor inheritance of the Markov-core
many-body Jordan amplification by the causal pole EP is asserted.

\section*{Code and data availability}
The numerical scripts and tabulated data used to generate the main and
Supplemental figures are included in the accompanying reproducibility archive.
The archive contains independent checks of the finite-$U$ contact algebra,
the rotating-drive two-cut EP and SOC continuation, and the complete-basis
finite-chain topology used in Fig.~\ref{fig:finiteU_fullk_exact}.  It also
contains the equation-of-motion identity, kept-space calibration, branch
tracking, no-truncation regression, the $N=10\to12$ common-contour Rouch\'e
audit, and scripts needed to reproduce the main and Supplemental figures.
A version-controlled copy of the code, data, figure outputs, and manuscript
sources is available in the GitHub repository
\href{https://github.com/VMKPHYSMATH/emergent-pt-dirac-impurity}{\texttt{VMKPHYSMATH/emergent-pt-dirac-impurity}}~\cite{kulkarni2026code}.
The exact public snapshot used for this work is Release \texttt{v2.1.0},
archived at
\href{https://doi.org/10.5281/zenodo.21888925}{doi:10.5281/zenodo.21888925};
the all-versions archival record is
\href{https://doi.org/10.5281/zenodo.21434682}{doi:10.5281/zenodo.21434682}.
Release \texttt{v2.1.0} includes the full-momentum certificate, SOC
continuation, finite-$U$ block-Wilson outputs, and finite-chain convergence
gates reported in this revision.

\section*{Supplemental Material}
Detailed derivations of the auxiliary-mode influence functional, Keldysh
mean-field equations, physical fluctuation--dissipation construction, complete
biorthogonal and finite-$U$ charge-resolvent identities, causal matrix-RG
validation, the finite-$U$ two-electron contact algebra, and parameter tables
are provided in the Supplemental Material~\cite{suppmat}.
The main text remains self-contained by stating the effective projected kernel,
physical spectrum, and claim boundary explicitly.

\begin{acknowledgments}
The author thanks colleagues and workshop participants for useful
discussions on driven quantum impurity systems, non-Hermitian
many-body physics, and Keldysh methods.
\end{acknowledgments}

%

\end{document}


\title{Supplemental Material for ``Emergent $\PT$ Symmetry and Exceptional Points in a Driven Dirac Impurity''}
\author{Vinayak M.~Kulkarni}
\affiliation{Theoretical Sciences Unit,
  Jawaharlal Nehru Centre for Advanced Scientific Research,
  Bangalore 560064, India}
\date{August 11, 2026}
\maketitle

This Supplemental Material provides the technical derivations and validation
checks that support the self-contained main article.  It includes the
auxiliary-mode influence functional, Keldysh mean-field equations, physical
fluctuation--dissipation relation (FDR) construction, number continuity,
shifted-resonance condition, the complete
biorthogonal and finite-$U$ charge resolvents, one-particle scattering-phase
checks, the controlled branchwise linearization, the exact finite-$U$
two-electron contact algebra, the full-momentum two-cut nonlinear-pole
exceptional-point (EP) certificate, its finite-\(U\) matrix block-Wilson
continuation, and numerical
validation figures.
Throughout, $\PT$ denotes parity--time symmetry.

\section*{Notation and parameter conventions}
\begin{description}[leftmargin=1.6cm,style=nextline]
\item[$t_{\rm av},\tau,T_{\rm b}$]
Wigner center time $(t+t')/2$, relative time $t-t'$, and reservoir temperature.
\item[$\eta,\zeta_\sigma$]
Chiral bath index $\eta=\pm$ and spin sign $\zeta_\uparrow=+1$, $\zeta_\downarrow=-1$.
\item[$\lambda,j_\alpha$]
Spin--orbit coupling (SOC) strength and eigenvalues of the dimensionless exchange matrix $\rho_bJ^{\rm bio}$; $j_{\rm scr}$ is the largest positive real part.
\item[$\alpha_{\rm b},k_{\rm max},D_{\rm uv}$]
Quadratic bath-dispersion coefficient, radial momentum cutoff, and ultraviolet
energy cutoff.  The normalized $k=0$ core figures contain neither cutoff;
$k_{\rm max}=\pi/4$ is used in the causal gate and
$D_{\rm uv}=\pi/4$ only in the sharp-cutoff reachability audit.
\item[$W_k,W,\Lambda_\eta^{\rm PV}$]
Static bare impurity--bath hybridization and principal-value bath shift.
The drive does not set $W_k$; at the saddle the vertex is $rW_k$.
\item[$\beta_0,\tilde\beta$]
Signed frozen drive-control amplitude and nonnegative slave-boson mean-field
(SBMF) projection $\tilde\beta=|\beta_0||b_c|$.  On the audited path
$D=0.10$ and $G=0.20\beta_0$, so $\beta_0$ is not itself a decay rate.
The local core and the SOC-degenerate $k=0$ bath-dressed condition coincide
exactly at $\beta_{0,{\rm core}}^{\rm EP}=0.50$.  The nonzero-$k$ root search
is boundary-limited at $\beta_0=0.55$ and establishes no interior projected
EP.
\item[$s$]
Impurity pole splitting; $\zeta_\sigma$ is used for the spin sign.
\item[$\mathsf{s}_a,\delta_a$]
Eigenvalues $\mathsf{s}_a=e^{2i\delta_a}$ and eigenphases of the complete
unitary scattering matrix.  They are distinct from the EP pole splitting
$s$ and from the fixed chiral labels $\eta=\pm$.
\item[$D,G,\Gamma_{\rm PT},\bar\Gamma$]
Constant-core coherent splitting, constant-core relative channel, projected
relative decay imbalance, and common bath width.  The symbols
$\tilde\Gamma_\eta$ and $\Gamma_\sigma^{\rm net}$ denote, respectively, the
chiral frozen-kernel width and the physical net channel width.
\item[$\omega,\Omega,\omega_0$]
Fourier frequency, drive frequency, and a selected spectral pole or resonance frequency; none of these denotes temperature.
\item[$\mathcal E_n,E_{\rm EP},\Lambda(\omega)$]
Non-Hermitian eigenvalues, EP eigenvalue, and real bath self-energy.
\item[$\mathcal P,\mathfrak a,\mathfrak b$]
Full-kernel Schur polynomial and the two local contact denominators; these are distinct from the bath coefficient $\alpha_{\rm b}$.
\item[$\mathcal F_{\rm dist},\mathcal F_{\rm MF}$]
Distribution matrix and mean-field free-energy functional.
\end{description}

For clarity, the same constant two-component matrix is written in three
conventions:
\begin{center}
\begin{tabular}{@{}lll@{}}
\toprule
representation & constant matrix & identification \\
\midrule
spin basis & $D\sigma_x+iG\sigma_z$ & $D=\Deff$, $G=\Gpt$ \\
fixed chiral basis & $D\sigma_z+iG\sigma_x$ & Hadamard rotation \\
companion algebra & $\gamma\sigma_x+i\beta\sigma_z$ & $\gamma=D$, $\beta=G$ \\
\bottomrule
\end{tabular}
\end{center}
The frozen drive coordinate is instead $\beta_0$, with
$G=0.20\beta_0$ on the audited path; it must not be identified with the
matrix coefficient $G$ or with the companion symbol $\beta$.
The normalized geometry plots use $D=0.50$ and $G=|\beta_0|$, whereas the
small-width causal/renormalization-group (RG) audit uses $D=0.10$ and
$G=0.20\beta_0$.  Both place
the dimensionless core crossing at $|\beta_0|=0.50$ because all constant-core
coefficients are rescaled together; their displayed $\beta_0$ axes are a
shared control-coordinate normalization, not identical physical decay rates.

\appendix

\section{Impurity-block EP condition and condition-number scaling}
\label{app:harmonic}
\label{app:EP_scaling}

This appendix records the impurity-block algebra used in the main text.
The Hadamard rotation itself is shown in
Eqs.~(\MainEqImpRotated)--(\MainEqVRotated) of the main text; here we keep only
the EP and condition-number scaling that underlies the diagnostic
interpretation.

The frozen impurity block is
$h_d=\bigl[\begin{smallmatrix}\teps+D&iG\\iG&\teps-D
\end{smallmatrix}\bigr]$, with eigenvalues $E_\pm=\teps\pm s$
and splitting $s=\sqrt{D^2-G^2}$.  The regimes are
$\PT$-unbroken ($D>G$), exceptional ($D=G$), and
$\PT$-broken ($G>D$) for positive parameters.  Right eigenvectors may be
chosen as $|r_\pm\rangle\propto[G,\,i(D\mp s)]^T$, so
\begin{equation}
\det R=2iG s,
\qquad R=(|r_+\rangle,|r_-\rangle).
\label{eq:detR}
\end{equation}
Thus, near the bare EP,
\begin{equation}
\keff=\|R\|\|R^{-1}\|\sim\frac{C}{|s|},
\label{eq:kappa_scaling}
\end{equation}
with a nonsingular constant $C$.  The standard local \(U\) does not cut off
this algebraic divergence on the equal-velocity, scalar-hybridization
linearized submanifold because it preserves the complexified pseudospin
symmetry.  Physical observables built from the complete resolvent remain
finite.  Curvature, unequal velocities, frequency dependence, or
component-dependent vertices can lift the exact symmetry and must be
analyzed separately.

\subsection{Slave-boson scaling and the core EP}

The bare impurity--bath hybridization $W$ in Eq.~(3) of the main article is
time independent.  At the saddle it becomes $rW$.  Since the projected
impurity self-energy is $\Sigma_d=W^\dagger G_cW$, every induced impurity
channel carries the same factor $r^2$:
\begin{equation}
\Gamma_{\rm PT}=r^2\Gamma_{\rm PT}^{(0)}(\beta_0),\quad
\Delta_{\rm eff}=r^2\Delta_0^{(0)},\quad
\bar\Gamma=r^2\Gamma_0 .
\label{eq:r2_channels}
\end{equation}
Consequently
\begin{equation}
h_{\rm rel}^2
=r^4\left\{[\Delta_0^{(0)}]^2
-[\Gamma_{\rm PT}^{(0)}(\beta_0)]^2\right\}I .
\label{eq:r4_identity_app}
\end{equation}
The common $r^4$ cancels from the local core condition for every $r>0$.
At $r=0$ the result is an ordinary decoupled degeneracy.  A bath-dressed
finite-$k$ minimum can drift because the SOC-resolved Schur terms do not
remain a common scalar shift; the audited search reaches the scan boundary
and is therefore not classified as an EP.
At $k=0$, by contrast,
$\varepsilon_{c,+}(0)=\varepsilon_{c,-}(0)=0$, and the Schur polynomial
factorizes exactly as
\begin{align}
\mathcal P(z,0)
&=\left(z-\tilde\epsilon_\xi-\frac{r^2W^2}{z}\right)^2
\nonumber\\[-0.2em]
&\quad-r^4\!\left\{[\Delta_0^{(0)}]^2
-[\Gamma_{\rm PT}^{(0)}(\beta_0)]^2\right\}.
\label{eq:k0_factorization_app}
\end{align}
The impurity-related $k=0$ double-root condition is therefore exactly the
local core condition for arbitrary $W$ and $r>0$.

\section{Angular Harmonic Decomposition and 2D$\to$1D Reduction}
\label{app:angular}

We show how the 2D Dirac bath with trigonal anisotropy reduces to
an effective 1D problem after integrating out all angular-momentum
harmonics $m\neq0$.
This Gaussian reduction is the microscopic origin of the
1D BA structure in Sec.~\MainSecBA\ of the main text.

\subsection{Angular Fourier Decomposition}

In 2D polar momentum coordinates $(k,\theta)$, the bath fermion
is decomposed into angular-momentum channels:
\begin{equation}
c_{k,\theta,\sigma}
=\frac{1}{\sqrt{2\pi}}\sum_{m=-\infty}^{\infty}
c_{k,m,\sigma}\,e^{im\theta}.
\label{eq:ang_decomp}
\end{equation}
The impurity at position $\mathbf{r}=0$ (the lattice site) is local,
so only the isotropic ($m=0$) channel hybridizes with it:
\begin{equation}
c_{\mathbf{r}=0,\sigma}
=\sum_k\frac{1}{\sqrt{2\pi}}\,c_{k,0,\sigma}.
\label{eq:local_bath}
\end{equation}
Higher harmonics $m\neq0$ are decoupled from the impurity in
the absence of anisotropy.

\subsection{$C_3$ Selection Rule from $\cos3\theta$}

The trigonal anisotropy $\beta_3(k)k^3\cos3\theta$ is expanded as
$\cos3\theta=\frac{1}{2}(e^{3i\theta}+e^{-3i\theta})$.
Substituting Eq.~(\ref{eq:ang_decomp}) and performing the
$\theta$ integral:
$\int_0^{2\pi}\frac{d\theta}{2\pi}\,e^{i(m'-m)\theta}\,e^{\pm3i\theta}
=\delta_{m',m\mp3}$.
The trigonal term therefore couples only channels differing by
$\Delta m=\pm3$:
\begin{equation}
V_{\rm trig}\propto\sum_{k,\sigma,m}
c^\dagger_{k,m,\sigma}c_{k,m-3,\sigma}+\mathrm{h.c.}
\label{eq:Vselect}
\end{equation}
The auxiliary fermions $O_{k,m,\sigma}$ represent precisely the
$m=\pm3,\pm6,\ldots$ channels activated by this selection rule.

\subsection{Gaussian Integration and Passive Relative Kernel}

Since $H_O+H_{\rm hyb}$ is bilinear in $O_{k,m,\sigma}$, integrating out
all $m\neq0$ modes is a Gaussian path integral within the quadratic
auxiliary sector.  For fixed $(k,\theta,\sigma)$ the auxiliary action has
exactly the form
\begin{equation}
\begin{split}
S_O+S_{cO}=&
\int_{\mathcal C}dt dt'\,\bar O(t)A_O(t,t')O(t') \\
&+\int_{\mathcal C}dt\,
\left[\bar c(t)V_\sigma(t)O(t)
+\bar O(t)V_\sigma^\dagger(t)c(t)\right].
\end{split}
\label{eq:app_grassmann_action}
\end{equation}
The Grassmann integral gives the exact quadratic influence action
\begin{equation}
S_{\rm infl}
=-\int_{\mathcal C}dt dt'\,
\bar c(t)V_\sigma(t)g_O(t,t')V_\sigma^\dagger(t')c(t'),
\label{eq:app_influence_action}
\end{equation}
so the explicit retarded vertex contraction is
\begin{align}
\Sigma_\sigma^{O,R}(t,t')
&=\sum_{m,m'\neq0}
V_{m\sigma}(t)\,
 g_{mm'}^{O,R}(t,t')\,
V^*_{m'\sigma}(t') .
\label{eq:app_vertex_contraction}
\end{align}
No anomalous auxiliary propagator is used.  For a number-conserving
Hermitian auxiliary Hamiltonian this contraction contains the phase
difference fixed by the vertex and its Hermitian conjugate.  The Wigner
step connecting this two-time expression to the frozen bath self-energy can
be displayed explicitly.  Set $T=t_{\rm av}=(t+t')/2$ and $\tau=t-t'$, and
write $\mathcal V_{mm'}=V_mV_{m'}^*e^{i3(m-m')\theta}$.  To first adiabatic
order,
\begin{align}
&V_{m\sigma}(T+\tau/2)V_{m'\sigma}^*(T-\tau/2)
\notag\\
&\quad=\mathcal V_{mm'}
e^{i\zeta_\sigma[\phi(T+\tau/2)-\phi(T-\tau/2)]}
\notag\\
&\quad=\mathcal V_{mm'}e^{i\zeta_\sigma\tau\dot\phi(T)}
\left[1+O(\tau^3\partial_T^3\phi)\right].
\label{eq:app_wigner_phase}
\end{align}
For a stationary auxiliary propagator, its Wigner transform is consequently
\begin{align}
\Sigma_\sigma^{O,R}(\omega,T)
&=\Sigma_0^{O,R}[\omega+\zeta_\sigma\dot\phi(T)]
\notag\\
&=\Sigma_0^{O,R}(\omega)
+\zeta_\sigma\dot\phi(T)\partial_\omega\Sigma_0^{O,R}(\omega)
\notag\\
&\quad+O[(\Omega/\epsilon_O)^2].
\label{eq:app_wigner_shift}
\end{align}
Here the remainder denotes the relative gradient correction in the off-shell
window $|\omega-\epsilon_O^{(m)}|\sim\epsilon_O$.  Defining
\begin{equation}
\Sigma_0^{\rm bath,R}=\Sigma_{\rm bg}^R+\Sigma_0^{O,R},
\qquad
\Delta_{\rm aux}=\dot\phi(T)\partial_\omega
\operatorname{Re}\Sigma_0^{O,R},
\label{eq:app_aux_definitions}
\end{equation}
therefore gives
\begin{equation}
\begin{split}
\Sigma_{\sigma}^{\rm bath,R}(\omega,T)
=&\,\Sigma_0^{\rm bath,R}(\omega)
+\zeta_\sigma\Delta_{\rm aux}(\omega,T) \\
&+O[(\Omega/\epsilon_O)^2],
\qquad \Delta_{\rm aux}\in\mathbb{R} .
\end{split}
\label{eq:app_aux_shift}
\end{equation}
For diagonal remote modes,
\begin{align}
\Sigma_0^{O,R}(\omega)
&=\sum_{m\ne0}\frac{|V_m|^2}
{\omega-\epsilon_O^{(m)}+i0^+},
\notag\\
\Delta_{\rm aux}
&=-\dot\phi(T)\sum_{m\ne0}|V_m|^2
\mathcal P\frac{1}{[\omega-\epsilon_O^{(m)}]^2}
\notag\\
&=-\dot\phi(T)\sum_{m\ne0}|V_m|^2
\partial_{\epsilon_O^{(m)}}
\mathcal P\frac{1}{\omega-\epsilon_O^{(m)}}.
\label{eq:app_aux_pv}
\end{align}
Thus the off-shell auxiliary sector produces a real principal-value
correction in the low-energy window.  If an auxiliary pole enters that
window, the $-i\pi\delta(\omega-\epsilon_O^{(m)})$ part must instead be
retained as ordinary on-shell bath damping.  The same frozen shift follows
after the unitary rotation
$O_{k,m,\sigma}\to e^{-i\zeta_\sigma\phi(t)}O_{k,m,\sigma}$, which removes
the drive phase from the vertices and shifts the auxiliary levels by
$-\zeta_\sigma\dot\phi(t)$.
The passive non-Hermitian component appears only after this shifted
channel is transmitted through the retarded Dirac propagator.  The point
is clearest from the quadratic $(d,c,O)$ inverse Green-function kernel,
\begin{equation}
\mathcal G^{-1}=
\begin{pmatrix}
 g_d^{-1} & -W^\dagger & 0\\
 -W & g_c^{-1} & -V_\sigma\\
 0 & -V_\sigma^\dagger & g_O^{-1}
\end{pmatrix}.
\label{eq:app_full_inverse}
\end{equation}
Here $W$ is static and all explicit drive dependence resides in
$V_\sigma(t)$.  Integrating out $O$ and then $c$ gives the two exact
Schur complements
\begin{equation}
G_{c,\sigma}^{-1}=g_c^{-1}-V_\sigma g_OV_\sigma^\dagger,
\qquad
\Sigma_{d,\sigma}=W^\dagger G_{c,\sigma}W .
\label{eq:app_two_schur}
\end{equation}
Equivalently,
\begin{equation}
\Sigma_{d,\sigma}=W^\dagger g_cW
+W^\dagger g_c(V_\sigma g_OV_\sigma^\dagger)g_cW+\cdots,
\label{eq:app_virtual_path}
\end{equation}
which explicitly displays the drive-dependent virtual path
$d\xrightarrow{W}c\xrightarrow{V_\sigma(t)}O
\xrightarrow{V_\sigma^\dagger(t')}c\xrightarrow{W^\dagger}d$.
The order of the two Gaussian integrations is immaterial.

For $\Xi_\sigma^R=\Xi_0^R+\zeta_\sigma\delta\Xi^R$, the general linear
correction is
\begin{equation}
\delta\Sigma_d^R
=W^\dagger G_{c,0}^R\,\delta\Xi^R\,G_{c,0}^R W .
\label{eq:app_exact_delta_sigma}
\end{equation}
It is complex because $G_{c,0}^R$ is a retarded continuum propagator,
even when the off-shell $\delta\Xi^R$ is real.  In the rigid-shift limit
$\delta\Xi^R\simeq\Delta_{\rm aux}$, with weak frequency dependence of
the spin-even auxiliary background,
\begin{equation}
\Sigma_{d,\sigma}^R(\omega,t_{\rm av})
\simeq
\Sigma_{d,0}^R(\omega-\zeta_\sigma\Delta_{\rm aux})
\simeq
\Sigma_{d,0}^R-\zeta_\sigma\Delta_{\rm aux}
\partial_\omega\Sigma_{d,0}^R .
\label{eq:app_dirac_conversion}
\end{equation}
Writing $\Sigma_{d,0}^R=\Lambda-i\bar\Gamma$, the real derivative gives
$\Dcoh=-\Delta_{\rm aux}\partial_\omega\Lambda$ in the $\sigma_z$
channel---it does \emph{not} renormalize the spin-mixing splitting
$\Deff\sigma_x$ at this order---while the imaginary derivative gives
\begin{equation}
\Gamma_{\rm PT}=\Delta_{\rm aux}\,\partial_\omega\bar\Gamma(\omega_0).
\label{eq:app_gamma_pt_dirac}
\end{equation}
Equation~(\ref{eq:app_gamma_pt_dirac}) is a scalar rigid-shift estimate, not
the exact matrix identity for the SOC bath.  Writing
$a=(\omega^+-\varepsilon_{k+})^{-1}$,
$b=(\omega^+-\varepsilon_{k-})^{-1}$,
$g_0=(a+b)/2$, and $g_x=(a-b)/2$, one instead has
\begin{equation}
-\partial_\omega g_0=g_0^2+g_x^2,\qquad
g_b\sigma_zg_b=(g_0^2-g_x^2)\sigma_z=ab\,\sigma_z .
\label{eq:app_matrix_shift_identity}
\end{equation}
Thus, in the sign convention of main-text Eq.~(\MainEqDeltaAux),
\begin{equation}
\begin{aligned}
\Gamma_{\rm PT}
&=\Delta_{\rm aux}\!\left[
\partial_\omega\bar\Gamma
-2\,\operatorname{Im}\sum_k|W_k|^2g_{x,k}^2\right]\\
&=\Delta_{\rm aux}\operatorname{Im}\sum_k|W_k|^2a_kb_k .
\end{aligned}
\label{eq:app_matrix_gamma_pt}
\end{equation}
An opposite convention for the spin-odd auxiliary shift reverses the full
equation.  In particular, when the exact positive-energy total width is
flat, $\partial_\omega\bar\Gamma=0$, the relative decay is generated
entirely by the SOC $g_x^2$ term.  The full-$k$ calculation below uses the
matrix resolvent itself and therefore retains this term; it does not use
Eq.~(\ref{eq:app_gamma_pt_dirac}) as its operative mechanism.
Here $\omega_0$ denotes the low-energy evaluation point of the frozen
kernel; at the compensation energy where $\partial_\omega\Lambda$
changes sign (of order $D_{\rm uv}/e$ for the linear Dirac width with a sharp
cutoff), the Kramers--Kronig detuning vanishes, $\Dcoh(\omega_0)=0$,
while $\partial_\omega\bar\Gamma$ remains finite.
For completeness, take the idealized symmetric Dirac width
$\bar\Gamma(\epsilon)=g|\epsilon|\Theta(D_{\rm uv}-|\epsilon|)$.  Its
principal-value Kramers--Kronig partner is
\begin{align}
\Lambda(\omega)
&=\frac{1}{\pi}\mathcal P\!\int_{-D_{\rm uv}}^{D_{\rm uv}}
d\epsilon\,\frac{\bar\Gamma(\epsilon)}{\omega-\epsilon}
\notag\\
&=-\frac{g\omega}{\pi}
\ln\left|\frac{D_{\rm uv}^2-\omega^2}{\omega^2}\right|.
\label{eq:app_dirac_kk}
\end{align}
Hence, for $|\omega|\ll D_{\rm uv}$,
\begin{align}
\partial_\omega\Lambda
&=-\frac{2g}{\pi}\left[\ln\frac{D_{\rm uv}}{|\omega|}-1\right]
+O(\omega^2/D_{\rm uv}^2),
\notag\\
\partial_\omega\bar\Gamma
&=g\,\operatorname{sgn}\omega .
\label{eq:app_dirac_kk_derivative}
\end{align}
This shows explicitly both the compensation scale
$|\omega_0|\simeq D_{\rm uv}/e$ in the low-energy form and the logarithmic
growth of $|\Dcoh/\Gamma_{\rm PT}|$ near the Dirac node.  For the exact
sharp-cutoff expression with $D_{\rm uv}=\pi/4$, numerical solution of
$\partial_\omega\Lambda=0$ gives $|\omega_c|=0.24608$, rather than the
low-energy estimate $D_{\rm uv}/e=0.28893$.  A smooth cutoff shifts this
number but not the mechanism.
The SOC split Dirac bath provides the needed helicity channels and
channel-resolved widths
\begin{equation}
\bar\Gamma_\eta(\omega)=
\pi\sum_k |W_{k\eta}|^2
\delta[\omega-\varepsilon_{c,\eta}(k)] .
\label{eq:app_helicity_width}
\end{equation}
After subtracting the common damping $-i\bar\Gamma I$, the remaining
co-decaying kernel contains the passive-$\PT$ core
$\Deff\sigma_x+i\Gamma_{\rm PT}\sigma_z$ of
main-text Eq.~(\MainEqPassivePTCore) together with the $\PT$-odd
Kramers--Kronig detuning $\Dcoh\sigma_z$; exact passive $\PT$ symmetry
holds at the compensated point $\Dcoh=0$, and a finite detuning acts as
the causal floor of main-text Eq.~(\MainEqDetunedEPConditions).  Indeed,
\begin{equation}
|s|^4=
\left(\Deff^2+\Dcoh^2-\Gamma_{\rm PT}^2\right)^2
+4\Dcoh^2\Gamma_{\rm PT}^2 .
\label{eq:app_causal_floor_objective}
\end{equation}
Minimizing over the real control $\Deff^2\ge0$ gives
\begin{equation}
|s|_{\min}=
\begin{cases}
\sqrt{2|\Dcoh\Gamma_{\rm PT}|},
& |\Dcoh|\le|\Gamma_{\rm PT}|,\\[2pt]
\sqrt{\Dcoh^2+\Gamma_{\rm PT}^2},
& |\Dcoh|>|\Gamma_{\rm PT}|.
\end{cases}
\label{eq:app_causal_floor_piecewise}
\end{equation}
The square-root expression quoted for the near-compensated regime therefore
has the necessary domain $|\Dcoh|\le|\Gamma_{\rm PT}|$; outside it the
minimum occurs at $\Deff=0$.  This is the
sense in which the strictly Hermitian driven model generates the
relative non-Hermitian kernel used in the main text.  Compensation is local
in frequency: for the audited values
$\bar\Gamma=0.12$, $\Gamma_{\rm PT}=0.10$, the full sharp-cutoff kernel one
common half-width from $\omega_c$ gives
$|s|_{\min}/\Gamma_{\rm PT}=1.0028$ on one side and $0.9004$ on the other.
Thus the finite-band problem has an order-$\Gamma_{\rm PT}$ avoided-
coalescence floor across a linewidth.  A structureless scalar flat bath,
for which both $\partial_\omega\bar\Gamma=0$ and $g_x=0$, produces only a
coherent spin splitting.  Flatness of the \emph{total} radial width alone
does not remove the passive relative decay: when the helicity channels are
nondegenerate, Eq.~(\ref{eq:app_matrix_gamma_pt}) leaves the SOC $g_x^2$
contribution.

The reduction relies on the following conditions:
\begin{enumerate}[label=(\roman*)]
\item Impurity at a $C_3$-invariant site ($\theta_{\rm imp}=0$
  mod $2\pi/3$), so the angular projection leaves a single radial
  bath channel.
\item Off-shell auxiliary harmonics, $\epsilon_O^{(m)}\gg D_{\rm uv}$ and
$|V|/\epsilon_O^{(m)}\ll1$, so the
  auxiliary contribution is a controlled principal-value shift.  Residual
  on-shell weight gives ordinary bath damping rather than the relative
  $\PT$ channel.
\item Slow drive on the auxiliary memory time, $\Omega\ll\epsilon_O^{(m)}$,
  so the correction is organized in powers of $\Omega/\epsilon_O$.
\item A retarded bath conversion channel: either an energy-dependent scalar
  width with $\partial_\omega\bar\Gamma\neq0$, as in the cone-patch estimate,
  or nonzero SOC matrix dressing $g_x$, as in the exact radial model.
\item $H_O$ bilinear: no many-body approximation is made in the
  auxiliary-mode integration.
\end{enumerate}

These conditions do not guarantee reachability of a chosen normalized core.
With $x=|V|/\epsilon_O$, phase amplitude $\pi/4$, and benchmark ratio
$\Gamma_{\rm PT}/\bar\Gamma=0.10/0.12$, the leading mapping requires
\begin{equation}
\frac{\Omega}{\epsilon_O}\gtrsim\frac{1.061}{x^2}.
\label{eq:app_reachability_bound}
\end{equation}
For $x\ll1$ this is incompatible with the off-shell/quasistatic hierarchy.
Accordingly the derivation fixes the form and symmetry of the passive
relative kernel, while the exact benchmark EP remains a normalized
constant-core construction rather than a demonstrated microscopic operating
point.

\subsection{Stationary rotating drive and full-momentum nonlinear-pole EP}
\label{app:full_k_causal_ep}

The preceding obstruction concerns the oscillatory quasistatic protocol.  A
stationary alternative follows from the same microscopic vertex by taking
$\phi(t)=\Omega t$.  After the unitary transformation of
main-text Eq.~(\MainEqAuxGaugeRotation), the remote levels are shifted by the
constant $-\zeta_\sigma\Omega$, so the rotated quadratic Hamiltonian is
exactly time independent.  For one remote Hermitian level with vertex $V_O$,
its spin-even and spin-odd contributions to the bath inverse are
\begin{align}
B_{\rm raw}(z)
&=V_O^2\frac{z-\epsilon_O}
{(z-\epsilon_O)^2-\Omega^2},
\notag\\
A(z)
&=-V_O^2\frac{\Omega}
{(z-\epsilon_O)^2-\Omega^2}.
\label{eq:app_rotating_aux_terms}
\end{align}
The constant spin-even principal-value shift is absorbed into the bath
chemical potential, $B(z)=B_{\rm raw}(z)-B_{\rm raw}(0)$.  This subtraction
does not affect the spin-odd term or the pole multiplicity.

For $\alpha_{\rm b}=1$, scalar momentum-independent impurity hybridization
$W$, and radial cutoff $K$, the physical-sheet impurity self-energy is
\begin{equation}
\Sigma_d^{\rm I}(z)=W^2\int_0^K\frac{k\,dk}{2\pi}
\left[(z-B-k^2)\mathbb I-\lambda k\sigma_x-A\sigma_z\right]^{-1}.
\label{eq:app_full_k_selfenergy}
\end{equation}
The integral is evaluated analytically and independently checked by adaptive
quadrature.  Every open helicity channel must be continued.  If $k_j$ is an
on-shell radial root on the real cut, define
\begin{align}
P_j&=\frac12\left[\mathbb I+
\frac{\lambda k_j\sigma_x+A\sigma_z}
{z-B-k_j^2}\right],
\notag\\
\rho_j&=\frac{k_j}{2\pi|v_j|}P_j,
\qquad
v_j=2k_j+\frac{\lambda^2k_j}{z-B-k_j^2}.
\label{eq:app_open_channel_residue}
\end{align}
Analytic continuation of the roots and projectors then gives the outgoing
resonance sheet
\begin{equation}
\Sigma_d^{\rm II}(z)=\Sigma_d^{\rm I}(z)
-2\pi iW^2\sum_{j\in{\rm open}}\rho_j(z).
\label{eq:app_two_cut_sheet}
\end{equation}
At the solution below there are two open channels.  Continuing only one cut
misses the EP.

With $F$ defined in main-text Eq.~(\MainEqFullKPoleDeterminant), the nonlinear
EP conditions are
\begin{equation}
F(z_{\rm EP})=0,\qquad
\partial_zF(z_{\rm EP})=0,\qquad
\partial_z^2F(z_{\rm EP})\neq0,
\label{eq:app_nonlinear_ep_conditions}
\end{equation}
together with geometric multiplicity one of the inverse kernel.  To certify
the zero without evaluating a singular eigenbasis, for each point on a
closed control loop in $(\epsilon_d,V_O)$ we branch-track the nearby critical
point $z_c$ satisfying $\partial_zF(z_c)=0$ and form
\begin{equation}
{\cal D}_{\rm loc}(\epsilon_d,V_O)=F(z_c;\epsilon_d,V_O).
\label{eq:app_local_discriminant}
\end{equation}
Because $\partial_z^2F$ is nonzero on the loop, the implicit-function theorem
keeps $z_c$ on one smooth critical-point branch.  Provided
${\cal D}_{\rm loc}$ is nonzero on the boundary, its integer
\begin{equation}
W_C
=\frac{1}{2\pi}\Delta_C\arg {\cal D}_{\rm loc}
\label{eq:app_local_degree}
\end{equation}
is the Brouwer degree of the smooth map
$(\epsilon_d,V_O)\mapsto(\operatorname{Re}{\cal D}_{\rm loc},
\operatorname{Im}{\cal D}_{\rm loc})$.  A nonzero degree forces at least one
enclosed zero; regular zeros contribute the sign of the corresponding real
$2\times2$ Jacobian.
At every such zero, $F=\partial_zF=0$; the retained lower bound on
$|\partial_z^2F|$ makes the zero quadratic, while the rank-one inverse-kernel
test below excludes a scalar or diabolic degeneracy.  Explicit tracking of
the two poles supplies the equivalent EP2 monodromy test: one circuit
exchanges their labels and their separation has half-integer winding.

For
\begin{equation}
\begin{aligned}
\lambda&=0.05,&W&=0.10,&K&=\pi/4,\\
\epsilon_O&=120,&\Omega&=60,&T_{\rm b}&=0.1,
\end{aligned}
\label{eq:app_full_k_witness_parameters}
\end{equation}
the solution is
\begin{align}
z_{\rm EP}&=0.011972834346-0.002807216274i,
\notag\\
V_O&=1.213422009614,\qquad
\epsilon_d=0.015170349783.
\label{eq:app_full_k_witness_solution}
\end{align}
The algebraic residuals are
\begin{align}
|F|&=1.44\times10^{-21},\qquad
|\partial_zF|=1.09\times10^{-15},
\notag\\
\partial_z^2F&=1.686448434-0.190336728i.
\label{eq:app_full_k_witness_residuals}
\end{align}
The two singular values of the inverse kernel are
$1.207\times10^{-3}$ and $1.22\times10^{-18}$, establishing geometric
multiplicity one.  On the reference ellipse
$\delta\epsilon_d=10^{-4}$ and $\delta V_O=10^{-2}$,
$W[\mathcal D_{\rm loc}]=+1$; the two poles exchange and their separation has
half-winding $+1/2$.  The minimum boundary modulus is
$2.10\times10^{-9}$, the largest phase step is $0.0882<\pi/2$, and
$|\partial_z^2F|$ remains above $1.697$.  The maximum pole-tracking step is
only $0.103$ of the local half-separation, and the maximum pole residual is
$8.62\times10^{-21}$.  Halving and doubling both ellipse radii leave the
winding and pole exchange unchanged.

The causal and microscopic controls are independent of the root solve.  The
analytic radial integral agrees with direct quadrature to
$7.93\times10^{-19}$.  Matching the upper rim of the physical sheet to the
continued lower rim gives a two-cut residual $4.02\times10^{-9}$ at
$\eta=10^{-8}$.  The physical rate-matrix eigenvalues are positive,
$0.00190368$ and $0.00309632$, so $\bar\Gamma=0.00250000$ and
$\Gamma_{\min}/\bar\Gamma=0.76147$.  The driven positive
$k=0$ threshold is $+0.00817959$; the certified pole is
$0.00379325=1.351|\operatorname{Im}z_{\rm EP}|$ above it.
Thus both open cuts, the threshold proximity, and the lower-branch van Hove
structure are already included in the continuation.  Finally,
\begin{equation}
\begin{aligned}
\frac{V_O}{\epsilon_O-\Omega}&=0.02022,&
\frac{\Omega}{K}&=76.39,\\
\frac{\Omega}{\bar\Gamma}&=2.40\times10^4,&
\frac{T_{\rm b}}{\bar\Gamma}&=40,&
\frac{\epsilon_O-\Omega}{T_{\rm b}}&=600.
\end{aligned}
\label{eq:app_full_k_control_ratios}
\end{equation}
Thus the auxiliary sector is off shell and thermally empty, while its exact
frequency dependence has been retained rather than expanded in
$\Omega/\epsilon_O$.  The large $\Omega/\bar\Gamma$ ratio would invalidate
an oscillatory quasistatic interpretation but is irrelevant to the exactly
stationary rotating-frame Hamiltonian.  The certificate itself is retarded
and temperature independent; linewidth-resolved equilibrium observables for
this witness require $T_{\rm b}\lesssim10^{-3}$.  Figure~\ref{fig:S_full_k_ep}
displays the local winding and pole exchange.

The same equations can be continued in SOC with $W$ held fixed at $0.10$
while retuning the real impurity gate and auxiliary vertex.  We obtain a
regular one-dimensional locus for
$0.02\leq\lambda\leq0.1578$.  Representative points are
\begin{center}
\begin{tabular}{@{}ccccc@{}}
\toprule
$\lambda$ & $\operatorname{Re}z_{\rm EP}$ & $\operatorname{Im}z_{\rm EP}$
& $\epsilon_d$ & $V_O$ \\
\midrule
$0.05$ & $0.01197283$ & $-0.00280722$ & $0.01517035$ & $1.21342201$ \\
$0.10$ & $0.02035701$ & $-0.00292855$ & $0.02323413$ & $1.83318448$ \\
$0.15$ & $0.03265022$ & $-0.00303645$ & $0.03515182$ & $2.41589290$ \\
\bottomrule
\end{tabular}
\end{center}
At all three points ${\cal D}_{\rm loc}$ has winding $+1$, the endpoint pole
permutation is $(12)$, and the pole-separation half-winding is $+1/2$.  The
maximum tracking-step/half-separation ratios are $0.103$, $0.100$, and
$0.119$, with no failed pole solves.  The corresponding threshold gaps in
pole widths are $1.351$, $0.577$, and $0.076$ and the passivity margins are
$0.761$, $0.614$, and $0.515$, respectively.  Throughout the continued segment
$V_O/(\epsilon_O-\Omega)<0.042$ and the physical rate matrix remains
positive.  As shown in Fig.~\ref{fig:S_full_k_soc_family}, the branch
approaches the upper open-channel threshold near the high-SOC end while its
passivity margin remains positive.  This threshold termination is distinct
from a passivity failure in a separate normalized sharp-cutoff audit.  We do
not classify its endpoint and do not infer a two-dimensional EP phase region.

This calculation establishes a microscopic possibility for a nonlinear
causal pole EP in the full curved momentum-summed retarded kernel.  It does
not show that the normalized $\lambda=W=0.75$ benchmark is a causal EP, and
it does not identify the nonlinear tangency condition
$F=\partial_zF=0$ with the nilpotent constant-matrix condition $M^2=0$.
Consequently no full-momentum Yang--Baxter equation (YBE), $RLL$, or
$RTT$ structure or many-body Jordan
amplification is inferred from this certificate.

This completes the microscopic link asserted at the start of the
appendix.  The key point is that the Gaussian auxiliary projection gives
a real spin-odd shift, and the retarded SOC matrix bath converts that shift
into a passive relative gain--loss component in the co-decaying frame,
together with its Kramers--Kronig companion detuning.  The same SOC
off-diagonal dressing supplies the $\PT$-even coherent mixing channel through
main-text Eq.~(\MainEqDeltaEffSOCOrigin) and, through its $g_x^2$ contribution in
Eq.~(\ref{eq:app_matrix_gamma_pt}), the relative decay even when the total
positive-energy radial width is flat.

\paragraph{Result: 2D bath $\to$ 1D effective bath.}
After the Gaussian integration, the impurity is coupled to
only the $m=0$ bath channel, labeled by radial momentum $k$
alone.
For the full dispersions
$\varepsilon_{k\pm}=\alpha_{\rm b}k^2\pm\lambda k$, define
$\Delta(\omega)=\sqrt{\lambda^2+4\alpha_{\rm b}\omega}$.  When both
positive-energy radial roots lie inside the cutoff, their exact densities are
\begin{equation}
\begin{aligned}
\rho_+(\omega)&=\frac{\Delta-\lambda}{4\pi\alpha_{\rm b}\Delta},&
\rho_-(\omega)&=\frac{\Delta+\lambda}{4\pi\alpha_{\rm b}\Delta},\\
\rho_++\rho_-&=\frac{1}{2\pi\alpha_{\rm b}}.&&
\end{aligned}
\label{eq:app_exact_radial_dos}
\end{equation}
Thus the complete radial bath has a constant positive-energy floor, not a
pseudogap.  For $-\lambda^2/(4\alpha_{\rm b})<\omega<0$, the lower branch
instead gives
\begin{equation}
\rho_-(\omega)=
\frac{\lambda}{2\pi\alpha_{\rm b}
\sqrt{\lambda^2+4\alpha_{\rm b}\omega}},
\label{eq:app_lower_van_hove}
\end{equation}
with a radial van Hove edge at $-\lambda^2/(4\alpha_{\rm b})$.  The
$|\omega|/\lambda^2$ Dirac law is the small-$k$ cone contribution of one
branch (or a cone-patch model), not the density of states (DOS) of the complete quadratic-plus-SOC
radial problem.

The Gaussian reduction leaves a single radial ($m=0$) bath carrying the
exact DOS above, hybridizing with the impurity through the
channel-preserving vertex of Eq.~(\ref{eq:local_bath}).  After the
spin rotation of main-text Sec.~\MainSecRotation\ this radial bath splits into
the two chiral channels $c_{k\pm}$, each coupled diagonally to a single
impurity component $\tilde{d}_\pm$.  This diagonal bare vertex does not,
however, diagonalize the complete scattering problem: the internal
$i\Gamma_{\rm PT}$ impurity matrix element mediates conversion between the
two asymptotic bath channels.  Main-text Sec.~\MainSecBA\ therefore first uses
eigenphases of the full \(2\times2\) unitary scattering matrix only for exact
one-particle phase counting.  Multiparticle factorization is asserted only
after the additional equal-velocity branchwise linearization and folding
stated explicitly there.

The same helicity decomposition also makes clear that the channel-odd
self-energy is generally complex:
\begin{equation}
\begin{aligned}
g_x&=\frac12\left[(\omega^+-\epsilon_{k+})^{-1}
-(\omega^+-\epsilon_{k-})^{-1}\right],\\
\Sigma_x&=\frac12(\Sigma_+-\Sigma_-)
=\frac{\Lambda_+-\Lambda_-}{2}
-i\frac{\Gamma_+-\Gamma_-}{2}.
\end{aligned}
\label{eq:app_complex_channel_odd_self_energy}
\end{equation}
The normalized Markov core denotes only the first, coherent term by
$\Delta_{\rm eff}$; the full-$k$ causal certificate retains both terms and
their frequency dependence.

\section{Finite-\texorpdfstring{$U$}{U} Matrix Block-Wilson Continuation}
\label{app:finiteU_fullk}

This section tests the interaction stability of the nonlinear causal pole EP
of Sec.~\ref{app:full_k_causal_ep}.  It is separate from the
scalar-hybridization Markov-core residue audit in
Sec.~\ref{app:nrg_jordan}.  The present calculation retains the
SOC-resolved \(2\times2\) physical-bath hybridization, the analytic two-cut
continuation, and the exact frequency dependence of the remote auxiliary
level.  The low-energy physical bath is represented by a finite matrix Wilson
chain; the remote level remains in the analytic one-body kernel.

\subsection{Matrix chain and exact finite-chain response}

For the \(\lambda=0.05\), \(W=0.10\) witness of
Eq.~(\ref{eq:app_full_k_witness_parameters}), the positive matrix spectral
measure is logarithmically discretized with Wilson parameter
\(\Lambda=1.5\) and mapped by block Lanczos to a chain with \(2\times2\)
onsite and hopping blocks.  We use 12 chain sites for the endpoint center and
pointwise certificate, then repeat the complete-basis calculation at $N=10$
and $12$ for a common-contour chain-length test.  Each length averages four
independent logarithmic discretization shifts.  Star-to-chain reconstruction errors lie
between \(3.5\times10^{-15}\) and \(2.3\times10^{-14}\).

The impurity interaction is
\begin{equation}
H_U=U n_{d\uparrow}n_{d\downarrow}.
\end{equation}
Each finite chain is solved in untruncated fixed-charge sectors for its
ground state and \(N\pm1\) correction vectors.  Writing
\(q_\alpha=n_{d\bar\alpha}d_\alpha\), we use
\begin{align}
G_{\alpha\beta}(z)
&=\langle\!\langle d_\alpha;d_\beta^\dagger\rangle\!\rangle_z,
\notag\\
F_{\alpha\beta}(z)
&=\langle\!\langle q_\alpha;d_\beta^\dagger\rangle\!\rangle_z.
\end{align}
The four chain responses are averaged before forming the exact finite-chain
equation-of-motion self-energy
\begin{equation}
\Sigma_U(z)=U F(z)G^{-1}(z).
\label{eq:app_finiteU_eom}
\end{equation}
An independent Dyson construction agrees with
Eq.~(\ref{eq:app_finiteU_eom}) to maximum absolute residual
\(4.67\times10^{-16}\) and maximum relative residual
\(1.32\times10^{-11}\).  At \(U=0\), the exact chain response agrees with the
direct one-body Green function to \(2.02\times10^{-14}\); anticommutator sum
rules are satisfied at machine precision.

The determinant \(F_U\) is obtained by adding Eq.~(\ref{eq:app_finiteU_eom})
to the same analytic outgoing-sheet one-body kernel used in
Eq.~(\ref{eq:app_full_k_selfenergy}).  Simultaneously solving
\(F_U=\partial_zF_U=0\) while continuing from \(U=0\) gives
\begin{center}
\begin{tabular}{@{}ccccc@{}}
\toprule
\(U\) & \(\operatorname{Re}z_{\rm EP}\) &
\(\operatorname{Im}z_{\rm EP}\) & \(\epsilon_d\) & \(W\)\\
\midrule
0      & 0.01197283 & -0.00280722 & 0.01517035 & 0.10000000\\
0.0010 & 0.01189782 & -0.00255800 & 0.01483973 & 0.09577738\\
0.0015 & 0.01185780 & -0.00244927 & 0.01468704 & 0.09385894\\
0.0020 & 0.01181724 & -0.00234964 & 0.01454259 & 0.09205708\\
0.0025 & 0.01177670 & -0.00225815 & 0.01440607 & 0.09036305\\
\bottomrule
\end{tabular}
\end{center}
The \(U=0.0025\) endpoint is reached both by direct initialization and by
stepping from \(U=0.0020\).  All four discretization shifts retain ground
charge two.  In the units of the article,
\(U/D_{\rm uv}=3.18\times10^{-3}\); with the retuned local width
\(\bar\Gamma_{\rm self}=W^2/4=0.002050\), however,
\(U/\bar\Gamma_{\rm self}=1.225\).  The endpoint is therefore finite and
moderate relative to the local width but remains far from a bandwidth-scale
or deep-Kondo strong-coupling limit.

\subsection{Pointwise topological gate}

At the interacting center we evaluate a 64-point ellipse with half-widths
\(\delta\epsilon_d=10^{-4}\) and \(\delta W=10^{-3}\).  At every point the
nearby critical point satisfying \(\partial_zF_U=0\) and the two associated
poles are independently branch tracked.  The resulting gates are
\begin{center}
\begin{tabular}{@{}lr@{}}
\toprule
Diagnostic & Result\\
\midrule
Local-discriminant winding & \(-1.00000000000012\)\\
Endpoint pole permutation & \((12)\)\\
Pole-separation half-winding & \(-0.50000000000005\)\\
Minimum boundary modulus & \(1.5060\times10^{-9}\)\\
Maximum phase increment & \(0.27725\)\\
Minimum \(|F_U''|\) & \(1.71153\)\\
Maximum \(|F_U'|\) & \(1.1150\times10^{-15}\)\\
Minimum pole separation & \(8.3767\times10^{-5}\)\\
Maximum pole residual & \(2.5452\times10^{-21}\)\\
Assignment/separation ratio & \(0.42415<1/2\)\\
Maximum sum-rule error & \(3.7682\times10^{-15}\)\\
\bottomrule
\end{tabular}
\end{center}
Thus the contour is nondegenerate, its phase sampling is resolved, the
critical-point branch remains quadratic, and the pole assignment satisfies
the explicit separation guard.  The winding and pole exchange in
main-text Fig.~\MainFigFiniteUFullK\ certify one interacting double zero
inside the ellipse without diagonalizing at the defective point.

As a discretization-center check, increasing the average from four to eight
shifts moves \(\operatorname{Re}z_{\rm EP}\),
\(\operatorname{Im}z_{\rm EP}\), \(\epsilon_d\), and \(W\) by
\(0.022\%\), \(0.71\%\), \(0.14\%\), and \(0.33\%\), respectively.
This is a center-convergence check, not an infinite-chain extrapolation.

\subsection{Common-contour chain-length gate}

The $N=10$ and $N=12$ interacting centers are solved independently rather
than inherited from one another.  Their maximum absolute coordinate shift is
$8.737\times10^{-5}$ and their maximum relative coordinate shift is
$3.746\times10^{-3}$.  We then evaluate both chains on the same ellipse,
centered on the reoptimized $N=12$ point, with half-widths
$\delta\epsilon_d=2\times10^{-4}$ and $\delta W=2\times10^{-3}$.
At 256 contour points the branch-tracked local discriminant $D_N$ gives the
complete-basis gates
\begin{center}
\begin{tabular}{@{}lcc@{}}
\toprule
Diagnostic & $N=10$ & $N=12$\\
\midrule
Winding & $-1$ & $-1$\\
Half-winding & $-1/2$ & $-1/2$\\
Pole permutation & $(12)$ & $(12)$\\
$\min_C|D_N|$ & $2.417\times10^{-9}$ & $2.897\times10^{-9}$\\
Assignment ratio & $0.1388$ & $0.1352$\\
Sum-rule error & $1.256\times10^{-15}$ & $6.280\times10^{-15}$\\
\bottomrule
\end{tabular}
\end{center}
Here ``assignment'' is the maximum branch-assignment step divided by the
local half-separation.  It remains well below $1/2$ for both chains.  The
pairwise sufficient inequality is
\begin{equation}
\begin{aligned}
\sup_C|D_{12}-D_{10}|&=1.638\times10^{-9}\\
&<\min_C|D_{10}|=2.417\times10^{-9},
\end{aligned}
\label{eq:app_finiteU_rouche}
\end{equation}
or a conservative Rouch\'e ratio $0.6775$.  Consequently the two finite
chains have the same number of enclosed zeros on this common contour.  The
128-point calculation gives the same winding, exchange, and half-winding,
with ratio $0.6758$; doubling the resolution halves the assignment ratios
and changes the Rouch\'e ratio by only $0.25\%$.

Equation~(\ref{eq:app_finiteU_rouche}) is a finite-$N$ stability statement.
It does not bound the omitted Wilson-chain tail and therefore does not prove
an infinite-chain or thermodynamic limit.

\subsection{Kept-space calibration and scope}

To quantify what an ordinary truncated run would resolve, we compare
kept-space responses with the exact response on the identical four
12-site chains:
\begin{center}
\begin{tabular}{@{}ccccc@{}}
\toprule
\(N_{\rm keep}\) & max. \(G\) & max. \(F\) &
max. \(\Sigma_U\) & sum-rule deficit\\
\midrule
320 & \(11.67\%\) & \(47.35\%\) & \(38.00\%\) & \(2.61\%\)\\
640 & \(11.76\%\) & \(32.90\%\) & \(24.74\%\) & \(1.51\%\)\\
960 & \(3.71\%\) & \(4.57\%\) & \(4.18\%\) & \(1.40\%\)\\
\bottomrule
\end{tabular}
\end{center}
The retained deviations decrease at \(N_{\rm keep}=960\), but this comparison
is not used to certify the topology:
the load-bearing result is the untruncated complete-basis response on the
specified finite chains.

Two limitations are explicit.  First, the $N=10\to12$ Rouch\'e pass is a
controlled comparison of two complete-basis finite chains, but chain-length
and \(\Lambda\) convergence to the continuum still require a true complete-basis spectral numerical
renormalization group (NRG) or density-matrix NRG (DM-NRG)
implementation; at 12 sites the omitted hopping is not parametrically below
the pole scale.  Second, a short-chain four-band discretization including the
remote auxiliary pair selects ground charges \(1,2,2,1\) across the four
shifts.  Its center is contaminated by finite-size charge steps and is
reported neither as a pass nor as a failure.  The certified calculation
therefore establishes a zero-temperature interacting EP of the finite
low-energy matrix chains dressed by the exact analytic auxiliary kernel.  It
does not establish a thermodynamic full-density-matrix NRG (FDM-NRG) fixed
point, finite-temperature
observability, a phase boundary, or universal interaction scaling.

\section{Keldysh Mean-Field Theory}
\label{app:keldysh}

The slave-boson representation $d^\dagger_\sigma=f^\dagger_\sigma b$
introduces spinon fermions $f_\sigma$ (the Kondo quasiparticles)
and a slave boson $b$ (enforcing no double occupancy) with
condensate $b_c=\langle b\rangle$ and Lagrange multiplier $\delta_c$
enforcing the constraint $r^2+\sum_\sigma f^\dagger_\sigma f_\sigma=Q$
(where $r=|b_c|$).  Here $Q=1$ is the physical spin-$1/2$ constraint,
while $Q=Nq_{\rm occ}$ denotes the large-$N$ or channel-rescaled normalization
used in the corresponding numerical saddle.
The quadratic spinon action can be organized either with an explicit bath
or with its complete influence kernel.  In the latter representation the
retarded impurity kernel has the schematic form
\begin{equation}
\mathcal G_{ff}^{R,-1}
=\omega-\epsilon_d-\delta_c
-r^2D_0(\beta_0)\sigma_x-r^2K_\Gamma^R(\omega,\beta_0),
\label{eq:spinon_inverse_influence}
\end{equation}
and its lesser partner contains $r^2K_\Gamma^<$.  Therefore
\begin{align}
\partial_r[rW]&=W,\qquad
\partial_r[r^2D_0]=2rD_0,\nonumber\\
\partial_r[r^2K_\Gamma]&=2rK_\Gamma .
\label{eq:r_derivative_bookkeeping}
\end{align}
These three derivatives respectively generate the hybridization,
coherent-channel, and dissipative-influence forces below.
For the $U\to\infty$ Coleman saddle, variation of the complete quadratic
contour action gives the basis-independent equations
\begin{align}
0&=r^2+n_f-Q,\qquad
n_f=-i\int\frac{d\omega}{2\pi}{\rm Tr}_sG_{ff}^<(\omega),
\label{eq:saddle2_app}\\
0&=2r\delta_c+\mathcal F_W+2rD_0\,\mathcal X
+2r\mathcal I_\Gamma .
\label{eq:saddle1_app}
\end{align}
Here
\begin{align}
\mathcal F_W
&=-i\int\frac{d\omega}{2\pi}\sum_k{\rm Tr}_s
\!\left[V_HG_{cf,k}^<+V_H^\dagger G_{fc,k}^<\right],\\
\mathcal X
&=-i\int\frac{d\omega}{2\pi}{\rm Tr}_s[\sigma_xG_{ff}^<],\\
\mathcal I_\Gamma
&={\rm Re}\!\left\{-i\int\frac{d\omega}{2\pi}{\rm Tr}_s
\!\left[K_\Gamma^RG_{ff}^<+K_\Gamma^<G_{ff}^A\right]\right\}.
\label{eq:saddle_forces}
\end{align}
The last line is the Langreth projection of the derivative of the
dissipative influence action.  Dropping $K_\Gamma^<G^A$ generally produces a
complex force and is not a valid thermal saddle.  If the relevant bath is
retained explicitly instead, $\mathcal I_\Gamma$ is absent as a separate term
because it is already contained in $\mathcal F_W$; the two descriptions must
not be combined.
Equation~(\ref{eq:saddle1_app}) is divided by $2r$ only after imposing a
finite-$r$ guard.  At the $r=0$ boundary the undivided equation is used, so
the numerical update never introduces a spurious $1/r$ singularity.
Under the Hadamard rotation,
$\sigma_x\mapsto\sigma_z$, $\sigma_z\mapsto\sigma_x$, and
$V_H\mapsto W_kI$.  Hence in the chiral basis
\begin{align}
\mathcal F_W^\pm
&=-i\int\frac{d\omega}{2\pi}\sum_{k,\eta}W_k
\left(G_{c_\eta f_\eta,k}^<+G_{f_\eta c_\eta,k}^<\right),\\
\mathcal X_\pm
&=-i\int\frac{d\omega}{2\pi}(G_{++}^<-G_{--}^<),
\end{align}
while the relative dissipative force depends on the off-diagonal chiral
components because $\widetilde K_\Gamma\propto iG\sigma_x$.
The full matrix lesser Green function must therefore be retained.
The lesser Green's functions are evaluated from the bath sandwich
$G^<=G^R\Sigma_{\rm bath}^<G^A$, with
$\Sigma_{\rm bath}^<=if\Gamma_{\rm bath}$.  When the complete thermal
influence kernel is retained this is algebraically equivalent to
$G^<=f(\omega)[G^A-G^R]$~\cite{kamenev2023field}; the sandwich form is used
in the solver so that the channel matrix and charge-continuity test remain
explicit.  This is an instantaneous-equilibrium closure at leading
adiabatic order.  Corrections of order $O[(\Omega/\epsilon_O)^2]$ and the
corresponding Wigner gradients would encode drive-induced redistribution;
no energy-pumping result is asserted at the retained order.
For the distinct stationary rotating-drive witness, no Wigner freezing is
used.  The impurity and physical bath fields are untransformed, while the
nearest shifted auxiliary level lies $(\epsilon_O-\Omega)/T_{\rm b}=600$
above the chemical potential.  Its lesser occupation is therefore bounded by
$e^{-600}$ within the discrete off-shell closure; the thermal bath lesser
kernel used for that retarded witness is not an adiabatic assumption.
The causal constraint on the Lagrange multiplier is implemented
by projecting $\delta_c\leftarrow\mathrm{Re}(\delta_c)$ at each
iteration, which is equivalent to enforcing the retarded
(causal) boundary condition.

\paragraph{Scope of the saddle equations.}
Equations~(\ref{eq:saddle2_app})--(\ref{eq:saddle_forces}) specify the
consistent $U\to\infty$ saddle.  The finite-$U=2$ two-denominator
Schrieffer--Wolff expressions used later are a separate frozen analytic
estimate and are not inserted into this saddle.  The principal spectral
figures use frozen $r$ and $\delta_c$ and therefore do not claim a
finite-temperature self-consistent enhancement of the condensate.  The phase
of $b_c$ is a $U(1)$ gauge choice throughout.

The physical projected channels are those of Eq.~(\ref{eq:r2_channels}), with
$V_{\rm eff}=rW$.  Main-text Fig.~\MainFigFrozenComparison\ sets $r=1$
and uses a normalized geometry scan to display
the bifurcation.  Supplemental Fig.~\ref{fig:S_causal_rg} instead uses fixed
$\Gamma_0=\pi\rho W^2=0.12$ and
$\Delta_{\rm eff}^{(0)}=0.10$,
$\Gamma_{\rm PT}^{(0)}=0.20\beta_0$.  Hence $W$ is independent of
$\beta_0$ in the quantitative small-width validation.

\subsection{Low-temperature self-consistent gate}
\label{app:lowT_gate}

The corrected $U\to\infty$ saddle was solved for the fixed-$W$ small-width
path with the complete thermal Markov influence kernel.  At
$T_{\rm b}=10^{-6}$, all real, negative-real, and complex initial boson seeds
converge to the same gauge-fixed solution over
$0.35\leq\beta_0\leq0.55$.  The amplitude decreases smoothly from
$r=0.0106078$ to $0.00949995$; at the core point $\beta_0=0.50$,
$r=0.00980741$ and $\delta_c=1.00000012$.  Across the scan the maximum
constraint residual is $6.54\times10^{-8}$, the maximum divided
stationarity residual is $4.86\times10^{-10}$, the maximum grid-refinement
error is $1.71\times10^{-6}$, and the charge-continuity residual is
$3.66\times10^{-23}$.

Because both $D=r^2D_0$ and $G=r^2G_0$, the local core condition remains
$D_0^2=G_0^2$ and its root is exactly $\beta_0=0.50$ for every finite
$r$.  At $k=0$ the two SOC bath energies coincide, so the Schur dressing is a
common scalar shift and the dressed double-root condition is also exactly the
core condition.  For every tested nonzero $k$, the minimizing $\beta_0$ lies
at the upper boundary $0.55$ of the scan; no interior finite-$k$ EP is
therefore inferred.

At the reference temperature $T_{\rm b}=0.1$, the same closure has no
finite-$r$ solution on this parameter path and reaches the $r=0$ boundary.
This is why the $T_{\rm b}=0.1$ spectra remain frozen diagnostics and why no
finite-temperature condensate or Kondo-scale enhancement is inferred.
Figure~\ref{fig:S_lowT_saddle} reports both the converged low-temperature
branch and the numerical validation checks.

\paragraph{Scope of main-text Fig.~\MainFigFrozenComparison.}
Main-text Fig.~\MainFigFrozenComparison\ is a normalized $k=0$
constant-core calculations with fixed $W=0.75$; the SOC term vanishes at
that momentum, and no physical DOS, cutoff, temperature, or interaction
parameter enters those panels.  The exact
finite-\(U\) statement belongs instead to the controlled linearized contact
model of Sec.~\ref{app:integrable_reduction}; there the standard local
interaction preserves the symmetry-protected EP.

\section{Bath-Controlled FDR and Physical Spectral Function}
\label{app:FDR}

The complete reservoir-coupled problem uses the Keldysh lesser Green's
function fixed by the thermal bath.  No independent mode-by-mode occupation
prescription enters the physical occupations or saddle-point equations.

\paragraph{Physical bath construction.}
With $G^<_{ij}(t,t')=i\langle d_j^\dagger(t')d_i(t)\rangle$ and hence
$n_f=-i\,{\rm Tr}\,G^<(t,t)$,
\begin{align}
\Sigma^<_{\rm bath}&=if\Gamma_{\rm bath},\nonumber\\
G^<_{\rm bath}&=G^R\Sigma^<_{\rm bath}G^A
=f(G^A-G^R)
\label{eq:less_sign_app}
\end{align}
follows directly from the Keldysh identity for a system bilinearly coupled
to a thermal reservoir~\cite{kamenev2023field}.
It satisfies the bath-controlled FDR for this bilinear reservoir setup,
independent of whether $H$ is Hermitian.  In the driven problem this is the
leading instantaneous adiabatic closure; finite-frequency pumping requires
the omitted higher-gradient corrections.
The need for a complete doubled-contour construction is consistent with the
exact high-bias Anderson-impurity solution of Oguri and
Sakano~\cite{oguri2013highbias}, where a finite non-Hermitian
Liouville--Fock Hamiltonian reproduces physical Keldysh Green functions only
with the appropriate paired initial and final states.  Their high-energy
limit is cited as a structural precedent, not as a low-temperature
approximation to the present driven kernel.
In the diagonal-hybridization basis of $\Htilde$, the two
channels $\tilde{d}_\pm$ are independently coupled to the
$c_{k\pm}$ baths, and each satisfies the FDR:
$[\Gless_{\rm bath}]_{\pm\pm}=f(\omega)[\Gadv-\Gret]_{\pm\pm}$.

\paragraph{Number continuity.}
The same complete self-energy satisfies the stationary continuity identity
\begin{equation}
\mathcal I_N=\int\frac{d\omega}{2\pi}\,
{\rm Tr}\!\left[\Sigma^<G^>-\Sigma^>G^<\right]=0 .
\label{eq:number_continuity_app}
\end{equation}
This check uses the full channel matrix, including the thermal partner of the
relative retarded width, and independently validates the bath construction.

\paragraph{Physical DOS identity.}
For the complete physical kernel, $G^A=(G^R)^\dagger$ and
\begin{equation}
\mathcal A=i(G^R-G^A)=G^R\Gamma_{\rm tot}G^A,
\qquad \Gamma_{\rm tot}\succeq0 .
\label{eq:physical_spectral_matrix_app}
\end{equation}
Consequently,
\begin{equation}
\rho_{\rm imp}=\frac{1}{2\pi}\mathrm{Tr}(P_{\rm imp}\mathcal A)
=-\frac{1}{\pi}\mathrm{Im}\,\mathrm{Tr}(P_{\rm imp}G^R).
\label{eq:physical_dos_app}
\end{equation}
The two expressions are identical, not competing density-of-states (DOS)
prescriptions.  We verify their equality and the positivity of the complete
physical response numerically.  A signed response of the compensated
relative kernel after common damping is removed is only a kernel diagnostic
and is not interpreted as a physical density of states.

\section{Shifted Friedel-Type Resonance Condition}
\label{app:friedel}

This appendix gives the conservative form of the shifted resonance
condition used in main-text Sec.~\MainSecFSR.  The purpose is not to claim a
new exact sum rule, but to identify the reference position of the
low-energy peak within the SBMF Green's function.

\subsection{Resonance Position in the Lorentzian Regime}

The spinon retarded Green's function at the SBMF saddle point may be
written as
\begin{equation}
G^R_{f,\sigma}(\omega)
=\frac{1}{\omega-\teps + i\Gamma_\sigma^{\rm net}},
\qquad
\Gamma_\sigma^{\rm net}=\Gamma_\sigma-x_\sigma,
\label{eq:GRf_friedel}
\end{equation}
where $x_\sigma=\zeta_\sigma\Gpt=\pm\Gpt$ is the
spin-selective imaginary shift from the coarse-grained drive and
$\Gamma_\sigma^{\rm net}>0$ in the stable low-energy regime.  The corresponding
physical impurity spectral function is
\begin{equation}
A_{\rm imp}(\omega)
=\frac{1}{\pi}\sum_\sigma
\frac{\Gamma_\sigma^{\rm net}}{(\omega-\mathrm{Re}\,\teps)^2+(\Gamma_\sigma^{\rm net})^2}.
\label{eq:DOS_friedel}
\end{equation}
When the line shape is approximately Lorentzian, both spin channels
are centred at the same reference frequency
\begin{equation}
\omega_{\rm res}\simeq\mathrm{Re}(\teps).
\label{eq:omegaK_exact}
\end{equation}
This is the shifted Friedel-type spectral resonance condition used in the
figures.  In the strongly non-Lorentzian regime, the observed maximum
can deviate from this reference; those deviations are treated as
line-shape effects rather than as a breakdown of a sum rule.

\subsection{Origin of the Shift}

In the Hermitian particle-hole-symmetric SBMF problem,
$\epsilon_\xi=-U/2$ and the spin occupations are equal, so the saddle
point pins $\mathrm{Re}(\teps)$ near zero.  In the driven problem the
coarse-grained self-energy gives unequal effective widths
$\Gamma_\uparrow^{\rm net}\neq\Gamma_\downarrow^{\rm net}$ and a real projected shift
$\dSig$, which together move the saddle value through
\begin{equation}
\mathrm{Re}(\teps)
=\epsilon_\xi+\mathrm{Re}(\delta_c)+\dSig .
\label{eq:omegaK_shift}
\end{equation}
The shift is therefore a gauge-invariant consequence of the
spin-selective self-energy and of the SBMF saddle point.  It is not
attributed to the phase of $b_c$.

\paragraph{Numerical reference.}
For the frozen channel-projected kernel used in the figures,
$\mathrm{Re}(\teps)$ is the renormalized impurity level generated by
the same plotted data pipeline.  In the local-moment reference with
$\epsilon_\xi=-U/2=-1.0$, the frozen closure gives
$\mathrm{Re}(\teps)\simeq -1.0$.  If a separate finite-temperature
self-consistent SBMF closure is used, Eq.~(\ref{eq:omegaK_shift})
should instead be evaluated with that closure and the figure should be
regenerated from the same saddle.

\section{Slave-Boson Mean-Field Kondo Scales}
\label{app:sbmf}

This appendix summarizes the slave-boson mean-field (SBMF) scale
estimate in the rotated spinon basis.  The emphasis is on the
spin-selective passive relative decay shifts, the channel-resolved
resonance width, and the way an EP-active Jordan residue can modify
the screening kernel without implying a universal exact Kondo
temperature.

\subsection{Effective Spinon Hamiltonian}

Using the slave-boson representation $d^\dagger_\sigma=
f^\dagger_\sigma b$ with gauge-invariant condensate amplitude
$r=|b_c|$, the quadratic effective Hamiltonian for the spinon
fermions $f_\sigma$ is:
\begin{equation}
H_{\rm eff}
=\sum_\sigma\tilde{\epsilon}_\sigma
 f^\dagger_\sigma f_\sigma
+\sum_{k\sigma}
 \bigl[r V_{k\sigma}\,c^\dagger_{k\sigma}f_\sigma+\mathrm{h.c.}\bigr],
\label{eq:Heff_sbmf}
\end{equation}
where $V_{k\sigma}$ is the hybridization and
$\tilde{\epsilon}_\sigma=\teps+ix_\sigma$ is
the complex spinon level.
The chiral spinon modes $\tilde{f}_\pm=(f_\uparrow\pm
f_\downarrow)/\sqrt{2}$ are the same rotated basis used in
App.~\ref{app:keldysh}.
The spin-dependent projected imaginary shift
$x_\sigma\equiv \zeta_\sigma\Gpt=\pm\Gpt$
(with $\zeta_\uparrow=+1$, $\zeta_\downarrow=-1$) encodes the
drive-induced passive relative gain/loss; the real part
$\epsilon_{r,\sigma}=\mathrm{Re}(\tilde\epsilon_\sigma)=\teps$
is the renormalized level common to both spins.
The renormalized hybridization width is:
\begin{equation}
\Gamma_\sigma(\omega)
=\pi r^2\sum_k|V_{k\sigma}|^2\delta(\omega-\varepsilon_{k\sigma})
\equiv r^2\Gamma^{(0)}_\sigma(\omega),
\label{eq:Gamma_sbmf}
\end{equation}
and the spinon retarded Green's function is:
\begin{equation}
G^R_{f,\sigma}(\omega)
=\frac{1}{\omega-\tilde{\epsilon}_\sigma+i\Gamma_\sigma}.
\label{eq:GRf_sbmf}
\end{equation}

\subsection{Saddle-Point Equations at $T_{\rm b}=0$}

The constraint $r^2+\sum_\sigma\langle f^\dagger_\sigma f_\sigma\rangle=Q$
and the stationarity condition $\partial\mathcal F_{\rm MF}/\partial\delta_c=0$
give at $T_{\rm b}=0$:
\begin{align}
Q-r^2 &=\sum_\sigma\int_{-D_{\rm uv}}^{0}
\frac{d\omega}{\pi}
\frac{\Gamma_\sigma}{(\omega-\epsilon_{r,\sigma})^2+\Gamma_\sigma^2},
\label{eq:SP1}\\
\delta_c &=\sum_\sigma\int_{-D_{\rm uv}}^{0}
\frac{d\omega}{\pi}
\frac{(\omega-\epsilon_{r,\sigma})\Gamma_\sigma}
     {(\omega-\epsilon_{r,\sigma})^2+\Gamma_\sigma^2}.
\label{eq:SP2}
\end{align}
These are the standard SBMF equations evaluated with the
complex spinon level $\tilde\epsilon_\sigma=\epsilon_{r,\sigma}
+ix_\sigma$.

\subsection{Kondo-scale estimate near an EP}
\label{app:TK_EP}

Within slave-boson mean-field theory the low-energy scale is set by
the renormalized impurity width.  With $r=|b_c|$ and
$V_{\rm eff}=rV$,
\begin{equation}
\tilde{\Gamma}_\sigma(0)=r^2\Gamma^{(0)}_\sigma(0),
\qquad
\Gamma^{(0)}_\sigma(0)=\pi V^2\rho_{b,\sigma}(0).
\label{eq:Gammatilde_sigma}
\end{equation}
Matching to the usual poor-man's-scaling form gives the channel
estimate
\begin{equation}
T_{K,\sigma}\simeq
D_{\rm uv}\exp\!\left[
-\frac{\pi|\tilde{\epsilon}_\sigma|}
{2\Gamma^{(0)}_\sigma(0)}
\right],
\qquad
|\tilde{\epsilon}_\sigma|=
\sqrt{\epsilon_{r,\sigma}^{\,2}+x_\sigma^{\,2}},
\label{eq:TKsigma}
\end{equation}
where $x_\sigma=\zeta_\sigma\Gpt$ is the projected spin-selective imaginary
shift.  The sign difference $x_\uparrow=-x_\downarrow$ is essential
for the $\PT$ structure, but it does not by itself split $T_K$ in
Eq.~(\ref{eq:TKsigma}), because only $x_\sigma^2$ enters.  Distinct
channel scales arise when the SOC-split bath gives
$\Gamma^{(0)}_\uparrow(0)\neq\Gamma^{(0)}_\downarrow(0)$, or when the
real detunings $\epsilon_{r,\sigma}$ differ.

Let $h=\Deff\sigma_x+i\Gpt\sigma_z$, $h^2=s^2I$, and
$a(z)=z-\epsilon_0-\Sigma_0^R(z)$.  Away from the EP, the biorthogonal
projectors and complete resolvent are
\begin{align}
P_\pm&=\frac{1}{2}\left(I\pm\frac{h}{s}\right),\\
\frac{P_+}{a-s}+\frac{P_-}{a+s}
&=\frac{aI+h}{a^2-s^2}.
\label{eq:projector_resolvent_app}
\end{align}
Although $P_\pm$ separately grow as $1/s$, their singular pieces cancel in
the sum.  At the EP, $h_{\rm EP}^2=0$ and the limit is
\begin{equation}
(aI-h_{\rm EP})^{-1}=\frac{I}{a}+\frac{h_{\rm EP}}{a^2}.
\label{eq:Jordan_EP}
\end{equation}
The double pole produces a $t e^{-iE_{\rm EP}t}$ transient; it does not
change contour time ordering or supply a free multiplicative coupling.

The same cancellation occurs in the finite-$U$ virtual charge sectors.  At
particle--hole symmetry with $A=U/2$,
\begin{align}
(AI-h)^{-1}+(AI+h)^{-1}
&=\frac{2A}{A^2-s^2}I,\\
J_{\rm SW}(s)&=\frac{4AW^2}{A^2-s^2}I,\nonumber\\[-0.2em]
J_{\rm SW}(0)&=\frac{8W^2}{U}.
\label{eq:complete_SW_app}
\end{align}
Thus no $1/s$ factor remains in the exchange.  Under the reciprocal
eigenvector rescaling
$|r_\nu\rangle\to a_\nu|r_\nu\rangle$ and
$\langle l_\nu|\to\langle l_\nu|/a_\nu$, both $P_\nu$ and every
consistently transformed vertex--propagator--vertex product are invariant.
Accordingly, no Petermann, condition-number, or $Z^{\rm bio}$ multiplier is
attached to $G^{R,A,<}$, a diagrammatic vertex, or the Kondo exponent.
For example, a Gaussian physical response is assembled from the full matrix
propagators,
\begin{align}
\chi_{AB}^{R}(\omega)
={}&-i\!\int\!\frac{d\nu}{2\pi}\bigl\{
\operatorname{Tr}[A G^R(\nu+\omega)B G^<(\nu)]\notag\\
&\hspace{20mm}+\operatorname{Tr}[A G^<(\nu+\omega)B G^A(\nu)]\bigr\},
\label{eq:physical_matrix_bubble}
\end{align}
so all eigenbranches, projectors, and consistently transformed vertices are
summed before a physical contraction is taken.  Equation~(\ref{eq:physical_matrix_bubble})
illustrates the cancellation but is not needed to prove it: the proof is the
resolvent identity in Eq.~(\ref{eq:projector_resolvent_app}).  Conversely,
that identity does not remove genuine dependence carried by pole motion,
matrix residues, interaction vertices, or relaxation rates.  The finite-$U$
matrix-chain and transition-residue calculations below are independent
physical-basis and sum-rule consistency tests; they are not presented as a
second algebraic proof.

If $\Deff=D_0$ and $\Gpt=g_G\beta_0$, then
\begin{equation}
\partial_{\beta_0}s^2
=-2\Gpt g_G.
\label{eq:slope_s2_app}
\end{equation}
Both amplitudes can increase while $s$ either increases or decreases.
When $s^2$ increases, Eq.~(\ref{eq:complete_SW_app}) can increase through
the ordinary charge denominators; this is not geometric amplification.
For the pre-EP path used in the direct RG test, $s^2$ decreases toward zero.
The complete causal matrix flow gives active/control scale ratios
$1.0567$, $1.0179$, $0.9893$, and $0.9166$ at thresholds
$g_\ast=0.30$, $0.40$, $0.50$, and $1.00$, respectively.  This verifies a
bounded early-flow ordering but not a threshold-robust Kondo-scale
enhancement.

\section{Full Scattering Matrix and Contact Algebra}
\label{app:Smatrix}

The Fisher--Lee and Wigner--Smith constructions used here are standard
scattering tools~\cite{fisherlee1981,wigner1955,smith1960}.  Their role is
to test causality and phase counting for the projected kernel; the novelty
claimed in the article is the microscopic origin of that kernel.

\subsection{Spin, chiral, and scattering-eigenchannel bases}

Let $\Psi_s=(d_\uparrow,d_\downarrow)^T$ denote the original spin basis and
$U_H=2^{-1/2}\bigl(\begin{smallmatrix}1&1\\1&-1\end{smallmatrix}\bigr)$.
For the compensated Markovian core,
\begin{align}
h_s&=\teps I+D\sigma_x+iG\sigma_z-i\bar\Gamma I,\nonumber\\
\Gamma_s&=i(\Sigma_s^R-\Sigma_s^A)
=2(\bar\Gamma I-G\sigma_z),
\label{eq:spin_kernel_rate}
\end{align}
where $D=\Delta_{\rm eff}$ and $G=\Gamma_{\rm PT}$.  With
$z=\omega-\teps+i\bar\Gamma$ and
$\mathcal D(z)=z^2-D^2+G^2$, direct inversion gives
\begin{equation}
G_s^R(\omega)=\frac{1}{\mathcal D(z)}
\begin{pmatrix}
z+iG&D\\ D&z-iG
\end{pmatrix}.
\label{eq:GR_spin_explicit}
\end{equation}
Writing $a=z-iG$ and $b=z+iG$ also gives the Schur forms
\begin{align}
[G_s^R]_{\uparrow\uparrow}
&=\frac{1}{a-D^2/b},&
[G_s^R]_{\downarrow\downarrow}
&=\frac{1}{b-D^2/a},\nonumber\\
[G_s^R]_{\uparrow\downarrow}
&=[G_s^R]_{\downarrow\uparrow}
=\frac{D}{ab-D^2}.
\label{eq:spin_schur_coefficients}
\end{align}
Thus the local denominators are useful Schur complements, but they do not
make the complete Green function or scattering matrix diagonal.

The fixed Hadamard/chiral fields are
$\Psi_\chi=U_H^\dagger\Psi_s=(\widetilde d_+,\widetilde d_-)^T$.
Every object, including the rate matrix, must be rotated:
\begin{align}
h_\chi&=U_H^\dagger h_sU_H
=\teps I+D\sigma_z+iG\sigma_x-i\bar\Gamma I,\nonumber\\
\Gamma_\chi&=U_H^\dagger\Gamma_sU_H
=2(\bar\Gamma I-G\sigma_x),\nonumber\\
G_\chi^R&=U_H^\dagger G_s^RU_H
=\frac{1}{\mathcal D(z)}
\begin{pmatrix}
z+D&iG\\ iG&z-D
\end{pmatrix}.
\label{eq:chiral_kernel_rate_green}
\end{align}
If $p=z-D$ and $q=z+D$, the corresponding chiral Schur coefficients are
\begin{align}
[G_\chi^R]_{++}
&=\frac{1}{p+G^2/q},&
[G_\chi^R]_{--}
&=\frac{1}{q+G^2/p},\nonumber\\
[G_\chi^R]_{+-}
&=[G_\chi^R]_{-+}
=\frac{iG}{pq+G^2}.
\label{eq:chiral_schur_coefficients}
\end{align}
Equations~(\ref{eq:spin_schur_coefficients}) and
(\ref{eq:chiral_schur_coefficients}) are the corrected coefficient
evaluation replacing the former block-diagonal contact argument.

\subsection{Contact matching and Fisher--Lee coefficients}

Choose an asymptotic-channel coupling matrix $\mathcal W_\nu$ such that
$2\pi\mathcal W_\nu\mathcal W_\nu^\dagger=\Gamma_\nu$, with
$\nu=s,\chi$.  For an incoming channel vector $A_\nu^{\rm in}$, the
impurity contact amplitude and outgoing vector satisfy
\begin{align}
e_\nu(\omega)&=G_\nu^R(\omega)\mathcal W_\nu A_\nu^{\rm in},\nonumber\\
A_\nu^{\rm out}
&=A_\nu^{\rm in}-2\pi i\mathcal W_\nu^\dagger e_\nu
=S_\nu(\omega)A_\nu^{\rm in}.
\label{eq:contact_matching}
\end{align}
For the positive square-root factorization this gives
\begin{equation}
S_\nu=I-i\Gamma_\nu^{1/2}G_\nu^R\Gamma_\nu^{1/2},
\qquad
S_\chi=U_H^\dagger S_sU_H .
\label{eq:Fisher_basis_covariance}
\end{equation}
In the spin representation define
$\gamma_\uparrow=2(\bar\Gamma-G)$ and
$\gamma_\downarrow=2(\bar\Gamma+G)$.  The explicit coefficients are
\begin{align}
[S_s]_{\uparrow\uparrow}
&=1-i\gamma_\uparrow\frac{z+iG}{\mathcal D(z)},\nonumber\\
[S_s]_{\downarrow\downarrow}
&=1-i\gamma_\downarrow\frac{z-iG}{\mathcal D(z)},\nonumber\\
[S_s]_{\uparrow\downarrow}
&=[S_s]_{\downarrow\uparrow}
=-i\sqrt{\gamma_\uparrow\gamma_\downarrow}
\frac{D}{\mathcal D(z)} .
\label{eq:S_spin_coefficients}
\end{align}
The chiral coefficients must be obtained from the same covariant rotation,
or equivalently from Eq.~(\ref{eq:Fisher_basis_covariance}) using
$\Gamma_\chi$ and $G_\chi^R$.  It is incorrect to rotate $G^R$ while
leaving $\Gamma$ diagonal.

For the validation point
$\bar\Gamma=0.12$ and $D=G=0.10$, at
$\omega=\teps$,
\begin{align}
S_s&=
\begin{pmatrix}
\mathsf a&i\mathsf b\\
i\mathsf b&\mathsf a
\end{pmatrix},\nonumber\\
S_\chi&=
\operatorname{diag}(\mathsf a+i\mathsf b,\mathsf a-i\mathsf b),
\nonumber\\
\mathsf a&=0.3888889,\qquad \mathsf b=0.9212847 .
\label{eq:S_basis_numeric_audit}
\end{align}
Consequently the earlier number $0.9213$ is a spin-representation
off-diagonal coefficient and, at the resonance center, the imaginary part
of the two unit-modulus scattering eigenvalues.  It is not a chiral
conversion amplitude.  In the same Markov gate,
$[S_\chi(\teps)]_{+-}=0$, while
$\max_\omega|[S_\chi]_{+-}|=0.8333333$ at
$|\omega-\teps|=0.12$.  These component values are basis dependent; the
eigenphases, $\det S$, and Wigner--Smith eigenvalues and trace are the
quantities used in the basis-independent checks.

\subsection{Four contact amplitudes and exact folded identity}
\label{app:four_contact_exact}

The four coefficients that occur when the two incoming channels are treated
simultaneously are most cleanly organized as the two columns of the impurity
contact Green function.  Let $e_{\eta'\eta}$ denote the impurity amplitude in
fixed chiral component $\eta'$ generated by an incoming asymptotic channel
$\eta$.  With
\begin{equation}
p=z-D,\qquad q=z+D,\qquad
\mathcal P_0=pq+G^2 ,
\label{eq:contact_pq}
\end{equation}
the exact quadratic contact equations are
\begin{equation}
\begin{pmatrix}
p&-iG&0&0\\
-iG&q&0&0\\
0&0&q&-iG\\
0&0&-iG&p
\end{pmatrix}
\begin{pmatrix}
e_{++}\\e_{-+}\\e_{--}\\e_{+-}
\end{pmatrix}
=\mathcal W
\begin{pmatrix}
\Phi_+\\0\\\Phi_-\\0
\end{pmatrix}.
\label{eq:four_contact_exact}
\end{equation}
Here $\mathcal W$ is the contact coupling and $\Phi_\eta$ is the incoming
amplitude.  Equation~(\ref{eq:four_contact_exact}) is simply the two-column
form of Eq.~(\ref{eq:chiral_kernel_rate_green}); it is therefore exact for
the quadratic projected kernel and introduces no two-particle assumption.

Rows 2 and 4 give
\begin{equation}
e_{-+}=\frac{iG}{q}e_{++},\qquad
e_{+-}=\frac{iG}{p}e_{--}.
\label{eq:cross_contact_elimination}
\end{equation}
Eliminating them leaves the two direct amplitudes with dressed contact
denominators
\begin{align}
\widehat p\,e_{++}&=\mathcal W\Phi_+,&
\widehat q\,e_{--}&=\mathcal W\Phi_-,
\nonumber\\
\widehat p&=p+\frac{G^2}{q}=\frac{\mathcal P_0}{q},&
\widehat q&=q+\frac{G^2}{p}=\frac{\mathcal P_0}{p}.
\label{eq:folded_contact_denominators}
\end{align}
Consequently,
\begin{equation}
\frac{\widehat p}{\widehat q}=\frac{p}{q}.
\label{eq:contact_ratio_identity}
\end{equation}
This identity is useful but must be interpreted correctly.  Both folded
denominators contain the same quadratic pole polynomial $\mathcal P_0$;
the ratio removes that common factor.  It therefore exposes the common
denominator responsible for the Jordan limit and for cancellation between
individual projectors.  It is not a basis-independent amplification factor.
For simultaneous equal sources one obtains
$e_{--}/e_{++}=p/q$, whereas changing the incoming superposition changes
that amplitude ratio.

\subsection{Controlled finite-\texorpdfstring{$U$}{U} Anderson contact
algebra}
\label{app:integrable_reduction}

The four amplitudes in Eq.~(\ref{eq:four_contact_exact}) are the two columns
of a \emph{one-particle} Green function.  They should not be confused with
the two-electron exchange matrix.  To obtain the latter, first impose the
controlled reduction stated in the main text: equal folded velocities,
linear dispersion, constant \(M=D\sigma_z+iG\sigma_x\),
momentum-independent scalar hybridization \(V\), and the standard local
interaction \(U n_{d+}n_{d-}\).  With momenta measured in energy units, the
two-electron bath wave function in the ordered sector \(x_1<x_2\) is
\begin{align}
\Psi_{ab}(x_1,x_2)={}&
\mathcal A_{12}^{ab}e^{ipx_1+iqx_2}
\notag\\
&+\mathcal A_{21}^{ab}e^{iqx_1+ipx_2},
\qquad x_1<x_2,
\label{eq:two_electron_ansatz_app}
\end{align}
supplemented by sectors with one bath electron and one impurity electron and
by a doubly occupied impurity amplitude \(D_{pq}\).  Substitution in the
Schr\"odinger equation, integration across the point contact, and elimination
of those impurity amplitudes separate into triplet and singlet sectors.  In
the normalization
\(\Gamma_A=V^2/(2v)\), the resulting contact equations are
~\cite{Wiegmann1980,KawakamiOkiji1982,WiegmannTsvelick1983,
ChaoPalacios2011}
\begin{align}
\mathcal A_{21}^{(t)}&=\mathcal A_{12}^{(t)},\label{eq:triplet_match_app}\\
\left[\Delta\mathcal B+i\,2U\Gamma_A\right]\mathcal A_{21}^{(s)}
&=\left[\Delta\mathcal B-i\,2U\Gamma_A\right]\mathcal A_{12}^{(s)},
\label{eq:singlet_match_app}\\
\Delta\mathcal B&=\mathcal B(p)-\mathcal B(q),\nonumber\\
\mathcal B(p)&=p(p-2\epsilon_d-U).
\label{eq:B_def_app}
\end{align}
Equations~(\ref{eq:triplet_match_app}) and
(\ref{eq:singlet_match_app}) display exactly where the virtual doublon
enters: it changes only the antisymmetric pseudospin channel.  With
\(\Pi_t=(I+P_{12})/2\) and \(\Pi_s=(I-P_{12})/2\),
\begin{align}
\mathcal A_{21}&=R_{12}(p,q)\mathcal A_{12},\nonumber\\
R_{12}(p,q)
&=\Pi_t+
\frac{\Delta\mathcal B-i\,2U\Gamma_A}
     {\Delta\mathcal B+i\,2U\Gamma_A}\Pi_s
\nonumber\\
&=\frac{\Delta\mathcal B\,I_{12}
       +i\,2U\Gamma_A P_{12}}
      {\Delta\mathcal B+i\,2U\Gamma_A}.
\label{eq:Anderson_R_app}
\end{align}
Thus the contact matrix follows directly from the two projector-resolved
matching equations; no rank-one Feshbach ansatz is added.

For \(U\ne0\), set
\begin{equation}
u(p)=\frac{\mathcal B(p)}{2U\Gamma_A},\qquad x=u(p)-u(q).
\label{eq:dressed_rapidity_app}
\end{equation}
Then
\begin{equation}
R_{12}(x)=\frac{xI_{12}+iP_{12}}{x+i}
=\frac{1}{x+i}
\begin{pmatrix}
x+i&0&0&0\\
0&x&i&0\\
0&i&x&0\\
0&0&0&x+i
\end{pmatrix}.
\label{eq:R4_explicit_app}
\end{equation}
The bare pair-energy factor is absorbed exactly because
\begin{equation}
\mathcal B(p)-\mathcal B(q)
=(p-q)(p+q-2\epsilon_d-U).
\label{eq:B_factor_app}
\end{equation}
The rapidity map is generically two-to-one,
\(u(p)=u(2\epsilon_d+U-p)\), and is singular at \(U=0\).  Global
injectivity is not required for the matrix algebra, while the \(U=0\)
quadratic benchmark is treated separately.

For the unnormalized matrix
\(\mathcal R_{12}(x)=xI_{12}+iP_{12}\), the permutation identities give
\begin{align}
&\mathcal R_{12}(u-v)\mathcal R_{13}(u-w)
\mathcal R_{23}(v-w)
\notag\\
&\qquad=
\mathcal R_{23}(v-w)\mathcal R_{13}(u-w)
\mathcal R_{12}(u-v),
\label{eq:YBE_app}
\end{align}
and, for every \(S\in GL(2,\mathbb C)\),
\begin{equation}
(S\otimes S)^{-1}\mathcal R_{12}(x)(S\otimes S)
=\mathcal R_{12}(x),
\label{eq:R_GL2_app}
\end{equation}
because \([P_{12},S\otimes S]=0\).  For
\(S=S_C=e^{\theta\sigma_y/2}\), \(\tanh\theta=G/D\),
\begin{equation}
S_C^{-1}(D\sigma_z+iG\sigma_x)S_C
=\sqrt{D^2-G^2}\,\sigma_z .
\label{eq:CPT_diag_app}
\end{equation}
The same canonical transformation on bath and impurity fields leaves the
kinetic term, scalar hybridization, and
\(n_{d+}n_{d-}=N_d(N_d-1)/2\) invariant.  It therefore preserves the
ordinary YBE rather than producing a new bulk \(R\)-matrix.

The non-diagonal pseudospin sector can be displayed without choosing
the diagonal positive-metric frame.  Define
\begin{equation}
\mathcal J^\mu
=\frac12\int_0^{\mathcal L}dx\,\bar\chi\sigma_\mu\chi
 \frac12\bar d\sigma_\mu d,\qquad
\mathcal J^\pm=\mathcal J^x\pm i\mathcal J^y .
\label{eq:global_ladders_app}
\end{equation}
For \(M=D\sigma_z+iG\sigma_x\), the conserved global charge is
\begin{equation}
Q_M
=2D\mathcal J^z+iG(\mathcal J^++\mathcal J^-).
\label{eq:global_charge_ladders_app}
\end{equation}
Accordingly, the nonzero ladder terms (often written \(S^\pm\)) are
not discarded.  For \(D^2\ne G^2\), the global canonical action
induced by \(S_C\) maps Eq.~(\ref{eq:global_charge_ladders_app}) to
\(2\sqrt{D^2-G^2}\,\mathcal J^z\), while
Eq.~(\ref{eq:R_GL2_app}) keeps the same Yang--Baxter algebra.

At \(D^2=G^2\), the diagonalizer in Eq.~(\ref{eq:CPT_diag_app}) is singular,
but \(M^2=0\).  On a finite ring the bounded gauge map moves the constant
matrix into
\begin{equation}
\mathcal G=e^{iM\mathcal L/v}=I+iM\mathcal L/v,
\qquad \mathcal G^{-1}=I-iM\mathcal L/v .
\label{eq:unipotent_twist_app}
\end{equation}
Because Eq.~(\ref{eq:R_GL2_app}) also holds for this non-diagonalizable
\(\mathcal G\), the monodromy
\begin{equation}
T_a(z)=\mathcal G_a
\mathcal R_{aN}(z-u_N)\cdots\mathcal R_{a1}(z-u_1)
\label{eq:twisted_monodromy_app}
\end{equation}
satisfies twisted RTT and generates commuting transfer matrices.  The
finite-volume domain proof, many-body Jordan theorem, projective
descendant-spin-root contraction, and four-state Fock lift are proved in the companion work
~\cite{KulkarniYB2026}.  There the EP value of
Eq.~(\ref{eq:global_charge_ladders_app}) is treated directly as a
nonzero nilpotent global ladder generator: it creates the
symmetry-protected Jordan chain rather than acting as an omitted
integrability-breaking perturbation.  These results are quoted here only to
identify the precise continuation of the physical contact algebra.

The supplied script \nolinkurl{verify_finiteU_contact_algebra.py} independently
checks the displayed \(4\times4\) matrix.  For generic complex rapidities and
a generic non-unitary \(S\), it gives Frobenius residuals
\(1.14\times10^{-15}\) for YBE,
\(4.48\times10^{-16}\) for \(GL(2,\mathbb C)\) covariance, and
\(1.66\times10^{-16}\) for the metric diagonalization; the singlet and triplet
eigenvalue residuals vanish to machine precision.

\subsection{One-particle contact-phase quantization}

Inside the passivity window, $S^\dagger S=I$.  Let
$\mathsf{s}_a=e^{2i\delta_a}$, $a=1,2$, denote its two eigenvalues.
These scattering eigenchannels are not identified with the fixed chiral
labels.  Periodic closure on a ring after the stated finite-window radial
linearization gives the exact one-particle phase condition
\begin{equation}
\det[I-e^{ik\mathcal L}S(k)]=0
\quad\Longleftrightarrow\quad
e^{ik_{ja}\mathcal L}\mathsf{s}_a(k_{ja})=1 .
\label{eq:S_full_quantization}
\end{equation}
This is exact phase counting for the complete causal one-particle matrix.
It is logically distinct from the multiparticle contact matrix
Eq.~(\ref{eq:Anderson_R_app}), which requires the additional controlled
linearization stated above.

\section{Biorthogonal Structure and Eigenvectors}
\label{app:biorth}

The frozen impurity block is
$h_d=\bigl[\begin{smallmatrix}
\teps+D&iG\\iG&\teps-D\end{smallmatrix}\bigr]$
with splitting $s=\sqrt{D^2-G^2}$ and eigenvalues
$E_\pm=\teps\pm s$.

\paragraph{Right eigenvectors.}
\begin{equation}
|r_+\rangle\propto\begin{pmatrix}G\\i(D-s)\end{pmatrix},\quad
|r_-\rangle\propto\begin{pmatrix}G\\i(D+s)\end{pmatrix}.
\label{eq:right_evecs}
\end{equation}

\paragraph{Left eigenvectors.}
Because $h_d$ is complex symmetric, the left eigenvectors may be taken
as the transposes of the right eigenvectors:
\begin{equation}
\langle l_+|\propto\begin{pmatrix}G & i(D-s)\end{pmatrix},\quad
\langle l_-|\propto\begin{pmatrix}G & i(D+s)\end{pmatrix}.
\label{eq:left_evecs}
\end{equation}
Direct verification gives $\langle l_\pm|r_\mp\rangle=0$ and
\begin{equation}
\langle l_+|r_+\rangle=2s(D-s),\qquad
\langle l_-|r_-\rangle=-2s(D+s).
\end{equation}
Thus the biorthogonal normalization diverges as $1/s$ at the EP:
\begin{equation}
\mathcal{M}_\pm\mathcal{N}_\pm
=\frac{1}{\langle l_\pm|r_\pm\rangle}
\sim \frac{1}{s}\xrightarrow{s\to0}\infty.
\label{eq:biorth_norm}
\end{equation}
This is the normalization of an individual eigenspace, not an additional
physical spectral weight.  The rescaling cancels within each normalized
projector, and the divergent pieces of the two projectors cancel in the
complete resolvent [Eq.~(\ref{eq:projector_resolvent_app})].  It is
consistent with $\keff\sim C/|s|$ [Eq.~(\ref{eq:kappa_scaling})].
At the EP ($s\to0$): $|r_+\rangle\propto|r_-\rangle$,
eigenvectors coalesce, and the Jordan block emerges with
$h_d|r_+\rangle=E_{\rm EP}|r_+\rangle$,
$(h_d-E_{\rm EP})|j\rangle=|r_+\rangle$.

\subsection{Explicit Contact Projector}

The normalized rank-one projector in the $\{\tilde d_+,\tilde d_-\}$
contact space is
\begin{equation}
P_+=\frac{|r_+\rangle\langle l_+|}{\langle l_+|r_+\rangle}
=\frac{1}{2s(D-s)}
\begin{pmatrix}
G^2 & iG(D-s)\\
iG(D-s) & -(D-s)^2
\end{pmatrix},
\label{eq:Pplus_contact}
\end{equation}
with $P_+^2=P_+$.
The full Hamiltonian $\Htilde(k)$ contains additional diagonal bath
dispersions $\varepsilon_{c,\pm}(k)$ that do not enter this local
projector.  The projector is used only to track biorthogonal normalization;
by itself it does not establish a two-body $R$-matrix for the full
energy-dependent channel-converting problem.  The distinct finite-\(U\)
contact derivation under controlled linearization is given in
Sec.~\ref{app:integrable_reduction}.

\section{Causal Fisher--Lee and Wigner--Smith Check}
\label{app:smith}

For the compensated Markovian small-width kernel in the spin
representation, define
\begin{align}
H_s&=D\sigma_x-i\bar\Gamma I+iG\sigma_z,\qquad
G_s^R=(\omega I-H_s)^{-1},\\
\Gamma_s&=2(\bar\Gamma I-G\sigma_z).
\end{align}
The passivity condition $\bar\Gamma\geq|G|$ makes
$\Gamma_s$ positive semidefinite.  The scattering and
delay matrices are
\begin{align}
S_s&=I-i\Gamma_s^{1/2}G_s^R\Gamma_s^{1/2},\\
Q_s&=-iS_s^\dagger\partial_\omega S_s .
\end{align}
The chiral matrices follow by the simultaneous rotation in
Eq.~(\ref{eq:Fisher_basis_covariance}).  Direct use of the Dyson identity
gives $S_\nu^\dagger S_\nu=I$ and $Q_\nu=Q_\nu^\dagger$ in either
representation $\nu=s,\chi$.
The trace also obeys
\begin{equation}
\operatorname{Tr}Q_\nu=-i\,\partial_\omega\log\det S_\nu .
\label{eq:smith_det}
\end{equation}
In the compensated unbroken Markov core, let
$x=\omega-\teps$ and $s=\sqrt{D^2-G^2}\in\mathbb R$.  Causality fixes
the scattering determinant entirely from the two poles and their
advanced zeros:
\begin{align}
\det S(\omega)
&=\prod_{\alpha=\pm}
\frac{x-\alpha s-i\bar\Gamma}
     {x-\alpha s+i\bar\Gamma},
\label{eq:detS_two_poles}\\
\operatorname{Tr}Q(\omega)
&=\sum_{\alpha=\pm}
\frac{2\bar\Gamma}
{(x-\alpha s)^2+\bar\Gamma^2}.
\label{eq:smith_two_pole_profile}
\end{align}
At the resonance center this gives the analytic pre-EP concentration
\begin{equation}
\operatorname{Tr}Q(\teps)
=\frac{4\bar\Gamma}{s^2+\bar\Gamma^2}.
\label{eq:smith_center_value}
\end{equation}
Relative to the matched Hermitian kernel,
\begin{equation}
\frac{\operatorname{Tr}Q_G(\teps)}
{\operatorname{Tr}Q_{G=0}(\teps)}
=\frac{D^2+\bar\Gamma^2}
{D^2-G^2+\bar\Gamma^2}>1
\label{eq:smith_preep_ratio}
\end{equation}
for $0<|G|<|D|$ at fixed $D$ and $\bar\Gamma$.  This is a
basis-independent analytic concentration of the scattering phase density
along the specified matched path: the two fixed-weight phase peaks move
together and concentrate at the center.  Its integral remains two, and
changing $D$ or the rate eigenvalues provides adverse controls, so
Eq.~(\ref{eq:smith_preep_ratio}) is not a universal enhancement theorem or
a Kondo-temperature formula.
Resolving the determinant into the two eigenvalues $\mathsf{s}_a$ of the complete
unitary matrix gives
\begin{equation}
\frac{1}{2\pi}\operatorname{Tr}Q_\nu
=\sum_{a=1}^{2}\frac{1}{2\pi i}\partial_\omega\log \mathsf{s}_a
\equiv K_{\rm ph} .
\label{eq:smith_ba_kernel_app}
\end{equation}
Thus the logarithmic contact-phase density of
main-text Eq.~(\MainEqContactPhaseCompact) is not an independent scale ansatz:
it is the causal scattering density-of-states shift of the same complete
retarded/advanced problem.  At $\beta_0=0.50$, the independent full-matrix
gate gives a maximum scattering-eigenvalue modulus error
$9.99\times10^{-16}$ and a finite-difference phase--Smith identity error
$5.53\times10^{-9}$.  The basis-resolved coefficients are given in
Eq.~(\ref{eq:S_basis_numeric_audit}); no individual off-diagonal matrix
element is used as a basis-independent diagnostic.

At the EP, $D=G$ and the pole pair coalesces at
$E_{\rm EP}=-i\bar\Gamma$.  The complete Jordan resolvent gives
\begin{equation}
\operatorname{Tr}Q_{\rm EP}(\omega)
=\frac{4\bar\Gamma}{\omega^2+\bar\Gamma^2},
\qquad
\int\frac{d\omega}{2\pi}\operatorname{Tr}Q_{\rm EP}=2 .
\label{eq:smith_ep_app}
\end{equation}
For $\bar\Gamma=0.12$, this predicts
$\operatorname{Tr}Q_{\rm EP}(0)=33.3333333$, reproduced numerically to
machine precision.  The finite-window numerical integral is $1.99236$; the
small deficit is the omitted Lorentzian tail.  Across the validation grid,
the maximum scattering-unitarity error is $1.33\times10^{-15}$, the maximum
Smith-Hermiticity error is $2.52\times10^{-14}$, and the maximum discrepancy
in Eq.~(\ref{eq:smith_det}) is $4.97\times10^{-14}$.

Figure~\ref{fig:S_smith} includes the essential adverse control.  A
rate-matched off-EP kernel with $D=0.05$ reaches a largest proper-delay peak
$63.77$, exceeding the EP value $30.56$.  Once normalized by the slowest
decay rate, the EP is not enhanced.  The delay profile is therefore used as
a causal coalescence and linewidth fingerprint, not as an independent
many-body amplification factor.

\section{Causal Channel Density and Matrix Poor-Man's Scaling}
\label{app:matrix_rg}

This appendix records the complete Green-function matrix analysis used for
Supplemental Fig.~\ref{fig:S_causal_rg}.  It is the standard one-loop
poor-man's-scaling construction~\cite{anderson1970poor} applied to the
energy-dependent positive channel density of the projected kernel.

\subsection{Positive matrix density}

For either the spin or chiral representation,
\begin{align}
\rho_\nu(\omega)
&=\frac{i}{2\pi}\left[G_\nu^R(\omega)-G_\nu^A(\omega)\right]\nonumber\\
&=\frac{1}{2\pi}G_\nu^R(\omega)\Gamma_\nu(\omega)
G_\nu^A(\omega),\qquad \nu=s,\chi .
\label{eq:rho_matrix_rg}
\end{align}
The second equality is the retarded/advanced Dyson identity.  It proves
positivity directly:
\begin{equation}
v^\dagger\rho_\nu v
=\frac{1}{2\pi}
\left\|\Gamma_\nu^{1/2}G_\nu^Av\right\|^2\geq0 .
\label{eq:rho_psd_proof}
\end{equation}
It also makes basis covariance explicit,
$\rho_\chi=U_H^\dagger\rho_sU_H$.  Consequently the two density
eigenvalues are independent of whether the calculation is displayed in
the spin or fixed chiral basis.

\subsection{Exchange matrix and flow}

The finite-$U$ Schrieffer--Wolff transformation fixes the reciprocal bare
exchange matrix,
\begin{equation}
J_0=2W^2\left(
\frac{1}{|\epsilon_d|}
+\frac{1}{|\epsilon_d+U|}
\right)I .
\label{eq:J0_matrix_rg}
\end{equation}
Let $\ell=\ln(D_{\rm uv}/\Lambda)$ and define the symmetric shell density
\begin{equation}
\rho_{\rm sh}(\Lambda)
=\rho(+\Lambda)+\rho(-\Lambda).
\label{eq:rho_shell}
\end{equation}
The noncommuting one-loop flow integrated in the numerical gate is
\begin{equation}
\frac{dJ}{d\ell}
=J(\ell)\rho_{\rm sh}(\Lambda)J(\ell).
\label{eq:matrix_poor_man_flow}
\end{equation}
No simultaneous diagonalization of $J$ and $\rho_{\rm sh}$ is assumed.
For Hermitian $J$ and positive Hermitian $\rho_{\rm sh}$,
$J\rho_{\rm sh}J$ is Hermitian, so the flow preserves Hermiticity.
Under a fixed unitary basis change, $J\mapsto U^\dagger JU$ and
$\rho\mapsto U^\dagger\rho U$, making
Eq.~(\ref{eq:matrix_poor_man_flow}) covariant.

To compare channels without assigning significance to a basis component,
the code forms
\begin{align}
\rho_{\rm av}(\Lambda)&=\frac{\rho_{\rm sh}(\Lambda)}{2},\nonumber\\
g(\Lambda)&=\rho_{\rm av}^{1/2}J(\Lambda)
\rho_{\rm av}^{1/2},\qquad
g_{\max}=\lambda_{\max}[g(\Lambda)] .
\label{eq:dimensionless_matrix_g}
\end{align}
The operational scale $T_\ast(g_\ast)$ is the cutoff at which
$g_{\max}(\Lambda)$ first reaches a specified threshold $g_\ast$.
It is deliberately called an operational weak-flow scale, not a
thermodynamic Kondo temperature.

\subsection{Coefficient and convergence audit}

For $k_{\rm max}=\pi/4$ the gate uses
$\rho_{\rm ref}=1/(2k_{\rm max})$, $\Gamma_0=\pi\rho_{\rm ref}W^2=0.12$,
and therefore
\begin{equation}
W^2=0.06,\qquad J_0=0.24 .
\label{eq:rg_numeric_coefficients}
\end{equation}
At $\beta_0=0.50$, $D_0=G_0=0.10$ and the two rate eigenvalues are
$0.04$ and $0.44$.  The selected-grid minimum density eigenvalue is
$0.0365034$, while the strict pre-EP-window minimum is
$0.0241015>0$.  The maximum difference between the two forms
of Eq.~(\ref{eq:rho_matrix_rg}) is $1.79\times10^{-15}$, the exchange
Hermiticity error is zero, and the maximum step-refinement drift is
$5.92\times10^{-10}$.

The active/control threshold-scale ratios at $\beta_0=0.50$ are
$1.0567$, $1.0179$, $0.9893$, and $0.9166$ for
$g_\ast=0.30$, $0.40$, $0.50$, and $1.00$, respectively.  The ordering
therefore reverses near $g_\ast=0.4627$.  The calculation establishes a
bounded early-flow ordering, but not a threshold-independent
enhancement of the strong-coupling Kondo scale.

\section{Finite-\texorpdfstring{$U$}{U} Biorthogonal NRG Residue Audit}
\label{app:nrg_jordan}

We next test the Jordan response by an operator-resolved non-Hermitian NRG
calculation in the scalar-hybridization Wilson-chain control of the projected
model~\cite{burke2025non}.  This is a weak-coupling structural control of the
local finite-$U$ residue in the early Wilson shells, not a low-energy fixed
point or a phase diagram for the full SOC-resolved bath.  It therefore
neither locates nor excludes the interacting continuation of the full-$k$
causal pole EP.  That distinct problem is addressed by the exact finite-chain
matrix block-Wilson calculation of Sec.~\ref{app:finiteU_fullk}.  The present
section instead asks whether the scalar Markov-core transition residue retains
the nilpotent matrix direction.  The
two transition operators are propagated independently as
$L^\dagger cR$ and $L^\dagger c^\dagger R$; a no-truncation regression checks
their noninteracting matrix elements before the interacting scan.

At the exact Jordan point a complete biorthogonal eigenbasis is singular.
We therefore calculate at finite coherent detuning
$\delta_{\rm coh}>0$, track the same two impurity-supported poles through
$U$ and $\delta_{\rm coh}$, and only then extrapolate to
$\delta_{\rm coh}\to0$.  If
\begin{equation}
G(z)=\frac{W_+}{z-z_+}+\frac{W_-}{z-z_-}+G_{\rm reg}(z),
\end{equation}
the coefficient that approaches the Jordan double-pole residue is
\begin{equation}
B=\frac{1}{2}(W_+-W_-)(z_+-z_-)
 =B_0 I+B_x\sigma_x+B_y\sigma_y+B_z\sigma_z .
\label{eq:nrg_double_pole_B}
\end{equation}
In the Hadamard/chiral basis used by the Wilson-chain adapter, the compensated
EP generator is
\begin{equation}
N_{\rm EP}=\Deff\sigma_z+i\Gpt\sigma_x,
\qquad N_{\rm EP}^2=0,
\label{eq:nrg_nilpotent_generator}
\end{equation}
we monitor the Frobenius-angle alignment, normalized trace, and nilpotent
component mismatch,
\begin{align}
A_J&=\frac{|\operatorname{Tr}(N_{\rm EP}^\dagger B)|}
{\|N_{\rm EP}\|_F\|B\|_F},\nonumber\\
f_{\rm tr}&=\frac{|\operatorname{Tr}B|}{\sqrt{2}\|B\|_F},
&
\epsilon_N&=\frac{|B_x-iB_z|}{|B_x|+|B_z|}.
\label{eq:nrg_jordan_diagnostics}
\end{align}

Supplemental Fig.~\ref{fig:S_NRG_Jordan} reports the validated part of the
audit.  The reference calculation uses
$\Lambda=3$, $z=0.5$, $N_{\rm keep}=320$, bath exponent $r=1$,
$\beta_0=0.50$, $\Deff=\Gpt=0.10$, $0\leq U\leq0.10$, and the
early/high-energy NRG iterations $n=4$--$6$ in charge sectors $q=\pm1$.
Five detunings between
$10^{-5}$ and $10^{-4}$ are used.  The reliable tracked-pair fraction is
$0.993939$; over the direct finite-detuning ensemble the median alignment is
$0.99999963$, the median trace fraction is $8.3567\times10^{-4}$, and the
median nilpotent mismatch is $7.3179\times10^{-7}$.  Thus the finite-$U$
calculation retains the nearly traceless nilpotent matrix direction.

The stronger pole-splitting claims do not pass the same convergence gate.
Across iteration, sector, fitting window, and detuning model, the complex
interaction-exponent ensemble has median $p=1.359$ with a $68\%$ spread of
$1.991$.  The proposed complex quartic coefficient also fails its plateau
test (median absolute log-slope $0.880$ and relative complex scatter
$1.118$).  These failures do not prove the absence of a scaling regime, but
they preclude extracting a universal exponent, quartic coefficient, or
interaction-induced EP floor from this calculation.  Accordingly only the
matrix-residue result is used in the main text.

\section{Scope and Division of Labor}
\label{app:YBE}

Three logically distinct statements are now kept separate.

\begin{enumerate}[label=(\roman*)]
\item For the complete frozen retarded kernel derived from the driven model,
the exact claims are the matrix Fisher--Lee relation, one-particle
contact-phase quantization, and the Wigner--Smith identity.  These require no
many-body factorization.

\item After equal-velocity branchwise linearization and folding, with
constant \(M\), scalar hybridization, and component-symmetric local \(U\),
the exact two-electron contact matrix is
Eq.~(\ref{eq:Anderson_R_app}).  Its dressed-rapidity difference form and
\(GL(2,\mathbb C)\) covariance prove YBE and the compatibility of the
metric similarity.  On this submanifold the standard interaction does
not lift the symmetry-protected EP.

\item Ref.~\cite{KulkarniYB2026} supplies the mathematical continuation:
the bounded finite-volume gauge equivalence, RLL and twisted RTT relations,
commuting transfer matrices, the explicit treatment of the nonzero global
pseudospin ladders, many-body Jordan blocks, and the compatible four-state
Fock lift.  Its descendant \emph{spin} roots meet at the
projective point \(\lambda=\infty\), with scaled coordinates given by a
generalized Laguerre polynomial; no finite Anderson charge-root coalescence
is asserted.  Those results are not rederived from the physical scattering
plots in this article.
\end{enumerate}

Curvature, unequal folded velocities, energy-dependent \(M(\omega)\),
nonscalar hybridization, or component-dependent interactions break one or
more hypotheses of item (ii).  They can shift or split the EP and may spoil
factorization.  No Yang--Baxter theorem or thermodynamic Bethe ansatz is
claimed for that complete driven kernel.  Conversely, no universal
interaction-induced avoided-coalescence scale is introduced: such a scale
would require a specified symmetry-breaking self-energy calculation.


\setcounter{figure}{0}
\renewcommand{\thefigure}{S\arabic{figure}}

\begin{figure*}[p]
\centering
\includegraphics[width=0.92\textwidth]{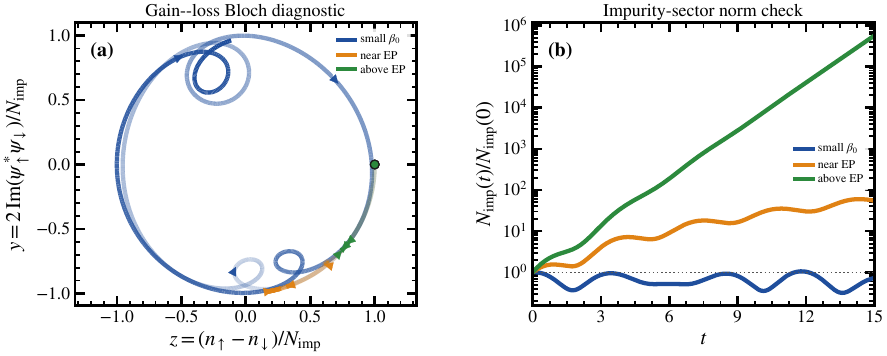}
\caption{%
\textbf{Gain--loss Bloch diagnostic for the impurity sector.}
This is the same normalized $k=0$ constant-core calculation as main-text
main-text Fig.~\MainFigFrozenComparison: $\teps=-1$, $D=0.50$ in panels
(a)--(c), $D=0$ in panels (d)--(f), $G=|\beta_0|$, and fixed
$W=0.75$.  It contains no physical temperature, interaction, DOS, or cutoff.
Panel (a) shows the normalized $(z,y)$ trajectory for representative values of
$\beta_0$ below, near, and above the EP.  Below the EP the normalized co-decaying trajectory remains close to the
unbroken relative-kernel orbit, while close to and beyond the EP the trajectory
is progressively distorted.  Panel (b) tracks the impurity-sector norm
$N_{\rm imp}(t)/N_{\rm imp}(0)$ and shows that the above-EP trajectory acquires a
strong late-time growth in the normalized diagnostic, whereas the below-EP and near-EP traces remain
well controlled over the plotted time window.
}
\label{fig:S_bloch}
\end{figure*}

\begin{figure*}[p]
\centering
\includegraphics[width=0.92\textwidth]{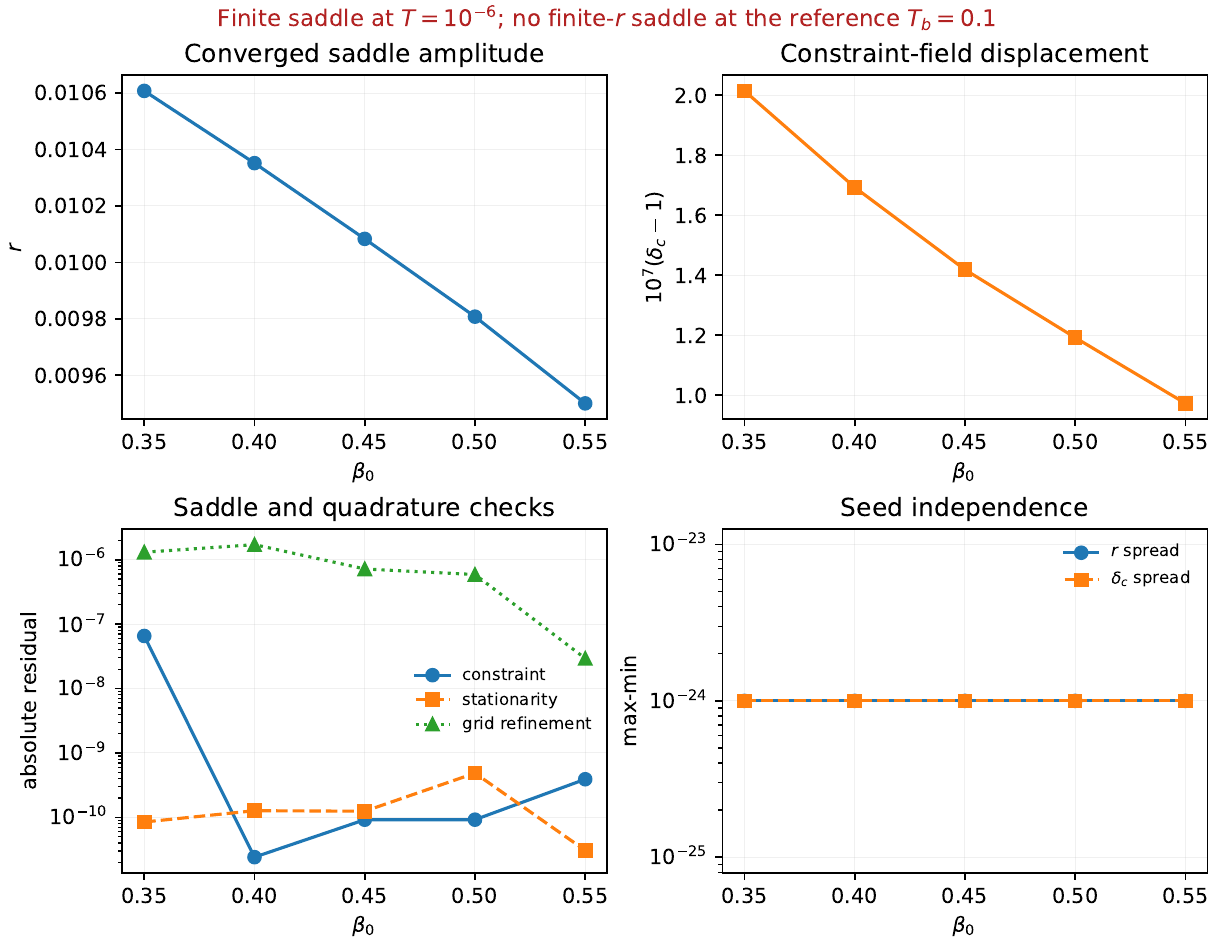}
\caption{%
\textbf{Low-temperature $U\to\infty$ saddle and numerical validation tests.}
At $T_{\rm b}=10^{-6}$, panel (a) shows the converged amplitude $r$ and
panel (b) the constraint-field displacement over
$0.35\leq\beta_0\leq0.55$.  Panel (c) reports the final constraint,
stationarity, and grid-refinement residuals; panel (d) shows that real,
negative-real, and complex initial boson seeds converge to the same
gauge-fixed solution.  The complete sandwich construction satisfies charge
continuity to $3.66\times10^{-23}$; the maximum constraint, stationarity,
and grid-refinement residuals are $6.54\times10^{-8}$,
$4.86\times10^{-10}$, and $1.71\times10^{-6}$, respectively.  The title
records the distinct result at
the manuscript reference temperature: the same closure has no finite-$r$
branch at $T_{\rm b}=0.1$.  The figure therefore validates the low-temperature
saddle and the exact core-EP scaling, but does not claim finite-temperature
condensate enhancement.
}
\label{fig:S_lowT_saddle}
\end{figure*}

\begin{figure*}[p]
\centering
\includegraphics[width=0.92\textwidth]{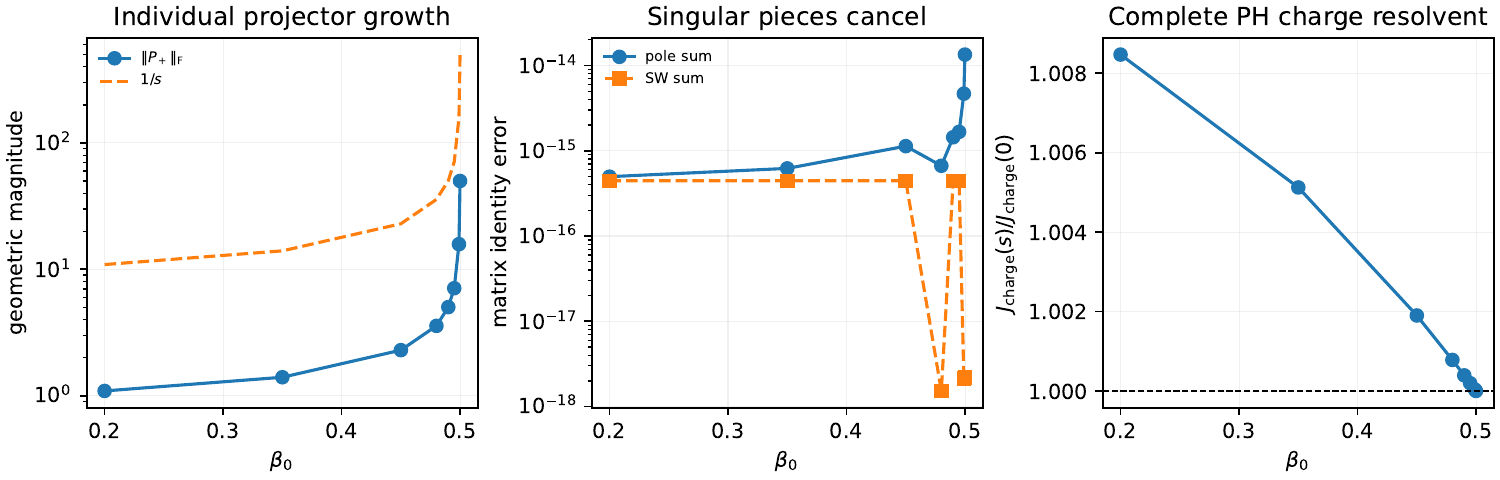}
\caption{%
\textbf{Biorthogonal-projector and finite-$U$ resolvent checks.}
Panel (a) shows the growth of an individual normalized projector and the
artificial $1/s$ factor on the pre-EP side.  Panel (b) verifies numerically
that the two-projector pole sum equals the direct matrix resolvent and that
the two virtual charge sectors equal the closed Schrieffer--Wolff resolvent;
the plotted discrepancies are numerical roundoff.  Panel (c) shows the
complete particle--hole charge factor normalized to its EP limit.  It remains
finite and approaches unity although each projector grows.  These checks
exclude an independent Petermann or $Z^{\rm bio}$ multiplier from the Green
function and screening estimate.
}
\label{fig:S_peakgap}
\end{figure*}

\begin{figure*}[p]
\centering
\includegraphics[width=0.92\textwidth]{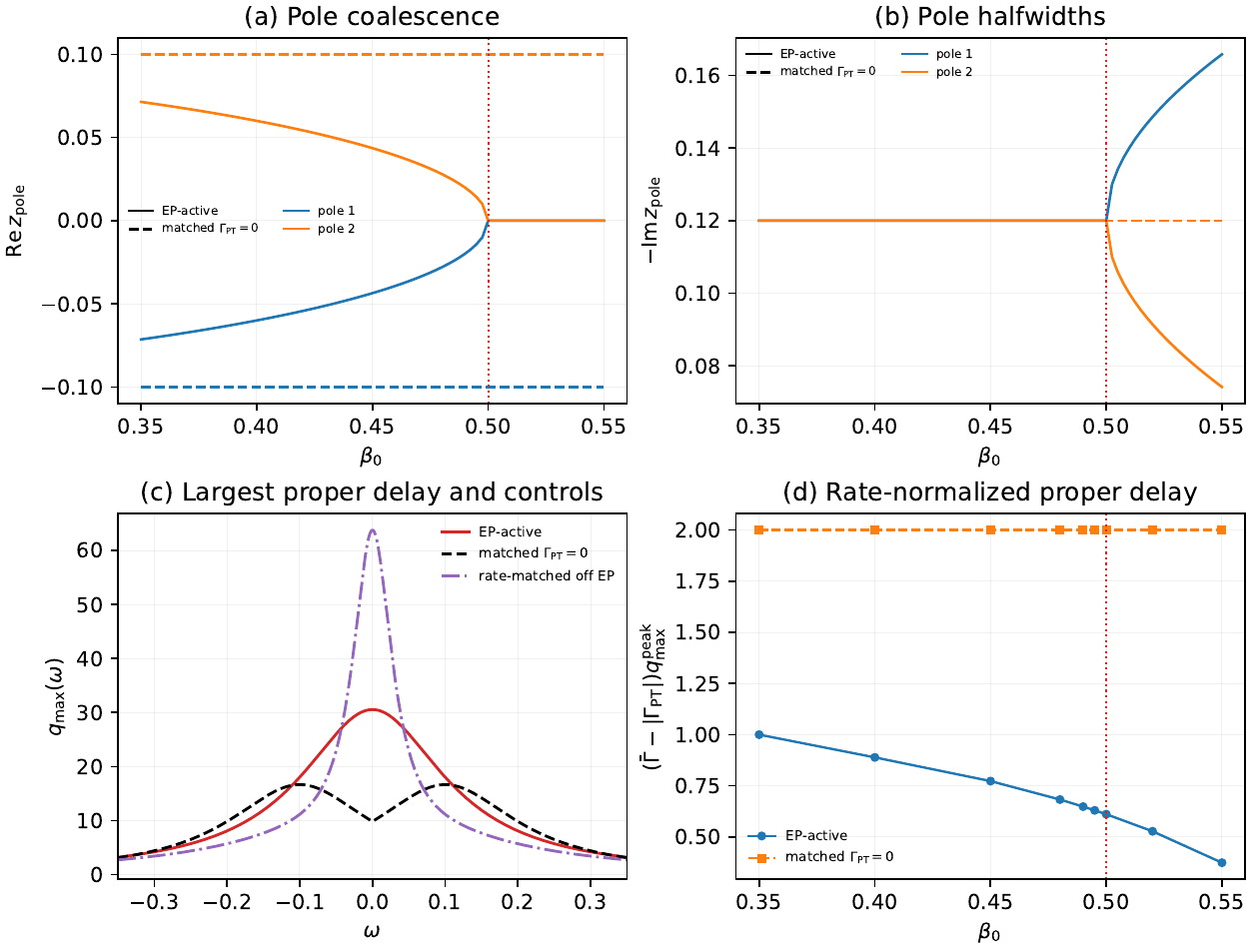}
\caption{%
\textbf{Causal Smith-delay validation and adverse controls.}
Panel (a) tracks the analytic pole pair to the exact core EP at
$\beta_0=0.50$.  Panel (b) shows the two passive halfwidths: they remain
positive throughout the retained window and split only on the relative-broken
side.  Panel (c) compares the largest proper-delay peak with a Hermitian
control matched by total rate and with an off-EP rate-matched kernel
($D=0.05$); the latter exceeds the EP peak.  Panel (d) normalizes the largest
proper delay by the slowest rate and likewise shows no independent EP
enhancement.  Together with the unitarity and trace-delay checks in
App.~\ref{app:smith}, these controls establish the Smith profile as a causal
coalescence/linewidth fingerprint, not a Kondo-scale multiplier.
}
\label{fig:S_smith}
\end{figure*}

\begin{figure*}[p]
\centering
\includegraphics[width=0.92\textwidth]{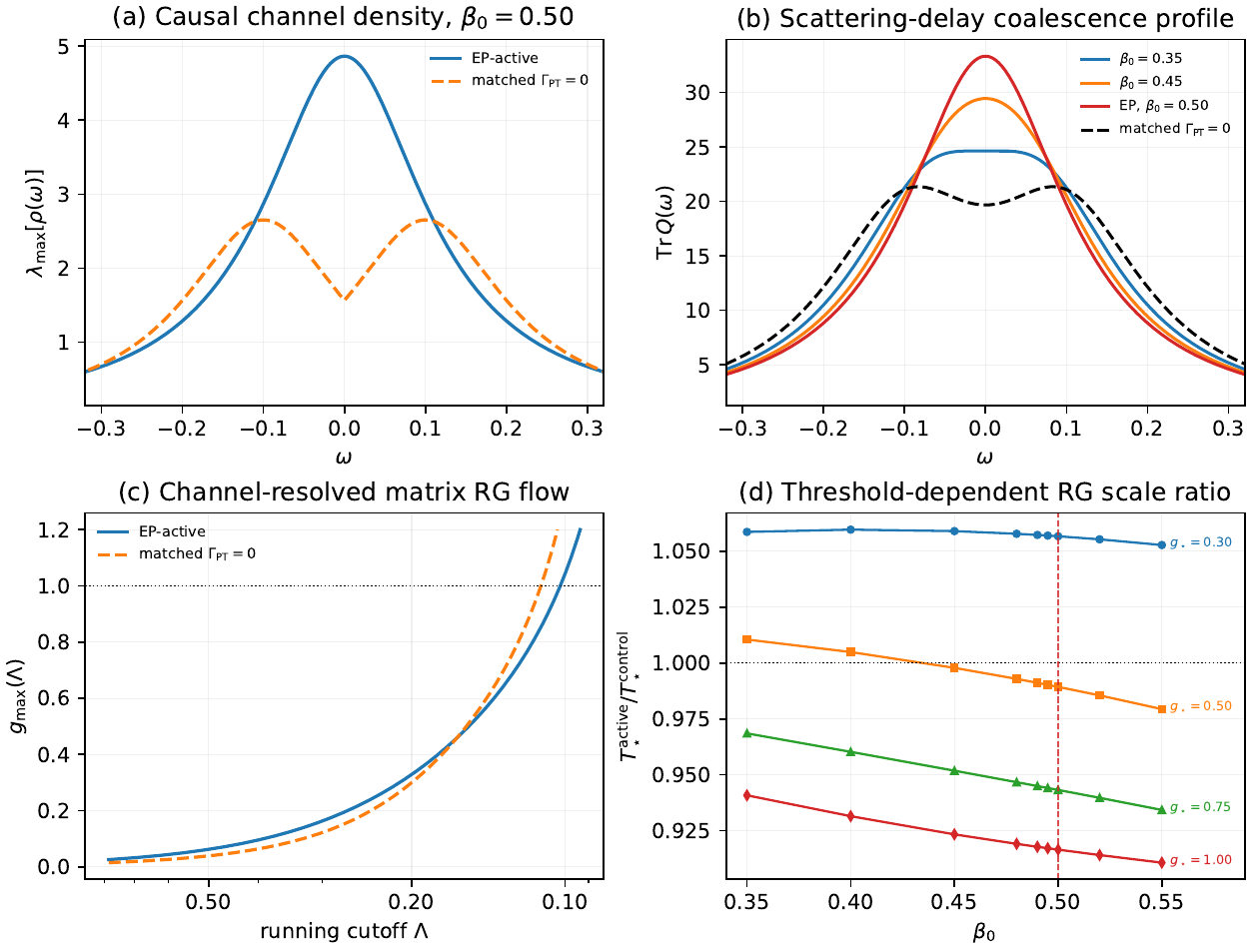}
\caption{%
\textbf{Causal scattering and channel-resolved RG support.}
The conservative fixed-$W$ validation uses the complete positive channel
density from $G^R$ and $G^A$, with $\Gamma_0=0.12$ and unit charge gap.
Panel (a) compares the largest causal channel-density eigenvalue at
$\beta_0=0.50$ with the matched $\Gamma_{\rm PT}=0$ control.
Panel (b) gives the Wigner--Smith trace; the passive Fisher--Lee matrix is
unitary to $1.33\times10^{-15}$ and
$\operatorname{Tr}Q(0)=4/\Gamma_0=33.3333$ at the EP.
Panels (c) and (d) show that the EP-active exchange eigenvalue is initially
larger, but its threshold-scale ratio crosses below the control before strong
coupling.  This supports a bounded early-flow ordering, not an
enhanced thermodynamic Kondo temperature.
}
\label{fig:S_causal_rg}
\end{figure*}

\begin{figure*}[p]
\centering
\includegraphics[width=0.92\textwidth]{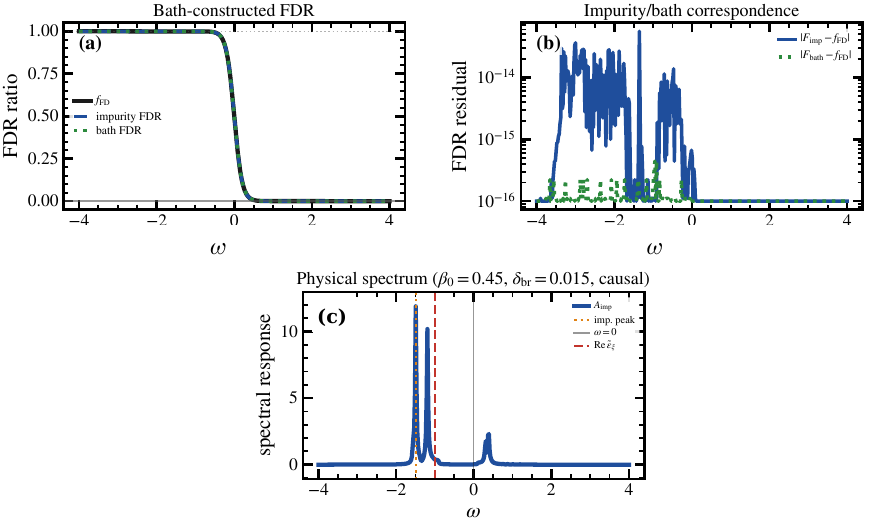}
\caption{%
\textbf{Bath-controlled fluctuation--dissipation and spectral checks.}
The uniform reference is $\epsilon_\xi=-U/2=-1$, $U=2$, and
$T_{\rm b}=0.1$.  Panels (a) and (b) verify the physical construction
$G^<_{\rm phys}=G^R\Sigma^<_{\rm bath}G^A
=f_{\rm FD}(G^A-G^R)$ with maximum residual
$5.61\times10^{-14}$ at $\delta_{\rm br,FDR}=10^{-3}$.
Panel (c) shows the physical impurity spectral response with
$\delta_{\rm br}=0.015$, evaluated from the complete retarded and advanced
kernel.  The two equivalent density-of-states evaluations in
Eq.~(\ref{eq:physical_dos_app}) agree numerically.
}
\label{fig:S_FDR}
\end{figure*}

\begin{figure*}[p]
\centering
\includegraphics[width=0.98\textwidth]{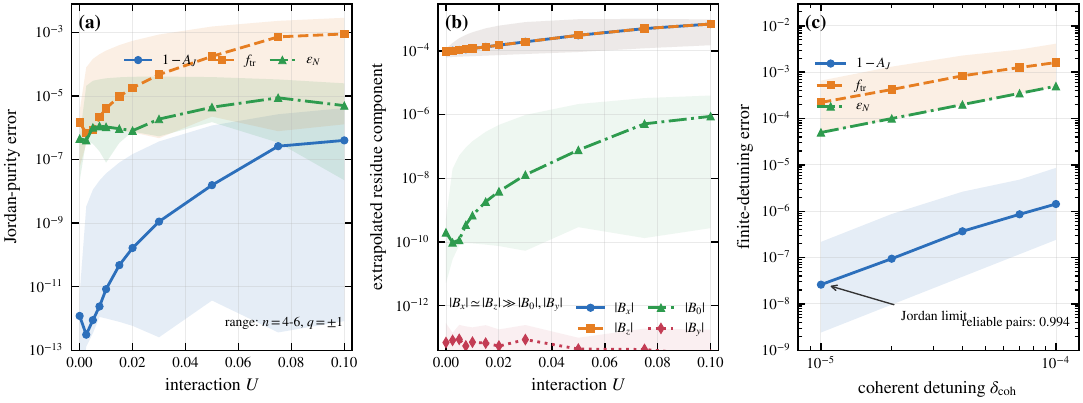}
\caption{%
\textbf{Finite-$U$ Jordan transition-residue audit for the
scalar-hybridization NRG control.}
(a) Zero-detuning extrapolations of the alignment error $1-A_J$, trace
fraction $f_{\rm tr}$, and nilpotent mismatch $\epsilon_N$.  Curves are
medians and bands are the full range over the early/high-energy iterations
$n=4$--$6$ and charge
sectors $q=\pm1$.  (b) Pauli components of the extrapolated double-pole
coefficient $B$: $|B_x|\simeq|B_z|\gg|B_0|,|B_y|$, as required by
$N_{\rm EP}=\Deff\sigma_z+i\Gpt\sigma_x$.  (c) Direct finite-detuning
convergence; curves are medians and bands are the 16th--84th percentile over
$U$, iteration, and tracked sector.  The reliable-pair fraction is $0.994$.
Parameters are $\Lambda=3$, $z=0.5$, $N_{\rm keep}=320$, $r=1$,
$\beta_0=0.50$, $\Deff=\Gpt=0.10$, and $0\leq U\leq0.10$.  The figure
establishes persistence of the nilpotent matrix-residue direction in this
weak-coupling structural window; it does not establish a low-energy fixed
point, pole coalescence, a universal interaction law, a thermodynamic Kondo
scale, or a phase boundary.
}
\label{fig:S_NRG_Jordan}
\end{figure*}

\begin{figure*}[t]
\centering
\includegraphics[width=0.96\textwidth]{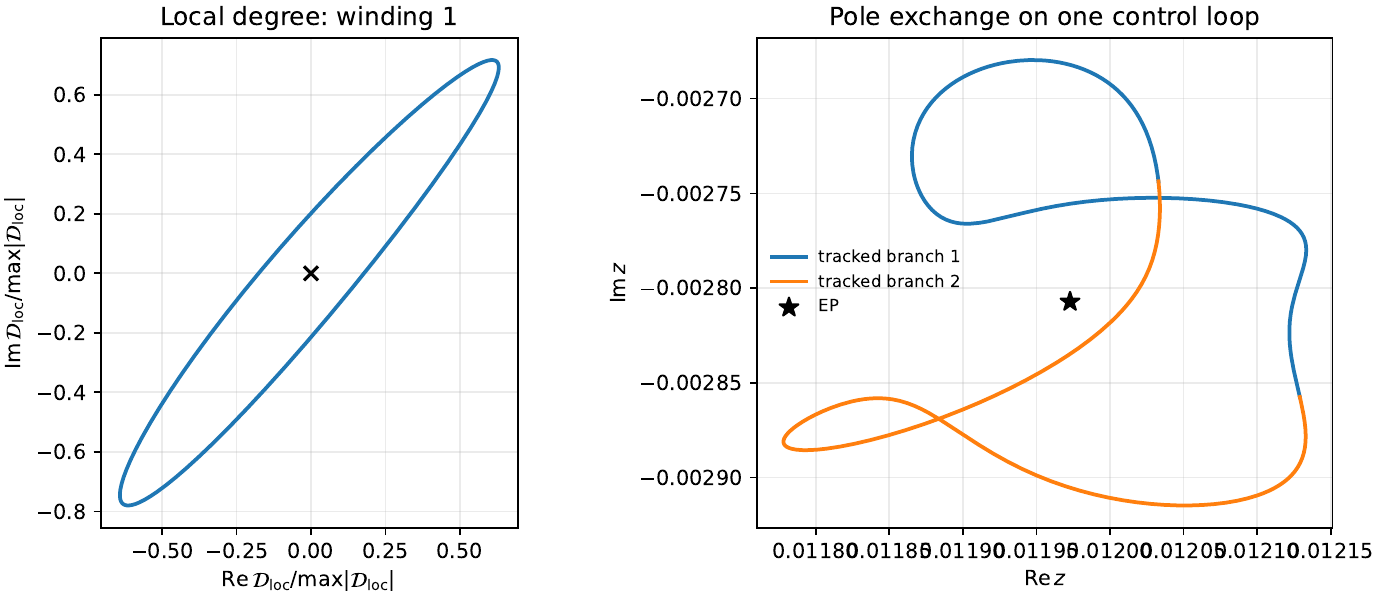}
\caption{%
\textbf{Full-momentum two-cut nonlinear causal-pole certificate.}
The stationary rotating-drive witness uses $\lambda=0.05$ and fixed
$W=0.10$,
$K=\pi/4$, $\epsilon_O=120$, $\Omega=60$, and the solution in
Eq.~(\ref{eq:app_full_k_witness_solution}).
Left: the normalized local discriminant ${\cal D}_{\rm loc}=F(z_c)$ winds
once around the origin on a nonsingular ellipse in $(\epsilon_d,V_O)$.
Right: the two resonance poles exchange after the same control loop; the star
marks $z_{\rm EP}$.  The winding is $+1$, the pole-separation half-winding is
$+1/2$, and all pole solves pass the separation and residual guards.  This
certifies a nonlinear pole EP of the stated full curved retarded kernel; it
is not a certificate for the nilpotent Markov-core EP or for full-momentum
Yang--Baxter integrability.
}
\label{fig:S_full_k_ep}
\end{figure*}

\begin{figure*}[t]
\centering
\includegraphics[width=0.94\textwidth]{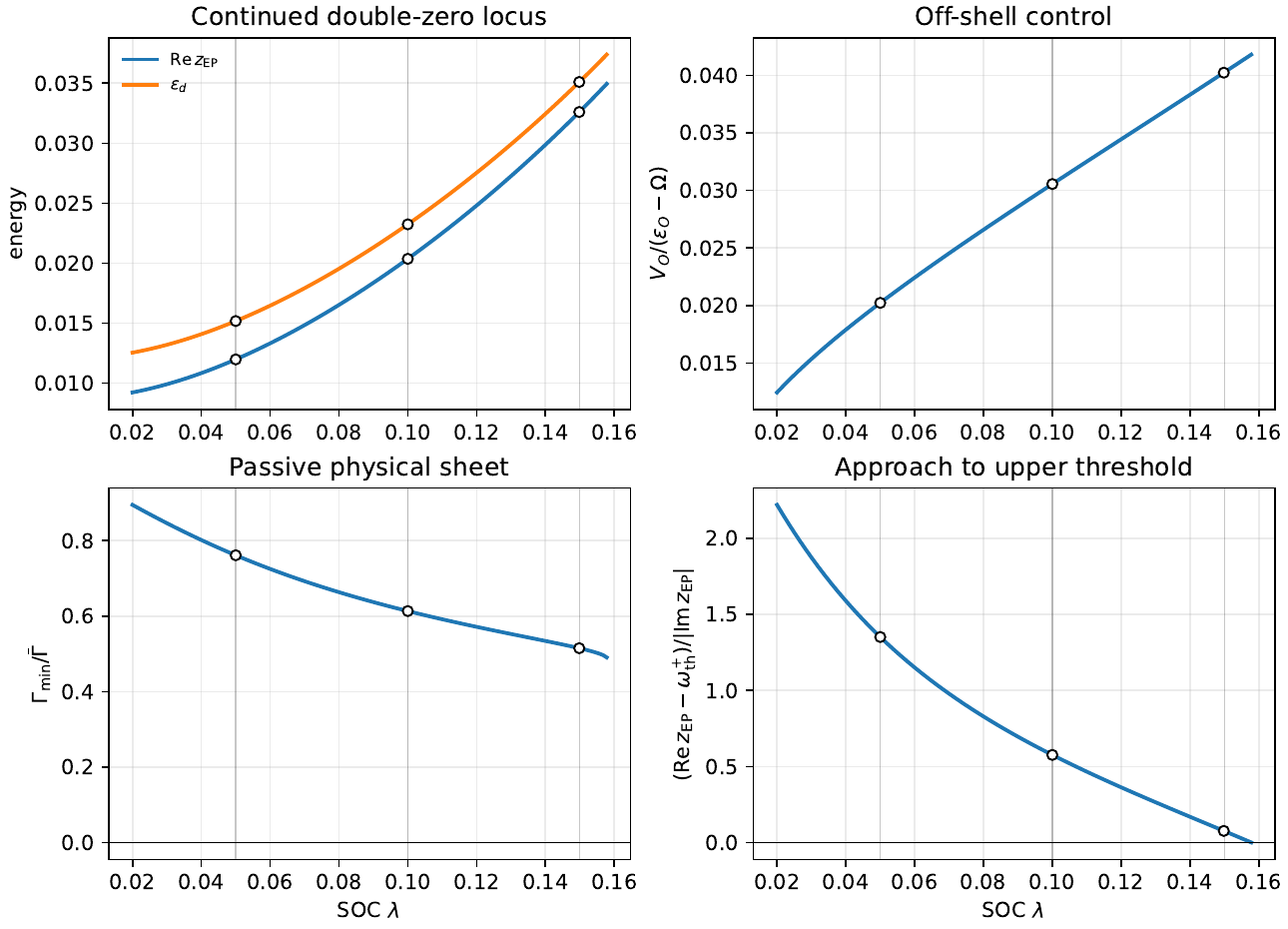}
\caption{%
\textbf{Finite SOC family of full-momentum nonlinear pole EPs.}
The double-zero equations are continued in $\lambda$ with $W=0.10$ fixed,
while the real impurity gate $\epsilon_d$ and off-shell auxiliary vertex
$V_O$ are retuned.
Vertical guides mark $\lambda=0.05$, $0.10$, and $0.15$, where independent
local-winding and pole-exchange certificates give winding $+1$, pole
permutation $(12)$, and half-winding $+1/2$.  The off-shell ratio remains
below $0.042$ and the physical rate matrix remains positive.  The last panel
shows the approach to the driven upper $k=0$ threshold.  This is a
one-dimensional locus in the enlarged parameter space; the approach to its
threshold endpoint occurs without loss of passivity, and neither that endpoint
nor any two-dimensional EP phase region is claimed.
}
\label{fig:S_full_k_soc_family}
\end{figure*}

\clearpage

\section*{Reproducibility}
The numerical scripts, parameter tables, and exported data used for the main
and Supplemental figures are included in the accompanying archive and in the
version-controlled GitHub repository
\href{https://github.com/VMKPHYSMATH/emergent-pt-dirac-impurity}{\texttt{VMKPHYSMATH/emergent-pt-dirac-impurity}}~\cite{kulkarni2026code}.
The exact public snapshot used for this work is Release \texttt{v2.1.0},
archived at
\href{https://doi.org/10.5281/zenodo.21888925}{doi:10.5281/zenodo.21888925};
the all-versions record is
\href{https://doi.org/10.5281/zenodo.21434682}{doi:10.5281/zenodo.21434682}.
Release \texttt{v2.1.0} includes the full-momentum certificate, SOC
continuation, finite-$U$ block-Wilson outputs, and finite-chain convergence
gates reported in this revision.
The accompanying source also includes
\nolinkurl{verify_finiteU_contact_algebra.py}, which reproduces the
finite-\(U\) \(R\)-matrix, metric covariance, and YBE residuals quoted above.
The script \nolinkurl{full_k_two_cut_ep_certificate.py} reproduces the exact
rotating-drive auxiliary closure, analytic full-momentum self-energy,
two-cut continuation, nonlinear double-zero solve, local winding, and pole
exchange of Fig.~\ref{fig:S_full_k_ep}.
The companion script \nolinkurl{full_k_ep_soc_family.py} continues the
double-zero locus in SOC, repeats the full topological certificate at three
representative points, and generates Fig.~\ref{fig:S_full_k_soc_family} and
its tabulated data.
The archived \nolinkurl{finiteU_fullk_*} JSON summaries record the exact
finite-chain interaction path, local winding, pole exchange,
equation-of-motion identity, \(N_z=4\to8\) center check, and kept-space
calibration of Sec.~\ref{app:finiteU_fullk}.
The archived \nolinkurl{fullk_n_scaling_gate_summary.json} and
\nolinkurl{FULLK_N_SCALING_AUDIT.md} record the independently optimized
$N=10$ and $12$ common-contour gate and Eq.~(\ref{eq:app_finiteU_rouche}).
The script \nolinkurl{make_finiteU_fullk_exact_figure.py} reads those summaries
and reproduces main-text Fig.~\MainFigFiniteUFullK.
The NRG archive additionally contains the independent transition-operator
propagation, no-truncation regression, branch-tracking and convergence
scripts, the gate summary, and the exported data used in
Fig.~\ref{fig:S_NRG_Jordan}.

%